\documentclass[fleqn,twoside,nofootinbib,unicode, hyperref=dvip]{revtex4}
\usepackage[dvips]{graphicx,psfrag}
\usepackage{amssymb}
\usepackage{amsmath}
\usepackage{latexsym}
\usepackage{graphicx}
\usepackage{epsfig}
\def\fnote#1#2{\begingroup\def\thefootnote{#1}\footnote{#2}\endgroup}
\input psfig.sty
\begin{document}
\title{Expanding Universe: slowdown or speedup?}
\author{Yu.\,L.\,Bolotin}
\email{ybolotin@gmail.com} \author{O.\,A.\,Lemets}\email{oleg.lemets@gmail.com}
\author{D.\,A.\,Yerokhin}\email{denyerokhin@gmail.com}
 \affiliation{A.I.Akhiezer Institute for
Theoretical Physics, National Science Center "Kharkov Institute of
Physics and Technology", Akademicheskaya Str. 1, 61108 Kharkov,
Ukraine}

\begin{abstract}
The purpose of the review - to provide affordable to a wider audience a description of  kinematics of the cosmological expansion, and dynamic interpretation of this process. The focus will be the accelerated expansion of the Universe. Will be considered by virtually all major opportunity to explain the accelerated expansion of the Universe, including an introduction to the energetic budget of the universe of so-called dark energy, a modification of Einstein's equations, and a new approach based on the holographic principle.
\end{abstract}
\maketitle

\section{Introduction} There are two fundamental reasons making the
question put in the title so principal. Firstly, in common sense,
the correct answer is necessary (though insufficient) to predict the
ultimate fate of the Universe. To be more specific, one must know
current values of the kinematic parameters (velocity, acceleration
etc.) in order to fix initial conditions necessary for solution of
the differential equations describing dynamics of the Universe.
Secondly, description of either decelerated or accelerated expansion
of the Universe requires absolutely distinct cosmological models.
Conventional substances (non-relativistic matter and radiation) from
one hand and general relativity from the other explain the
decelerated expansion. In order to obtain the accelerated expansion
one must either change the composition of the Universe rather
radically, or make even more responsible conclusion: the fundamental
physical theories lying in the basis of our understanding of the
World, are essentially wrong. Common experience tells us that it is
very unlikely to  answer the posed question in near future. Already
$80$ years passed after the discovery of the Universe expansion, and
we still cannot establish the source responsible for the initial
velocity distribution.  ''The Big Bang'' is nothing but euphemism
used to hide limitation of our knowledge. What does actually expand?
This question is still sharply controversial: ''How can vacuum
expand?'' \cite{RootofAllEvil}. Even more miraculous, how can vacuum
expand with acceleration or deceleration?  ``...how is it possible
for space, which is utterly empty, to expand? How can nothing
expand? The answer is: space does not expand. Cosmologists sometimes
talk about expanding space, but they should know better.''
\cite{Rees_Weinberg}.

No generation of homo sapiens can vanquish the temptation to believe
that it is chosen to acquire complete and final understanding of
Nature and Universe. In particular, the first half of the last
century, with discovery of Universe expansion by Hubble and due to
now fundamental physical theories, relativity and quantum mechanics
first of all, was marked by new cosmological paradigm --- the Big
Bang model. The main basis of the model is the celebrated Hubble
law, discovered in twenties of the last century and supported by all
types of up-to-date cosmological observations. The Big Bang model
was able to give satisfactory explanation for thermal evolution of
cooling Universe, predicted existence of relict radiation, correctly
described relative abundance of light elements and many other
features of the Universe.

Towards the end of the last century it was commonly hoped that the
Big Bang model supplemented by the Inflation Theory represented an
adequate (at least as a first approximation) model of Universe.
However this hope was never to come true. The cosmological paradigm
had to change for the reason of observations performed with ever
increasing precision. Right after the Hubble discovery the
cosmologists tempted to measure the deceleration of the expansion,
attributed to gravitation. They were so firmly confident to discover
exactly above stated effect, that the corresponding observable was
named the deceleration parameter. However, in the year $1998$ two
independent collaborations \cite{Riess,Perlmutter42}, studying
distant supernovae, presented convincing evidence for the fact that
the Universe expansion is accelerated. It turned out that the
brightness decreases in average much faster than it was commonly
believed according to the Big Bang model. Such additional dimming
means that the distance, attributed to given redshift, was somewhat
underestimated. And it in turn means that the universe expansion
accelerates: in the past the Universe expanded slower than nowadays.
The discovery of cosmological acceleration is probably one of the
most important observations not only in modern cosmology but in
physics in general. Accelerated expansion of Universe the most
directly demonstrates the fact that our fundamental theories are
either incomplete, or, even worse, misleading \cite{Trodden,Smolin}.

Physical origin of the cosmological acceleration still remains
greatest miracle. As we already mentioned above, if the Universe is
solely composed of matter and radiation, it must decelerate the
expansion. If the expansion in fact accelerates, we have to choices,
each of which forces us to revise our basic physical concepts.
\begin{enumerate}
  \item Up to $75\%$ of energy density in the Universe exists in form of unknown substance (commonly called the dark energy) with high negative pressure, providing the accelerated expansion.
  \item General Relativity theory must be revised on cosmological scales.
\end{enumerate}

We should remark that, besides the two above cited radical
possibilities to put the theory in agreement with the observations,
there still exist an obvious conservative way to get rid of the
problem: it implies more adequate utilization of available
theoretical possibilities. D.Wiltshire \cite{Wiltshire} expressed
this idea in the following way: I will take the viewpoint that
rather than adding further epicycles to the gravitational action
\footnote{generally speaking, ''epicycles'' here stand not only for
additional terms in the Hilbert-Einstein action, but also for such
new substances as dark energy and dark matter} the cosmological
observations which we currently interpret  in terms of dark energy
are inviting us to think more deeply about the foundations of
general relativity. The above mentioned term ''more adequate
utilization of the available possibilities'' need an explanation.
Let us give an example. In the context of cosmological models based
on homogeneous and isotropic Universe, in order to explain the
observed acceleration one have to involve a new form of matter with
negative pressure --- the dark energy. According to an alternative
interpretation \cite{Wiltshire,Buchert,Ishibashi,Backreaction}, the
accelerated expansion of the Universe follows from deviations from
the homogeneity. It is assumed that mass density in the Universe is
considerably inhomogeneous on the scales smaller than the Hubble
radius. In order to transit to effective description of homogeneous
and isotropic Universe, one should average and/or smooth out the
inhomogeneities up to some properly chosen scale of the averaging.
In such an averaged Universe one can define the ''effective
cosmological parameters''. After that it turns out that so obtained
equations of motion in general differ from the equations with the
same parameters in the models based on the cosmological principle of
uniform and isotropic Universe. If the difference looks like an
effective dark energy contribution, then one can try to explain the
accelerated expansion of Universe in frames of General Relativity
without any dark energy.

The present review considers all the three above listed
possibilities. It aims to present a comprehensive description for
both the kinematics of the cosmological expansion and the dynamical
description of the process. We concentrate our attention on the
acceleration of the cosmological expansion. It should be emphasized
that the equivalence principle directly links the acceleration with
the gravity nature and space-time geometry. Therefore its value is
extremely important for verification of different cosmological
models. It is the quantity which to great extent determines the
ultimate fate of the Universe.

\section{Cosmography --- the kinematics of expanding Universe}\label{SECII}
This section is devoted to a particular way of Universe description,
called the ''cosmography''\cite{Weinberg_GC}, entirely based on the
cosmological principle. The latter states that the Universe is
homogeneous and isotropic on the scales larger than hundreds
megaparsecs, and it allows, among the diversity of models for
description of the Universe, to select a specific class of uniform
and isotropic ones. The most general space-time metrics consistent
with the cosmological principle is the Friedmann-Robertson-Walker
(FRW) one:
\begin{equation}
    \label{FRW}
ds^2=dt^2-a^2(t)
\left\{\frac{dr^2}{1-kr^2}+r^2(d\theta^2+\sin^2\theta d\varphi^2)
\right\}.
\end{equation}

Here $a(t)$ is the scale factor and $r$ is coordinate of a point,
which does not take part in any motion except the global expansion
of the Universe. In some sense the main problem of cosmology is to
determine the dependence $a(t).$

The cosmological principle helps to construct the metrics of the
Universe and to make the first steps in interpretation of
cosmological observations. Recall that kinematics means the part of
mechanics that describes motion of bodies irrelative to forces
causing it. In that sense the cosmography is nothing but kinematics
of cosmological expansion. In order to construct the principal
cosmological quantity --- the time dependence of the scale factor
$a(t)$ --- one have to have the equations of motion (the Einstein
equations of general relativity) and an assumption on material
composition of the Universe, allowing to construct the
energy-momentum tensor. The efficiency of cosmography lies in the
fact that it allows to verify arbitrary cosmological models obeying
the cosmological principle. Modifications of general relativity or
introduction of new components (dark matter or dark energy) will
certainly change the dependence $a(t),$ but will not affect the
relations between the kinematic characteristics. Velocity of the
Universe expansion, determined by the Hubble parameter $H(t)\equiv
{\dot{a}(t)}/{a(t)},$ depends on time. The time dependence of the
latter is measured by the deceleration parameter $q(t).$ Let us
define it by Taylor expansion of the time dependence of the scale
factor $a(t)$ in vicinity of the current moment of time $t_0$:
\begin{equation}\label{a(t)_teylor}
    a(t)=a\left( {{t}_{0}} \right)+\dot{a}\left( {{t}_{0}} \right)\left[ t-{{t}_{0}} \right]+\frac{1}{2}\ddot{a}\left( {{t}_{0}} \right){{\left[ t-{{t}_{0}} \right]}^{2}}+\cdots
\end{equation}
Let us rewrite the relation in the following form
\begin{equation}\label{a(t)_teylor1}
   \frac{a(t)}{a\left( t_0\right)}=1+H_0\left[ t-t_0\right]-\frac{q_0}{2}H_{0}^{2}{{\left[ t-{{t}_{0}} \right]}^{2}}+\cdots,
\end{equation}
where the deceleration parameter reads
\begin{equation}\label{q(t)}
   q(t)\equiv
   -\frac{\ddot{a}(t)a(t)}{\dot{a}^2(t)}=-\frac{\ddot{a}(t)}{a(t)}\frac{1}{H^2(t)}.
\end{equation}
As will be shown below, accelerated growth of the scale factor
occurs at $q<0,$ while accelerated growth of the expansion velocity
$\dot{H}>0$ corresponds to $q<-1.$ When the sign of the deceleration
parameter was first defined, it seemed evident that gravitation
--- the only force governing the dynamics of Universe --- dumps its
expansion. A natural desire to deal with a positive parameter
predetermined the choice of the sign. Afterwards it turned out that
the choice made is inconsistent with the observed dynamics of
expansion and it is rather an example of historical curiosity.

For more detailed kinematical description of cosmological expansion
it is useful to consider the extended parameter set
\cite{Visser_Cosmography,Visser_Jerk}
\begin{eqnarray}\nonumber
  H(t) &\equiv& \frac{1}{a}\frac{da}{dt} ;\\\nonumber
  q(t) &\equiv&  -\frac{1}{a}\frac{{{d}^{2}}a}{d{{t}^{2}}}{{\left[ \frac{1}{a}\frac{da}{dt} \right]}^{-2}};\\\nonumber
  j(t)&\equiv& \frac{1}{a}\frac{{{d}^{3}}a}{d{{t}^{3}}}{{\left[ \frac{1}{a}\frac{da}{dt} \right]}^{-3}} ;\\\nonumber
  s(t) &\equiv& \frac{1}{a}\frac{{{d}^{4}}a} {d{{t}^{4}}}{{\left[ \frac{1}{a}\frac{da}{dt} \right]}^{-4}};\\\nonumber
  l(t) &\equiv& \frac{1}{a}\frac{{{d}^{5}}a} {d{{t}^{5}}}{{\left[ \frac{1}{a}\frac{da}{dt} \right]}^{-5}}.
\end{eqnarray}
Note that the latter four parameters are dimensionless. Derivatives
of lower parameters can be expressed through higher ones. So, for
example,
\begin{equation}\nonumber
   \frac{dq}{d\ln (1+z)}=j-q(2q+1).
\end{equation}
Let us Taylor expand the time dependence of the scale factor using
the above introduced parameters
\begin{equation}\label{a(t)_p}
   \begin{matrix}
   a(t)={{a}_{0}}\left[\right. 1+{{H}_{0}}\left( t-{{t}_{0}} \right)-\frac{1}{2}{{q}_{0}}H_{0}^{2}{{\left( t-{{t}_{0}} \right)}^{2}}+  \\
   \frac{1}{3!}{{j}_{0}}H_{0}^{3}{{\left( t-{{t}_{0}} \right)}^{3}}+\frac{1}{4!}{{s}_{0}}H_{0}^{4}{{\left( t-{{t}_{0}} \right)}^{4}}+\frac{1}{5!}{{l}_{0}}H_{0}^{5}{{\left( t-{{t}_{0}} \right)}^{5}}\ +{\mathrm O}\left( {{\left( t-{{t}_{0}} \right)}^{6}} \right)\left.\right].  \\
  \end{matrix}
\end{equation}
In terms of the same parameters the Taylor series for redshift reads
\begin{equation}\label{1+z_arr}
   1+z={{\left[ \begin{matrix}
  & 1+{{H}_{0}}(t-{{t}_{0}})-\frac{1}{2}{{q}_{0}}H_{0}^{2}{{(t-{{t}_{0}})}^{2}}+\frac{1}{3!}{{j}_{0}}H_{0}^{3}{{\left( t-{{t}_{0}} \right)}^{3}}+\frac{1}{4!}{{s}_{0}}H_{0}^{4}{{\left( t-{{t}_{0}} \right)}^{4}} \\
 & +\frac{1}{5!}{{l}_{0}}H_{0}^{5}{{\left( t-{{t}_{0}} \right)}^{5}}\ +{\mathrm O}\left( {{\left( t-{{t}_{0}} \right)}^{6}} \right)\\
\end{matrix} \right]}^{-1}};
\end{equation}
\begin{equation}
   z={{H}_{0}}({{t}_{0}}-t)+\left( 1+\frac{{{q}_{0}}}{2} \right)H_{0}^{2}{{(t-{{t}_{0}})}^{2}}+\cdots.
\end{equation}
Let us cite few useful relations for the deceleration parameter
\begin{eqnarray}\nonumber
  q(t)&=& \frac{d}{dt}\left( \frac{1}{H} \right)-1;\\\nonumber
  q(z) &=& \frac{1+z}{H}\frac{dH}{dz}-1;\label{q(z)} \\\nonumber
  q(z) &=& \frac{d\ln H}{dz}(1+z)-1; \\\nonumber
  q(a) &=& -\left( 1+\frac{\frac{dH}{dt}}{{{H}^{2}}} \right)=-\left( 1+\frac{a\frac{dH}{da}}{H} \right);\\\nonumber
  q &=& -\frac{d\ln (aH)}{d\ln a}.
\end{eqnarray}
For the case of single component liquid with density $\rho $ one has
\begin{equation}\label{q_rho}
    q(a)=-1-\frac{a\frac{d\rho }{da}}{2\rho }.
\end{equation}
The derivatives $\frac{dH}{dz},$ $\frac{{{d}^{2}}H}{d{{z}^{2}}},$
$\frac{d^{3}H}{dz^{3}}$ and $\frac{d^{4}H}{dz^{4}}$ can be expressed
through the parameters $q$ and $j$ solely
\begin{eqnarray}
\frac{dH}{dz}&=&\frac{1+q}{1+z}H;\nonumber \\
\frac{{{d}^{2}}H}{d{{z}^{2}}}&=&\frac{j-q^2}{\left(1+z \right)^2}H;\nonumber \\
\frac{d^{3}H}{dz^{3}} &=& \frac{H}{(1 + z)^{3}} \left(3 q^{2} + 3q^{3} -4 q j - 3 j -s \right); \nonumber \\
\frac{d^{4}H}{dz^{4}} &=& \frac{H}{(1 + z)^{4}} \left(- 12 q^{2} -
24 q^{3} - 15 q^{4} + 32 q j + 25 q^{2} j  + 7 q s + 12 j - 4 j^{2}
+ 8 s + l\right).\nonumber
\end{eqnarray}
Let us present for convenience the useful relations for
transformations from higher time derivatives to derivatives with
respect to red shift:
\begin{eqnarray}\label{eq:secondHt}
\frac{d^{2}}{dt^{2}} = (1+z) H \left[H+ (1+z) \frac{d H}{dz}\right]
\frac{d}{dz} + (1 + z)^{2} H^{2} \frac{d^{2}}{dz^{2}};
\end{eqnarray}
\begin{eqnarray}\label{eq:thirdHt}
\frac{d^{3}}{dt^{3}} &= &- (1 + z) H \left\{ H^{2} + (1 + z)^{2}
\left(\frac{dH}{dz}\right)^{2} + (1 + z) H \left[ 4 \frac{dH}{dz} +
(1 + z) \frac{d^{2}H}{dz^{2}}
\right] \right\} \frac{d}{dz} - 3 (1 + z)^{2} H^{2}  \nonumber \\
&\times& \left[ H + (1+z) \frac{dH}{dz} \right] \frac{d^{2}}{dz^{2}}
- (1 + z)^{3} H^{3} \frac{d^{3}}{dz^{3}};
\end{eqnarray}
\begin{eqnarray}\label{eq:fourthHt}
\frac{d^{4}}{dt^{4}} &=& (1 + z) H \left[ H^{2} + 11 (1+z) H^{2}
\frac{dH}{dz} + 11 (1+z) H \frac{dH}{dz} + (1+z)^{3}
\left(\frac{dH}{dz}\right)^{3} + 7 (1+z)^{2} H \frac{d^{2}H}{dz^{2}}  \right. \nonumber\\
&+& \left. 4 (1+z)^{3} H \frac{dH}{dz} \frac{d^{2}H}{d^{2}z} +
(1+z)^{3} H^{2} \frac{d^{3}H}{d^{3}z}\right] \frac{d}{dz}+ (1 +
z)^{2} H^{2} \left[ 7 H^{2} + 22 H \frac{dH}{dz} + 7 (1 +
z)^{2} \left(\frac{dH}{dz}\right)^{2} \nonumber \right. \\
&+& \left. 4 H \frac{d^{2}H}{dz^{2}} \right] \frac{d^{2}}{dz^{2}} +
6 (1 + z)^{3} H^{3} \left[ H + (1+z) \frac{dH}{dz} \right]
\frac{d^{3}}{dz^{3}} + (1 + z)^{4} H^{4} \frac{d^{4}}{dz^{4}} +  (1
+ z)^{4} H^{4} \frac{d^{4}}{dz^{4}};
\end{eqnarray}
Derivatives of the Hubble parameter squared with respect to the red
shift $\frac{{{d^{(i)}}{H^2}}}{{d{z^{(i)}}}},\;i = 1,2,3,4$
expressed through the cosmographic parameters take the form
\begin{equation}\nonumber
\frac{d(H^{2})}{dz} = \frac{2 H^{2}}{1 + z} (1 + q)
\end{equation}
\begin{equation}\nonumber
\frac{d^{2}(H^{2})}{dz^{2}} = \frac{2H^{2}}{(1 + z)^{2}} (1 + 2 q +
j);
\end{equation}
\begin{equation}\nonumber
\frac{d^{3}(H^{2})}{dz^{3}} = \frac{2H^{2}}{(1 + z)^{3}} (- q j - s);
\end{equation}
\begin{equation}\nonumber
\frac{d^{4}(H^{2})}{dz^{4}} = \frac{2H^{2}}{(1 + z)^{4}} ( 4 q j + 3
q s + 3 q^{2} j - j^{2} + 4 s + l).
\end{equation}
Time derivatives of the Hubble parameter can be expressed through
above defined cosmographic parameters $H,q,j,s,l$ as
\begin{eqnarray}\label{dotH_z}
  \dot{H}&=& -{{H}^{2}}(1+q);\\\nonumber
  \ddot{H} &=& {{H}^{3}}\left( j+3q+2 \right);\\\nonumber
 \dddot{H} &=& {{H}^{4}}\left[ s-4j-3q(q+4)-6 \right];\\\nonumber
  \dddot{H} &=& {{H}^{5}}\left[ l-5s+10\left( q+2 \right)j+30(q+2)q+24 \right].
 \end{eqnarray}
From the relations \eqref{dotH_z} one can easily see that the
accelerated growth of the acceleration velocity $\dot{H}>0$
corresponds to the case $q<-1.$

The Hubble parameter, as can be seen from the relation \eqref{q(z)},
is linked to the deceleration parameter with the integral relation
\begin{equation}\nonumber
    H={{H}_{0}}\exp \left[ \int_{0}^{z}{\left[ q({z}')+1 \right]d\ln (1+{z}')} \right].
\end{equation}
It immediately follows that in order to calculate the principal
characteristic of the expanding Universe $H(z)$ one needs the
information on dynamics of cosmological expansion, coded in the
quantity $q(z).$

\section{Dynamics of cosmological expansion briefly\label{SECIII}}
Dynamics of Universe in frames of General Relativity is described by
the Einstein equations
\begin{equation}\nonumber
    {{R}_{\mu \nu }}-\frac{1}{2}R{{g}_{\mu \nu }}=8\pi G{{T}_{\mu \nu
    }}.
\end{equation}
The energy-momentum tensor ${{T}_{\mu \nu }}$ describes the space
distribution of mass(energy), while the curvature tensor components
${{R}_{\mu \nu }}$ and its trace $R$ are expressed through the
metric tensor ${{g}_{\mu \nu }}$ and its derivatives of first and
second order. The Einstein equations generally are non-linear and
hard to analyze. The problem is simplified when considering mass
distribution with special symmetry properties provided by the
metric. For homogeneous and isotropic Universe described by the
FRW-metric, the Einstein equations reduce to the system of Friedmann
equations:
\begin{equation}
\label{Frid_1}
  \left( \frac{\dot a}{a} \right)^2 = \frac{8\pi G}{3}\rho  - \frac k{a^2},
\end{equation}
\begin{equation}
\label{Frid_2}
    \frac{\ddot a}a =  - \frac{4\pi G}3\left( \rho  + 3p \right).
\end{equation}

Here $\rho $ and $p$ are total energy density and pressure
respectively for all the components present in the Universe at the
considered time moment.

The system (\ref{Frid_1},\ref{Frid_2}) is sufficient for complete
description of the Universe dynamics. The conservation equation
\begin{equation}
\label{conser_eq}
    \dot\rho +3H(\rho+p)=0,
\end{equation}
follows from the Lorentz-invariance of the energy-momentum tensor
$T_{\nu,\mu}^\mu =0$ and, as can be easily seen, it is nothing but
the first principle of thermodynamics for ideal liquid with constant
entropy $dE+pdV=0.$ We remark, that this equation can be derived
from the Friedman equations (\ref{Frid_1},\ref{Frid_2}).

Let us now consider relative density for $i$th component of the
Universe:
    \begin{equation}\label{Omega}
    {{\Omega }_{i}}\equiv \frac{{{\rho }_{i}}}{{{\rho }_{c}}},\,\quad {{\rho }_{c}}\equiv \frac{3{{H}^{2}}}{8\pi G},\quad \Omega \equiv \sum\limits_{i}{{{\Omega }_{i}}.}
\end{equation}
Setting the relative curvature density by definition ${{\Omega
}_{k}}\equiv -\frac{k}{{{a}^{2}}{{H}^{2}}},$ the first Friedmann
equation can be presented in the form
    \begin{equation}\nonumber
    \sum\limits_{i}{{{\Omega }_{i}}=1}.
\end{equation}
In order to solve the Friedmann equations one has to define the
matter composition of the Universe and construct the state equation
for each component. In the simplest linear form the state equation
reads
\begin{equation}
\label{state_eq}
    p_{i} = w_{i}\rho_{i}.
\end{equation}
Solving the Friedmann equations for the case $w=const,$ $k=0$ one
obtains
    \begin{equation}\nonumber
    a(t)\propto {{\left( t/{{t}_{0}} \right)}^{\frac{2}{3(1+w)}}},\ \rho \propto {{a}^{-3(1+w)}}.
\end{equation}
(The scale factor is normalized by the condition $a\left( {{t}_{0}}
\right)=1.$) The above cited solutions exist only provided $w\ne
-1,$ the otherwise case will be considered separately. For the
Universe dominated by radiation (relativistic gas of photons and
neutrinos) one has $w=1/3,$ while for the matter dominated case
$w=0.$ As a result of such state equations, for the matter dominated
case one gets
\begin{equation}\nonumber
    a(t)\propto {{\left( t/{{t}_{0}} \right)}^{2/3}},\ \rho \propto {{a}^{-3}}.
\end{equation}
which is easily interpreted as the particle number conservation law.
For the radiation-dominated case the solution reads
\begin{equation}\nonumber
    \quad a(t)\propto {{\left( t/{{t}_{0}} \right)}^{1/2}},\quad \rho \propto {{a}^{-4}}.
\end{equation}
The latter result is a consequence of the fact that the energy
density of radiation decreases as ${{a}^{-3}}$ due to growth of
occupied volume (as Universe expands) and additionally as
${{a}^{-1}}$ due to the redshift. Note that from the equation
(\ref{conser_eq}) it follows that $\rho =const$ for $w=-1.$ In the
latter case the hubble parameter is constant leading to exponential
growth of the scale factor
\begin{equation}\nonumber
    a(t)\propto {{e}^{Ht}}.
\end{equation}
Therefore in the case of traditional cosmological components in form
of matter and radiation with $w=0$ and $w=1/3$ respectively, the
Universe expansion can only decelerate, i.e. $\ddot{a}<0.$

Using the definition of deceleration parameter, one can see that for
flat Universe filled by single component with state equation
$p=w\rho $
\begin{equation}\nonumber
    q=\frac{1}{2}(1+3w).
\end{equation}
In generic case $(k=(0,\pm 1),\,\quad \rho =\sum\limits_{i}{{{\rho
}_{i}}},\quad p=\sum\limits_{i}{{{\rho }_{i}}}{{w}_{i}})$ one gets
\begin{equation}\label{q_w}
    q=\frac{\Omega }{2}+\frac{3}{2}\sum\limits_{i}{{{w}_{i}}}{{\Omega }_{i}}.
\end{equation}
Using \eqref{Omega}, the latter relation can be rewritten in the
form
\begin{equation}\nonumber
    q=\frac{1}{2}\left( 1+\frac{k}{{{a}^{2}}{{H}^{2}}} \right)\left( 1+3\frac{p}{\rho }
    \right).
\end{equation}

Since the deceleration parameter $q$ is a slowly varying quantity
(e.g. $q=1/2$ for matter-dominated case and $q=-1$ in the Universe
dominated by dark energy in form of cosmological constant), then the
useful information is contained in its time average value, which is
very interesting to obtain without integration of the equations of
motions for the scale factor. Let us see how it is possible
\cite{LimaAge}. Let us define the time average $\bar{q}$ on the
interval $\left[ 0,t_0 \right]$ by the expression
\begin{equation}\nonumber
\bar{q}\left(t_0\right)=\frac{1}{t_0}\int_{0}^{t_0}{q(t)dt}.
\end{equation}
Substituting the definition of the deceleration parameter
    \begin{equation}\nonumber
    q(t)=-\frac{\ddot{a}a}{{{{\dot{a}}}^{2}}}=\frac{d}{dt}\left( \frac{1}{H} \right)-1,
\end{equation}
it is easy to see that
\begin{equation}\label{barq}
\bar{q}(t_0)=-1+\frac{1}{{{t}_{0}}{{H}_{0}}}
\end{equation}
or, equivalently,
\begin{equation}\label{t_0barq}
t_0=\frac{H_{0}^{-1}}{1+\bar{q}}.
\end{equation}
As was expected, the present age of universe scales as $H_{0}^{-1},$
but the proportionality coefficient depends only on the average
value of the deceleration parameter. It is worth to mention that
this purely kinematic result does not dependent on curvature of
space in the Universe, neither on number of component filling it,
nor on the particular type of the gravity theory. Let us somewhat
reformulate the obtained results for the average deceleration
parameter. For single-component flat Universe the Friedman equations
can be presented in the following form
\begin{eqnarray}
 8\pi G\rho &=& 3\frac{\dot{a}^2}{a^2};  \\
  8\pi Gp&=& -2\frac{\ddot{a}}{a}-\frac{\dot{a}^2}{a^2}.
\end{eqnarray}
For the single component Universe with the state equation $p=w\rho
,\quad w=const$ the equation for the scale factor
    \[a\ddot{a}+\left( \frac{1+3w}{2} \right){{\dot{a}}^{2}}=0,\]
has an exact solution
\begin{equation}\label{a(t)w}
a(t)={{a}_{0}}{{\left[ \frac{3}{2}\left( 1+w \right){{H}_{0}}t
\right]}^{\frac{2}{3(1+w)}}}.
\end{equation}
Now it immediately follows that
\begin{equation}\label{q(w)}
q=\frac{1+3w}{2}=const,\quad t_0=\frac{2H_{0}^{-1}}{3(1+w)}.
\end{equation}
In particular, one has for the cases of radiation $\left( w=1/3
\right),~q=1$ and non-relativistic matted $\left( w=0
\right),~q=1/2.$ From the other hand, accounting that in the
considered case $q=\bar{q},$ the latter relation can be rewritten as
$t_0=H_{0}^{-1}/\left( 1+\bar{q} \right),$ which coincides with the
relation \eqref{t_0barq}. The resulting expression \eqref{q(w)} can
be presented in the following form
\begin{equation}\label{TU}
T=\frac{H^{-1}}{1+\bar{q}},
\end{equation}
where $T,H,\bar{q}$ are age of Universe, Hubble parameter and
average deceleration parameter respectively. Since $\bar{q}$ is of
order of unity, then it immediately follows from \eqref{a(t)w} that
on any stage of Universe evolution the Hubble time $H_{0}^{-1}$
represents the characteristic time scale.

Further model-independent dynamical restrictions on the Universe
kinematics can be derived from the so-called energy conditions
\cite{HawkingEllis,CarrollSpacetime,VisserBarcelro,SantosAlcaniz,Ming-JianZhang}.
Those conditions based on rather general physical principles, impose
restrictions on the energy-momentum tensor components $T_{\mu \nu
}.$ Specifying a particular model for the medium\footnote{specific
model does not specify the state equation!} those conditions can be
transformed into inequalities, limiting possible values of pressure
and density in the medium. In the Friedmann model the medium is an
ideal fluid, where
\begin{equation}\label{Tmunu}
T_{\mu \nu }=\left( \rho +p \right){{u}_{\mu }}{{u}_{\nu
}}-p{{g}_{\mu \nu }},
\end{equation}
where $u_\mu$ is the fluid four-velocity, with total density $\rho $
and pressure $p$ given, respectively, by
\begin{equation}\label{rho(a)_p(a)}
\rho =\frac{3}{8\pi G}\left(
\frac{{{{\dot{a}}}^{2}}}{{{a}^{2}}}+\frac{k}{{{a}^{2}}} \right),~~
p=-\frac{1}{8\pi G}\left(
2\frac{{\ddot{a}}}{a}+\frac{{{{\dot{a}}}^{2}}}{{{a}^{2}}}+\frac{k}{{{a}^{2}}}
\right).
\end{equation}
The most common energy conditions (see, e.g.,
\cite{HawkingEllis,CarrollSpacetime,VisserBarcelro,SantosAlcaniz})
reduce to
\begin{eqnarray}\nonumber
  NEC &\Rightarrow& \rho +p\ge 0;\\\nonumber
  WEC &\Rightarrow& \rho \ge 0\quad and\quad \rho +p\ge 0; \\\nonumber
  SEC &\Rightarrow& \rho +3p\ge 0,and\quad \rho +p\ge 0; \\\nonumber
  DEC&\Rightarrow& \rho \ge 0\quad and\quad -\rho \le p\le \rho,
\end{eqnarray}
where NEC, WEC, SEC and DEC correspond, respectively, to the null,
weak, strong and dominant energy conditions. Because these
conditions do not require a specific equation of state of the matter
in the Universe, they provide very simple and model-independent
bounds on the behavior of the energy density and pressure.
Therefore, the energy conditions are among of many approaches to
understand the evolution of Universe. From Eqs. \eqref{rho(a)_p(a)},
one easily obtains that these energy conditions can be translated
into the following set of dynamical constraints relating the scale
factor
\begin{eqnarray}
  NEC &\Rightarrow& -\frac{{\ddot{a}}}{a}+\frac{{{{\dot{a}}}^{2}}}{{{a}^{2}}}+\frac{k}{{{a}^{2}}}\ge 0;\nonumber\\
  WEC &\Rightarrow& \frac{{{{\dot{a}}}^{2}}}{{{a}^{2}}}+\frac{k}{{{a}^{2}}}\ge 0; \label{EC} \\
  SEC &\Rightarrow& \frac{{\ddot{a}}}{a}\le 0;\nonumber \\
  DEC&\Rightarrow& \frac{{\ddot{a}}}{a}+2\left[ \frac{{{{\dot{a}}}^{2}}}{{{a}^{2}}}+\frac{k}{{{a}^{2}}} \right]\ge 0.\nonumber
\end{eqnarray}
For flat Universe the conditions \eqref{EC} can be transformed to
restrictions on the deceleration parameter $q$
\begin{eqnarray}
  NEC &\Rightarrow& q\ge -1;\nonumber\\
  SEC &\Rightarrow& q\ge 0; \label{ECq} \\
  DEC&\Rightarrow& q\le 2,\nonumber
\end{eqnarray}
while WEC is always satisfied for arbitrary real $a(t).$

The conditions \eqref{ECq}, considered separately, leave a principal
possibility for both decelerated $\left( q>0 \right)$ and
accelerated $\left( q<0 \right)$ expansion of Universe. The sense of
the restrictions for NEC in  \eqref{ECq} is quite clear. As it
follows from the second Friedmann equation, the condition of
accelerated expansion of Universe reduce to the inequality $\rho
+3p\le 0,$ i.e. accelerated expansion of Universe is possible only
in presence of components with high negative pressure  $p<-1/3\rho
.$ The energetic condition  $SEC$ excludes existence of such
components. As a result $q\ge 0$ in that case. At the same time the
conditions $NEC$ and $DEC$ are compatible with the condition
$p<-1/3\rho $ and therefore regimes with $q<0$ are allowed here. In
conclusion of the section let us pay attention to an interesting
feature of the expanding Universe dynamics. According to Hubble law,
the galaxies situated on the Hubble sphere, recede from us with
light speed. Velocity of the Hubble sphere itself equals to time
derivative of the Hubble radius ${{R}_{H}}=c/H,$
\begin{equation}\label{dRh}
\frac{d}{dt}\left( {{R}_{H}} \right)=c\frac{d}{dt}\left( \frac{1}{H}
\right)=-\frac{c}{{{H}^{2}}}\left(
\frac{{\ddot{a}}}{a}-\frac{{{{\dot{a}}}^{2}}}{{{a}^{2}}}
\right)=c(1+q).
\end{equation}
In the Universe with decelerated expansion $(q>0)$ the Hubble sphere
has superluminal velocity $c(q+1)$ and it outruns those galaxies.
Therefore the galaxies initially situated outside the Hubble sphere,
gradually enter inside it. Any observer in arbitrary point of the
Universe will see that the number of galaxies constantly increases.
In the case of accelerated expanding Universe   $(q<0)$ the Hubble
sphere has subluminal $c(1-|q|)$ and it lags behind the galaxies.
Therefore now the galaxies initially contained in the Hubble sphere
will gradually leave it and become unaccessible for observation.
Should we consider those unobservable galaxies as physically real?
The distinction between physics and metaphysics is the possibility
of experimental verification for the physical theories. Physics have
no deal with unobservable objects. However, the science constantly
expands its boundaries, thus including to consideration more and
more abstract concepts, once been metaphysical before: atoms,
electromagnetic waves, black holes... The list can be continued
further.

We are most likely inhabitants of accelerated expanding Universe.
Likewise in the decelerated expanding one, there are galaxies so
distant from us that no signal from them can be detected by
terrestrial observer. However, if the cosmic expansion accelerates,
then we recede from those galaxies with superluminal velocity.
Therefore, if their light have not reach us till now then it will
never come. Such galaxies are unaccessible for observation not only
temporarily, they are forever unobservable. Such ''never observable
galaxies'' descent from the same ''Big Bang'' like our Milky Way.
Should we attribute them to physical objects or to metaphysics?
Those who consider the science fiction as a realization of most
unbounded fantasy, are absolutely wrong. Compared to modern
cosmology, the science fiction is dull and lacks imagination.

Accelerated expansion of the Universe was first introduced in
cosmological models with creation of the inflation theory. It was
developed to fix multiple defeats of the Big Bang model. It turns
out that in order to resolve most of them it is sufficient to have
exponentially fast accelerated expansion of Universe at the very
beginning of its evolution, during just about $10^{-35}s.$ The
simplest way to obtain expansion of such type is to consider
dynamics of Universe with scalar field. The inflation theory is
formulated in many ways, starting from the models based on quantum
gravity and high-temperature phase transitions
theory with supercooling and exponential expansion in the false
vacuum state. To illustrate the main ideas of the theory let us consider
flat, homogeneous and isotropic Universe filled by the scalar field
$\varphi$ with the potential $V(\varphi),$ independent of the
coordinates. The first Friedmann equation (\ref{Frid_1}) in the
considered case takes the following form
\begin{equation}\nonumber
    H^2 = \frac{8\pi}{3M_{Pl}^2}\left(\frac{1}{2}\dot\varphi^2 + V(\varphi )\right).
\end{equation}
The conservation equation (\ref{conser_eq}), written in terms of the
scalar field, takes the form of Klein-Gordon equation, and in the
case of non-stationary background one obtains:
\begin{equation}
\label{K-G_eq}
    \ddot \varphi + 3H\dot \varphi + V'(\varphi ) = 0.
\end{equation}
In the fast expanding Universe the scalar field rolls down very
slowly, like a ball in viscous medium, while the effective viscosity
turns out to be proportional to the expansion velocity. In the
slow-roll regime
\begin{equation}\nonumber
    H\dot \varphi \gg \ddot \varphi,\quad V\left( \varphi \right) \gg \dot \varphi
    ^2.
\end{equation}
In the same limit the equations of motion take the form
\begin{eqnarray}\nonumber
3H\dot \varphi + V'(\varphi ) = 0;\\\nonumber H^2 = \frac{{8\pi
}}{{3M_{Pl}^2 }}V(\varphi ).
\end{eqnarray}
To be more specific, let us consider the simplest model of the
scalar field with mass $m$ and potential energy $V(\varphi) =
\frac{{m^2}}{2}\varphi^2.$ Right after the start of inflation $\ddot
\varphi \ll 3H\dot \varphi;~\dot \varphi^2 \ll m^2 \varphi^2,$
therefore
\begin{eqnarray}\nonumber
3\frac{{\dot a}}{a}\dot \varphi + m^2 \varphi = 0;\\\nonumber H =
\frac{{\dot a}}{a} = \frac{{2m\varphi }}{{M_{Pl} }}\sqrt {\frac{\pi
}{3}}.
\end{eqnarray}
Due to fast growth of the scale factor and slow variation of the
field because of strong friction, one has
\begin{equation}\nonumber
    a \propto e^{Ht},\quad H = \frac{2m\varphi}{M_{Pl}}\sqrt{\frac{\pi }{3}}.
\end{equation}
For better clarity let us obtain the state equation for the scalar
field in the slow-roll regime. For homogeneous scalar field in the
potential $V\left( \varphi \right)$ in the local Lorentz frame the
non-zero components of the energy momentum tensor are
\begin{equation}\nonumber
T_{00} = \frac{1}{2}\dot \varphi ^2 + V\left( \varphi \right) = \rho
_\varphi ;\;T_{ij} = \left( {\frac{1}{2}\dot \varphi ^2 - V\left(
\varphi \right)} \right)\delta _{ij} = p_\varphi \delta _{ij} .
\end{equation}
In the slow-roll regime $\dot \varphi ^2 \ll V(\varphi )$ and,
consequently, $p_\varphi \approx - \rho _\varphi .$ Therefore the
energy-momentum tensor in the slow-roll regime approximately
coincides with the vacuum one with $p = - \rho .$ Using (\ref{q_w})
and taking into account the fact that the inflation scenery with
exponential growth of the scale factor leads to the flat space case,
and therefore $\Omega = 1,$ one obtains that $q=-1$ during the
period of the inflation expansion.

Therefore the scalar field provides rather natural way to obtain the
accelerated expansion of Universe at least on early stages of its
evolution. As the field intensity decrease in the slow-roll regime,
viscosity falls down and the Universe comes out the inflation regime
with exponential growth of the scale factor.

We remark that the scalar field can provide decelerated expansion of
Universe as well as the accelerated one. Near the minimum of the
inflation potential the inflation conditions definitely violate and
the Universe leaves the inflation regimes. The scalar field start to
oscillate near the minimum. Assuming that period of the oscillations
is much less than the cosmological time scale, one can neglect the
expansion term in the equation (\ref{K-G_eq}), and it easy to come
up with effective state equation near the minimum of inflation
potential. Let us rewrite the scalar field equation(\ref{K-G_eq}) in
the form
\begin{equation}\nonumber
    \frac{d}{{dt}}\left( {\varphi \dot \varphi } \right) - \dot \varphi ^2 + \varphi V'_\varphi = 0.
\end{equation}
After averaging over the oscillation period the first term turns to
zero and then
\begin{equation}\nonumber
    \left\langle {\dot \varphi ^2 } \right\rangle \simeq \left\langle {\varphi V'_\varphi } \right\rangle .
\end{equation}
The effective (averaged) state equation now reads
\begin{equation}\nonumber
    w \equiv \frac{p}{\rho } \simeq \frac{{\left\langle {\varphi V'_\varphi } \right\rangle - \left\langle {2V} \right\rangle }}{{\left\langle {\varphi V'_\varphi } \right\rangle + \left\langle {2V} \right\rangle }}.
\end{equation}

For the case of quadratic potential $V \propto \varphi ^2$ one
obtains $w \simeq 0,$ which corresponds to the state equation of
non-relativistic matter.

We note also that the scalar field models are very wide-spread in
modern cosmology, allowing to describe not only accelerated
expansion, but much more complicated dynamics of Universe. Besides,
most of the scalar field models are well motivated in the particle
physics and concurring theories.

\section{Evidence for the accelerated expansion of Universe\label{SECIIII}}

The Hubble law tells nothing about the magnitude, sign or very
possibility of non-uniform expansion of Universe. The approximation
where it works is insensible to acceleration. Investigation of the
non-linear effects requires high red-shift data. If the observation
detect deviations from the linear law, then the magnitude and sign
of the deviation allow to judge about the sign of the cosmological
acceleration. If the detected deviation lies towards larger
distances at fixed redshift then the acceleration is positive. The
distance is estimated by brightness of the source in assumption that
the considered kind of sources represent the standard candles --- an
ensemble of object with practically equal luminosity. Therefore the
observed brightness of the objects depends only on the distance to
observer. The supernova outbursts of type Ia (exploding white
dwarfs) present an example of such kind of objects. As the white
dwarfs have very low mass dispersion, their luminosity is
practically the same. An additional advantage is the huge power of
$\left( \sim {{10}^{36}}W \right),$ released in explosion. Thanks to
that, they can be detected on distances compatible to size of the
observable Universe.

Given the emitted light intensity $L,$ or internal luminosity of the
object, and having measured the light intensity $F,$ which have
reached us, or observed flow, one can calculate the distance to the
object. The quantity defined this way is called the luminosity
distance ${d}_{L}$
\begin{equation}\label{d_L}
d_{L}^{2}=\frac{L}{4\pi F}.
\end{equation}
In order to determine the acceleration of the Universe expansion one
needs to express the luminosity distance in terms of the redshift
for the registered radiation. Let $E$ be internal (absolute)
luminosity of some source. A terrestrial observer registers the
photon flow $F.$ Increasing of the photon wavelength, and therefore
decrease of its energy, in the expanding Universe during its pass
from the source to the observer results in effective (apparent)
luminosity of the source $L=E/a(t).$ The conservation law for the
energy emitted on time interval $dt$ and absorbed on the interval
$d{{t}_{0}},$ implies
\begin{equation}\label{Ff}
F4\pi {{r}^{2}}d{{t}_{0}}=Edt=La(t)dt,
\end{equation}
where $r$ is the comoving distance between the source and the
observer at time moment $t_0,$ which coincides with the physical
distance in normalization of the form  $a\left(t_0\right)=1.$ Since
the comoving distance between source and observer does not change,
the conformal time interval $d\eta =dt/a$ between two light pulses
is the same at the point of emission and at the point observation
\begin{equation}\nonumber
    \frac{dt}{a(t)}=\frac{dt_0}{a_0}.
\end{equation}
Therefore from \eqref{Ff} follows that
\begin{equation}\label{F}
F=\frac{La^2(t)}{4\pi r^2}.
\end{equation}
Comparing this expression with the definition of the luminosity
distance \eqref{d_L}, one finds
\begin{equation}\label{d_Lrz}
d_L=\frac{r}{a(t)}=(1+z)r.
\end{equation}
The obtained result is physically clear. In the expanding Universe
the registered flow decreases by factor $(1+z)^2:$ first, due to
increase of the photon wave length $(1+z)$ times and, second, due to
increase of the time interval for arrival of the fixed energy
portion also $(1+z)$ times.

Let us now determine the comoving distance to the currently observed
light source as the function of its redshift. The equation of motion
for photon is $ds^2=0.$ Let us consider a radial trajectory with the
observer at the origin of the coordinates. In the case of spatially
flat metrics one has
\begin{equation}\label{ds_light}
ds^2=a^2(t)\left( d\eta^2-dr^2\right)=0.
\end{equation}
Accounting that
\begin{equation}\label{d_eta}
    d\eta =\frac{d\eta }{dt}\frac{dt}{da}\frac{da}{dz}dz=-\frac{dz}{H(z)},
\end{equation}
one finds
\begin{equation}\label{r(z)}
r(z)=\int_0^z{\frac{dz'}{H(z')}}.
\end{equation}
Therefore for spatially flat Universe
\begin{equation}\label{d_L(z)H}
    d_L=(1+z)\int_{0}^{z}{\frac{d{z}'}{H({z}')}}.
\end{equation}
In general case,
\begin{equation}\label{d_L(z)}
d_L(z)=c(1+z){{\left( 1-{{\Omega }_{0}}
\right)}^{-1/2}}H_{0}^{-1}S\left[ {{\left( 1-{{\Omega }_{0}}
\right)}^{1/2}}{{H}_{0}}\int_{0}^{z}{\frac{d{z}'}{H({z}')}} \right],
\end{equation}
where
\begin{equation}\label{S(x)}
S(x)=\left\{
  \begin{array}{c}
    \sin x,\quad {{\Omega }_{0}}>1;\\
    x,\quad \quad {{\Omega }_{0}}=1;  \\
   \sinh x,\ {{\Omega }_{0}}<1.\\
  \end{array}
\right.
\end{equation}
The quantities $H_0,\,\Omega_0\equiv \frac{\rho_0}{\rho_{cr}},\
\left( \rho_{cr}\equiv \frac{3H_{0}^{2}}{8\pi G} \right)$ refer to the
present moment of time. For the multi-component flat case
\begin{equation}\label{d_Lw}
d_L=\frac{1+z}{{{H}_{0}}}\int_{0}^{z}{\frac{d{z}'}{
\sqrt{\sum\limits_i{\Omega _{0i}{{\left( 1+{z}' \right)}^{3\left(
1+{{w}_{i}} \right)}}}}}}.
\end{equation}
The relation \eqref{d_L(z)}. can be rewritten in terms of the
deceleration parameter. For the flat case it is
\begin{equation}\label{d_L(z)q}
d_L(z)=(1+z)\int_0^z\frac{dz'}{H(z')}=(1+z)H_0^{-1}\int_0^z{du}\exp
\left[ -\int_0^u\left[ 1+q(v)d\ln (1+v) \right] \right].
\end{equation}
Let us use normal definition for the absolute value $\mu $ of the
distance to a standard candle (say, SNe1a):
\begin{equation}\label{mu}
    \mu (z)\equiv \left[ {{m}_{B}}(z)-{{M}_{B}} \right]=5\log \left( {{d}_{L}}/Mpc \right)+25.
\end{equation}
Using \eqref{d_L(z)q}, one can link that quantity to the
acceleration history $q(z)$ via
\begin{equation}\label{mu_z}
\mu (z)=25+5\log \left[ \frac{1+z}{H_0 Mpc} \right]\int_0^zdu\exp
\left( -\int_{0}^{u}{\left[ 1+q(u) \right]d\ln u} \right).
\end{equation}
Here $M_B$ and $m_B$ are absolute and apparent stellar magnitude of
the source respectively. The latter expression represents a
fundamental relation, linking the deceleration parameter history
with the $SNe 1a$ measurements.

We remark that the relation \eqref{d_L(z)q} is based solely on the
FRW-metrics. It means that the acceleration-deceleration dilemma can
be resolved without any assumption about applicability of general
relativity. However, supernova observations cannot help to determine
current deceleration parameter immediately. In order to interpret
the data, we must know  $H(z)$ or $q(z),$ which in turns requires to
specify the dynamical equations and material composition of the
Universe.

It is useful to cite the expression for the luminosity distance with
accuracy up to $z^2$ terms:
\begin{equation}\label{d_L(z)Tey2}
    d_L=\frac{z}{H_0}\left[ 1+\left( \frac{1-q_0}{2} \right)z+{\mathrm O}(z^2) \right],
\end{equation}
where in the flat case
$q_0=\frac{1}{2}\sum\limits_{i}\Omega_i(1+3w_i). $

It follows from \eqref{d_L(z)Tey2} that for small $z$ the luminosity
distance is proportional to the redshift, with the proportionality
coefficient equal to inverse value of the Hubble constant. For more
distant cosmological objects the higher order corrections for the
luminosity distance depend on the current value of the deceleration
parameter $q_0,$ ore, in other words, on quantity and type of
components, filling the Universe.

The expression for luminosity distance in the next order correction
over the redshift reads
  \begin{eqnarray}\label{d_L(z)Tey4}
   d_L(z)&=&\frac{cz}{H_0}[1+\frac{1}{2}\left( 1-q_0 \right)z-\frac{1}{6}\left( 1-q_0-3q_0^2+j_0 \right)z^2+\\\nonumber
   &+&\frac{1}{24}\left( 2-2q_0-15q_0^2-15q_0^3+5j_0+10j_0q_0+s_0\right)z^3+{\mathrm
   O}(z^4)].
\end{eqnarray}
Let us briefly discuss the methodology for verification of the
cosmological models using the type $SNe 1a$ supernova outbursts. Let
us consider more closely \cite{Perlmutter42,Riess} two supernovae
1992P at low redshift $z=0.026$ with $m=16.08$ and 1997ap at high
redshift $z=0.83$ with $m=24.32.$ As indicated earlier,
$d_L(z)\simeq z/H_0$ for $z\ll 1.$ Using \eqref{mu} we find
$M=-19.09.$ Then the luminosity distance of 1997ap is obtained by
substituting  $m=24.32$ and $M=-19.09$ for \eqref{mu}
\begin{equation}\label{H_0d_L}
H_0d_L\simeq 1.16\quad for\quad z=0.83.
\end{equation}
From the other hand, for the case of Universe filled by
non-relativistic matter, using \eqref{d_Lw}, one finds
\begin{equation}\nonumber
    {{H}_{0}}{{d}_{L}}\simeq 0.95.
\end{equation}
The latter result apparently contradicts the observations
\eqref{H_0d_L}.

Therefore, registration of outbursts for any kind of light sources
with the same internal luminosity, i.e. standard candles, allows to
determine the velocity of Universe expansion in different moments of
its evolution. Comparing the obtained results with the predictions
of different theoretical models, one can select the most adequate
ones. Despite of the conceptual simplicity, the problem have been
continuously facing many obstacles. Let us list just a few of them.
The supernova outbursts are rare and random. To collect sufficient
statistics one has to cover rather large part of the sky. The
outbursts takes place for a limited time, therefore it is important
to detect a supernova as soon as possible in order to follow the
dynamics of its brightness variation. And of course the main problem
is still connected to disputable applicability of the type Ia
supernova for a standard candle. In the early ninetieth the United
States created two groups for detection and analysis of the type
$SNeIa$ supernova outbursts: SuperNova Cosmology Project and High-Z
SuperNova Search. It were their results \cite{Perlmutter42,Riess}
which in the years 1998 and 1999 gave base for conclusion on
accelerated expansion of Universe, so dramatically affected not only
modern cosmology, but all physics as a whole. During the passed
decade the results \cite{Perlmutter42,Riess} were multiple times
reproduced with ever improving statistics. The main conclusion was
always reconfirmed: comparable recently (at $z\sim 0.5)$ our
Universe passed through the transition from decelerated expansion to
accelerated one. Let us present in more details the analysis carried
out by Riess  \cite{RiessIazl1}, where he used the so-called golden
set of $SNe 1a.$ The set contained $157$ well studied $SNe 1a$
supernova with redshifts ranged in $0.1<z<1.76.$ The analysis was
based on the relation \eqref{d_L(z)q} which is exact expression for
the luminosity distance in a geometrically flat Universe. In the
case of linear two-parameter decomposition
\begin{equation}\label{q(z)q_0q_1z}
q(z)=q_0+q_1z,
\end{equation}
the integral in \eqref{d_L(z)q} can be exactly evaluated and the
expression for the luminosity distance takes the form
\cite{CunhaLima}
\begin{equation}\label{d_L(z)q0q1}
    d_L(z)=\frac{1+z}{H_0}e^{q_1}q_1^{q_0-q_1}\left[ \gamma \left( q_1-q_0,(1+z)q_1\right)-\gamma \left( q_1-q_0,q_1\right) \right],
\end{equation}
where ${{q}_{0}},{{q}_{1}}$ are values of $q(z),\frac{dq(z)}{dz}$ at
$z=0,$ $\gamma $ is the reduced gamma-function. Using
\eqref{d_L(z)q0q1} one can obtain information about $q_0,q_1$ and,
therefore, about global behavior of $q(z).$ The dynamical ''phase
transition'' occurs at $q\left( {{z}_{t}} \right)=0$ or equivalently
at $z_t=-q_0/q_1.$

The second widely used parameterization is
\begin{equation}\label{d_L(z)q0q_1}
    q(z)=q_0+q_1\frac{z}{1+z}.
\end{equation}
Its advantage is in the fact that it is well-behaved at large $z,$
while the linear approximation suffers of divergences. In such
parameterization one has
\begin{equation}\label{d_L(z)q0q_1}
d_L(z)=\frac{1}{(1+z)H_0}e^{q_1}q_1^{-(q_0+q_1)}\left[ \gamma \left(
q_1+q_0,q_1\right)-\gamma \left( q_1+q_0,q_1/(1+z) \right) \right].
\end{equation}
Now the parameter $q_1$ determines the correction to $q_0$ in the
distant past: $q(z)=q_0+q_1$ at $z\gg 0.$ The likelihood for the
parameters $q_0$ and $q_1$ can be determined from $\chi^2$
statistic,
\begin{equation}\label{chi2}
\chi ^2\left( H_0,q_0,q_1\right) = \sum\limits_i {\frac{{{{\left(
{{\mu _{p,i}}\left( {{z_i};{H_0},{q_0},{q_1}} \right) - {\mu
_{0,i}}} \right)}^2}}}{{\sigma _{{\mu _{0,i}}}^2 + \sigma _v^2}}},
\end{equation}
where $\sigma_{\mu_{0,i}}$ is the uncertainty in the individual
distance modules, $\sigma_v$ is the dispersion in SNe redshifts due
to peculiar velocities. The obtained results give evidence in favor
of the Universe with recent acceleration $\left(q_0<0 \right)$ and
previous deceleration $(q_1>0)$ with $99.2\%$ and $99.8\%$
likelihood. In the case of linear decomposition of the parameter $q$
the transition from decelerated expansion in the past to current
accelerated expansion took place at the redshift $z_t=0.46\pm 0.13.$
Unfortunately one should not take this result too serious: the
linear approximation always leads to the transition, provided the
two parameters have opposite signs.

The consequent and statistically more reliable analysis
qualitatively supported the above cited results. Of course, the
quantitative results depend on the used set of supernova, but the
main result remains the same: we live in accelerated expanding
Universe, which passed from decelerated expansion to accelerated one
in the recent past.

Nowadays the method based on the supernova outbursts observations is
undoubtedly leading. But it turned out to have competitors. A
promising and completely independent alternative (and no way a
surrogate) is the observation of angular diameter distances $D_A(z)$
for a given set of distant objects. The combination of the
Sunyaev-Zeldovich effect with measurements of surface brightness in
X-ray range provides the method for determination of angular
diameter distances for galactic clusters \cite{Bonamente}. The
Sunyaev-Zeldovich stands for small perturbation of CMB spectrum
produced by inverse Compton scattering of relict photons passed
through clouds of hot electrons. Observing the temperature decrement
of the CMB spectrum towards galaxy clusters together with the X-rays
observations, it is possible to break the degeneracy between
concentration and temperature thereby obtaining $D_A(z).$ Therefore,
such distances are fully independent of the one given by the
luminosity distance.

For the case of spatially flat Universe described by FRW-metric, the
angular diameter distance is
\begin{equation}\label{d_A(z)FRW}
d_A(z)=\frac{1}{(1+z)H_0}\int_0^z\frac{du}{H\left( u
\right)}=\frac{1}{(1+z)H_0}\int_0^z\exp \left[ -\int_0^u\left[
(1+q({u}') \right]d\ln (1+{u}' \right]du,
\end{equation}
\cite{CunhaLimaHolanda} considered
the 38 measurements of angular diameter distances from galaxy
clusters as obtained through SZE/X-ray method by Bonamente and
coworkers \cite{Bonamente}  Lima et al used a maximum likelihood
determined by a $\chi^2$ statistics
\begin{equation}
\chi^2\left(z,p \right)=\sum\limits_i\frac{\left( {{D}_{A}}\left(
{{z}_{i}},p \right)-{{D}_{{{A}_{0,i}}}} \right)}{\sigma
_{{{D}_{A0i}}}^{2}+\sigma _{stat}^{2}},
\end{equation}
where $D_{A_{0,i}}$ is the observational angular diameter distance,
$\sigma_{D_{A0i}}$is the uncertainty in the individual distance,
$\sigma_{stat}$ is the contribution of the statistical errors added
in quadrature and the complete set of parameters is given by
$p\equiv \left( H_0,q_0,q_1\right).$ For the sake consistency, the
Hubble parameter $H_0$ has been fixed by its best it value
$H_{0}^{*}=80km/\sec \cdot Mpc^{-1}.$ In the case of linear parameterization
the results are the following: best its to the free parameters are
$q_0=-1.35,~q_1=4.2,~z_t=0.32.$ Such results favor a Universe with
recent  acceleration $\left(q_0<0 \right)$ and a previous
decelerating stage $\left( dq/dz>0 \right).$ In the case of the
parameterization
\begin{equation}\nonumber
    q(z)={{q}_{0}}+{{q}_{1}}\frac{z}{1+z},
\end{equation}
one gets $q_0=-1.43,~q_1=6.18,~z_t=0.3.$ In both cases the results
well agree with the ones obtained using the $SNe 1a$ data.

It was recently shown that the so-called luminous red galaxies
(LRG's) provide another possibility for direct measurements of the
expansion velocity \cite{JimenezLoeb,ZhangMa,MaZhangAstr}. Idea of
the method is to reconstruct the Hubble parameter from the time
derivative of the redshift
\begin{equation}\label{dz}
H(z)=-\frac{1}{1+z}\frac{dz}{dt}.
\end{equation}
The derivative can be found in measurements of the ''age difference
'' between two passively evolving galaxies at different but close
redshifts. The method was realized for $0.1<z<1.75.$ We stress that
the range includes the interesting for us transition region, where
$z\sim 0.5.$ The results of analysis for available data
\cite{CarvalhoAlkaniz_Cosmography} agree with the ones based on the
supernova and angular diameter. In nearest future a set of some
$2000$ passively evolving galaxies is going to be measured in the
range $0<z<1.5.$ Those observations will give about $1000$ values of
$H(z)$ with $15\%$ accuracy, provided the age of galaxies is
determined with $10\%$ precision.

Concluding the present section we present a novel and promising
direction to research the history of cosmological expansion of
Universe. We remind that astrophysicists initially used cepheids ---
the variable stars with intensity proportional to period of
brightness variation --- as the standard candles. A typical example
of cepheids is the  Polar Star, the brightest and closest to Earth
variable star with period of $3.97$ days. The cepheids are perfect
standard candles for galactic distances. They allowed to determine
size of our Galaxy and distance to our closest neighbor --- the
Andromeda galaxy. Research of the Universe dynamics involves
principally larger scales, and thus it requires to use essentially
more powerful standard candles. Recall, that the cosmological
principle, which postulates homogeneity and isotropy of the
Universe, and which all our above used equations of Universe
dynamics are based on, is valid on scales of $100\,Mpc$ and larger.
Using considerably more powerful radiation sources as the standard
candles, namely type Ia supernovae, already allowed us to advance
remarkably deeper to the history of Universe. However the
possibilities of the new standard candles are also limited. Up to
present time the type Ia supernovae were observed only up to the
redshift $z<2,$ though more reliable reconstruction of the
cosmological expansion history requires even higher redshifts and
therefore more powerful standard candles. It turns out that objects
with required properties are already available! It is the so-called
gamma-ray bursts(GRB) --- giant energy emissions of explosive type
with duration from tree to hundred second, observed in the hardest
part of the electromagnetic spectrum. The typical energy, emitted by
the gamma-ray burst is $\sim 10^{54} \, erg$ which is one order of
magnitude higher than at a supernova outburst. It is already
compatible to the Sun rest mass! The events producing the gamma-ray
bursts are so powerful that they can be observed even by naked eye,
though they occur on distances of billions light years from the
Earth! The energy yields in form of a collimated flow, called a jet.
Occurrence of the jets means that we can see only small number of
all bursts that happen in Universe. Distribution of GRB durations
has pronounced bimodal character.

\begin{center}
 \begin{figure}[tbhp]
 \centering
 \includegraphics[width=0.7\textwidth]{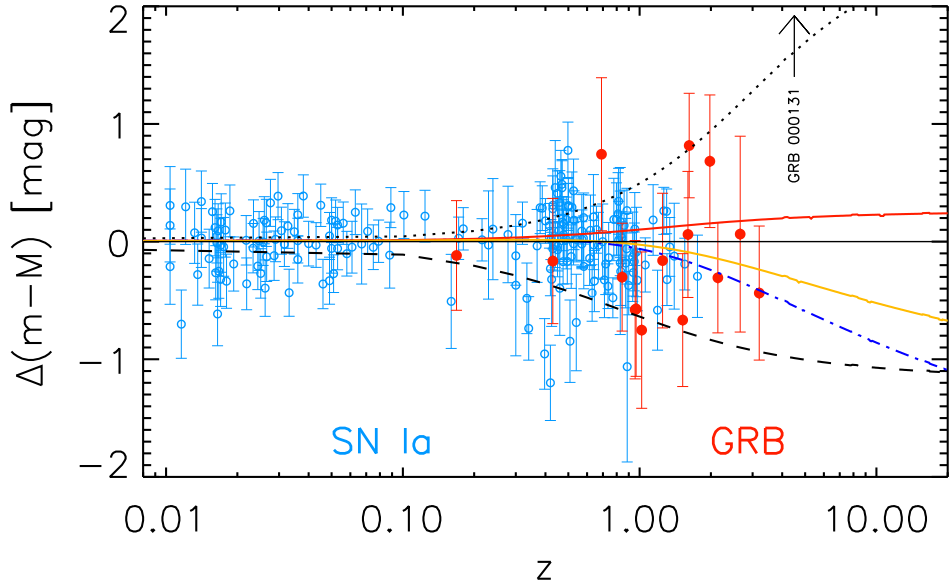}
 \caption{\label{GRBs15}  15 GRB with measured redshifts and collimation angle(compared to Ia supernova data in SCM)}.
 \end{figure}
 \end{center}

Origin of short GRB is probably linked to coalescence of two neutron
stars, or a neutron star and a black hole. The events with longer
duration are presumably due to creation of a black hole in collapse
of massive stellar nucleus (>25 Solar mass) with high angular
momentum --- it is the so-called collapsar model. The possibility to
use the GRB as the standard candles is based on occurrence of the
so-called ''Amati relation'', which relates the peak frequency of
the burst with its total energy. This relation is a direct analog of
the period-luminosity relation for cepheids. Strong dispersion (see
Fig.\ref{GRBs15}) yet limits applications of the GRB as the standard
candles, but the possibility to advance to the region of
considerably higher redshifts makes this direction potentially very
attractive.

\section{Scale factor dynamics in SCM}
\begin{figure}[tbhp]
 \centering
 \includegraphics[width=0.7\textwidth]{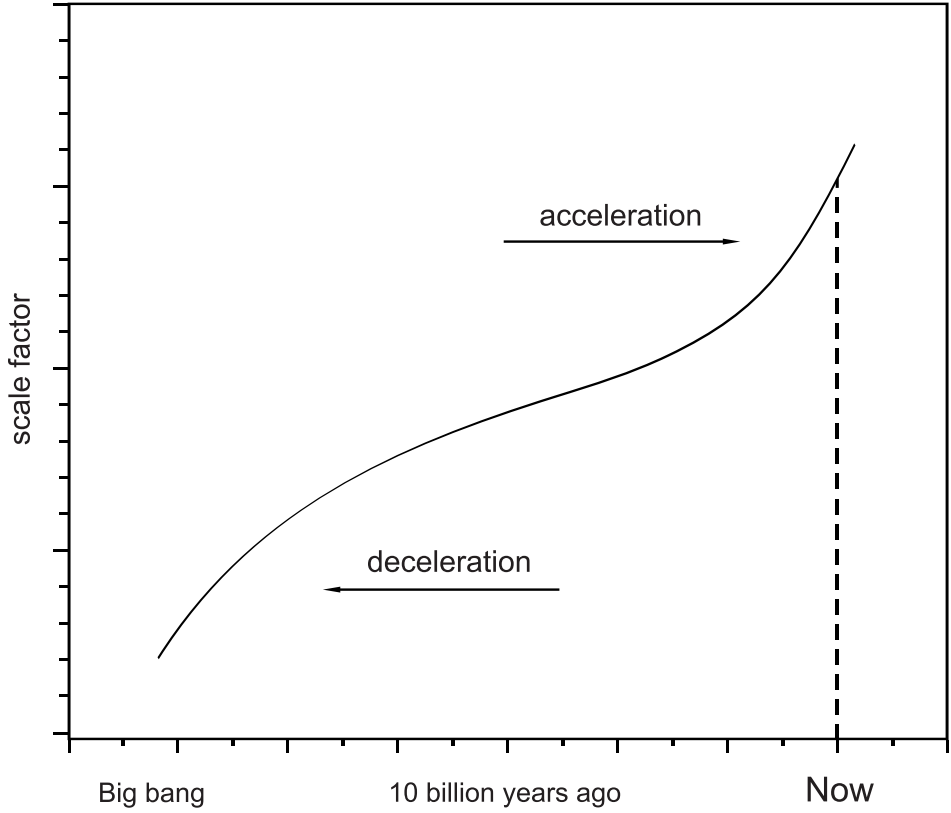}
 \caption{\label{a_SKM}  Time dependence of the scale factor in SCM.}
 \end{figure}

Now let us consider evolution of the deceleration parameter in
Standard Cosmological model (SCM). Recall that the Big Bang model
included only two substances -- matter and radiation -- which could
give only decelerated expansion of the Universe. According to SCM,
the Universe is presently dominated by the dark energy --- a
component with negative pressure. It is the component responsible
for the observed accelerated expansion of Universe. Let us determine
the redshift and time moment of the transition to the accelerated
expansion, i.e. find the inflection point on the curve (see Fig.
\ref{a_SKM}), describing the time dependence of the scale factor
The second Friedmann equation in the SCM \eqref{Frid_2} can be
transformed to the following
\begin{equation}\label{Fri_2SKM}
\frac{{\ddot{a}}}{a}=\frac{1}{2}H_{0}^{2}\left[ 2{{\Omega }_{\Lambda
0}}-{{\Omega }_{m0}}{{(1+z)}^{3}} \right],
\end{equation}
where $\Omega_{\Lambda 0}$ and ${{\Omega }_{m0}}$ are present values
of the relative density for dark energy in form of cosmological
constant and for matter respectively. Now it follows for the
redshift at the point of transition from decelerated expansion to
accelerated one
\begin{equation}\label{z_trSCM}
z^*={{\left( \frac{2{{\Omega }_{\Lambda 0}}}{{{\Omega }_{m0}}}
\right)}^{1/3}}-1.
\end{equation}
With the SCM parameters $\Omega_{\Lambda 0}\simeq 0.73,\ {{\Omega
}_{m0}}\simeq 0.27$ one gets $z^*\simeq 0.745.$ Note that the result
\eqref{z_trSCM} can be obtained using the fact that for a
multi-component Universe with state equations $p_i=w_i\rho_i$
respectively, the deceleration parameter \eqref{q_w} equals (recall
that $\Omega =\sum \limits_i \Omega_i =1$ in the flat case )
\begin{equation}\label{q_SCM}
    q=\frac{1}{2}-\frac{3}{2}\frac{{{\Omega }_{\Lambda 0}}}{{{\left( 1+z \right)}^{3}}{{\Omega }_{m0}}+{{\Omega }_{\Lambda 0}}}.
\end{equation}
The condition $q=0$ allows to reproduce the relation
\eqref{z_trSCM}. Let us discuss its asymptotes. For early Universe
$(z\to \infty ),$ filled by components with positive pressure ,
$q\left( z\to \infty  \right)=\frac{1}{2},$ i.e. as it was expected
the expansion is decelerated, while in the distant future with
domination of the cosmological constant the accelerated expansion is
observed: $q\left( z\to -1 \right)=-1.$ The latter result is a
trivial consequence of the exponential expansion $a\propto
{{e}^{Ht}}$ in the case of Universe dominated by dark energy in the
form of cosmological constant.

We remark that the dependence $q(t)$ can be immediately obtained
from the definition $q=-\ddot{a}/a{{H}^{2}},$ using the SCM
solutions for the scale factor:
\begin{eqnarray}\nonumber
  a(t)&=&{{A}^{1/3}}{{\sinh }^{2/3}}\left( t/{{t}_{\Lambda }} \right);\label{a(t)_SCM} \\
  A & \equiv & \frac{{{\Omega }_{m0}}}{{{\Omega }_{\Lambda 0}}},\quad {{t}_{\Lambda }}\equiv \frac{2}{3}H_{0}^{-1}\Omega _{\Lambda 0}^{-1/2}.
 \end{eqnarray}
It results in the following
\begin{equation}\label{q(t)_SCM}
q(t)=\frac{1}{2}\left[ 1-3\tanh^{2}\left( \frac{t}{t_{\Lambda
}}\right) \right],
\end{equation}
 \begin{figure}[tbhp]
 \centering
 \includegraphics[width=0.7\textwidth]{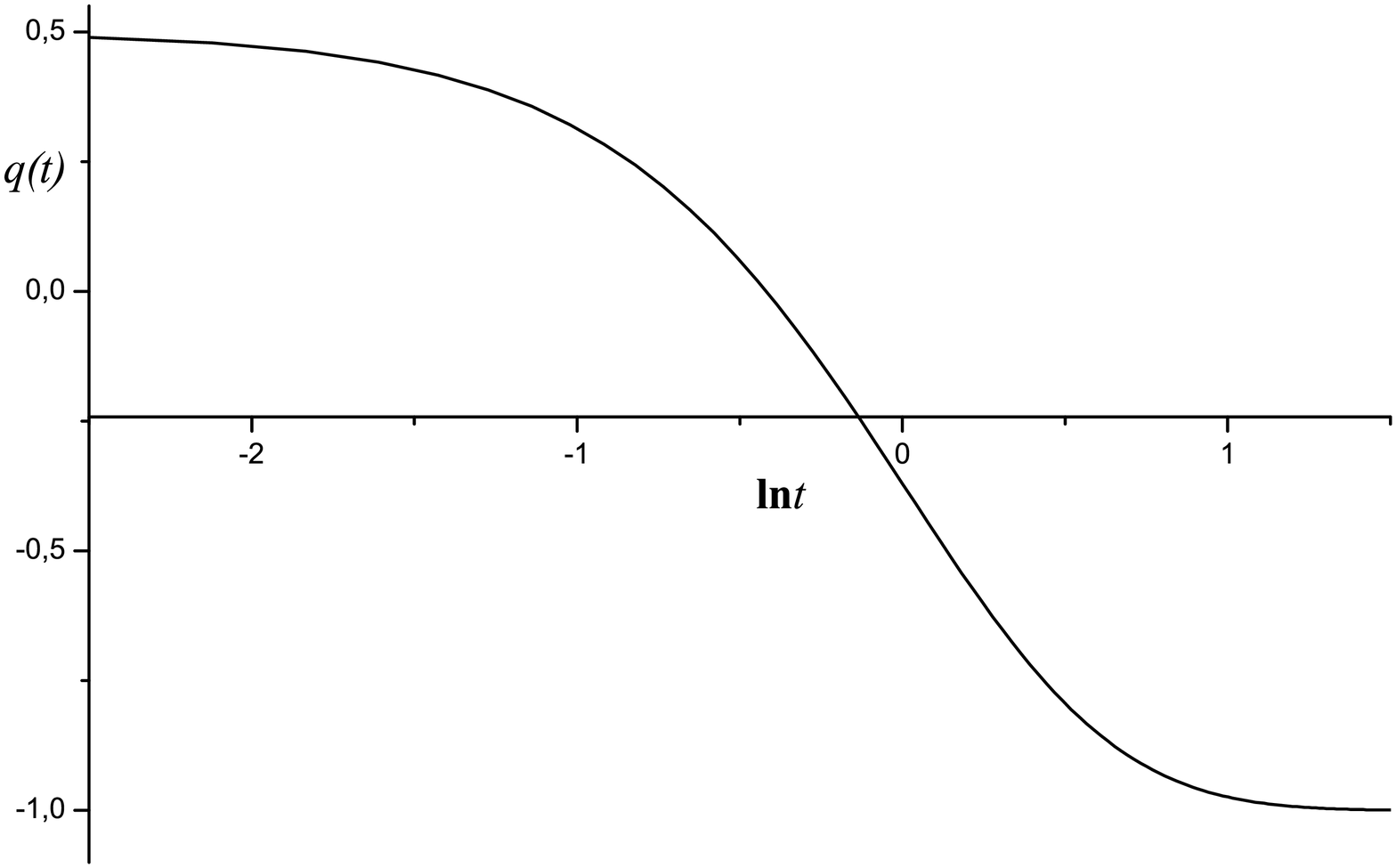}
 \caption{\label{qp_SKM} Time dependence of the deceleration parameter $q(t)$ in SCM.}
 \end{figure}
The dependence $q(t)$ is presented on Fig.\ref{qp_SKM}. Note that
the time asymptotes at $t\to 0$ and $t\to \infty $ correspond to the
above obtained ones at $z\to \infty $ and $z\to -1.$

Let us now determine the time of the transition to the accelerated
expansion. Inserting the relation \eqref{a(t)_SCM}, one obtains
\begin{equation}\label{t(a)_SCM}
t\left( a \right)=\frac{2}{3}\Omega _{\Lambda
0}^{-1/2}H_{0}^{-1}{arsh}\left[ {{\left( \frac{{{\Omega }_{\Lambda
0}}}{{{\Omega }_{m0}}} \right)}^{1/2}}{a}^{3/2} \right].
\end{equation}
Transforming from the redshift to the scale factor
$a^*={{(1+z^*)}^{-1}}={{\left( \frac{{{\Omega }_{m0}}}{2{{\Omega
}_{\Lambda 0}}} \right)}^{1/3}}$ , one gets
\begin{equation}\label{t_tr}
    t^*\equiv(a^*)=\frac{2}{3}\Omega _{\Lambda 0}^{-1/2}H_{0}^{-1}\sinh^{-1}\left( 1/2 \right)\simeq 5.25\,Gyr.
\end{equation}
Because of physical importance of the obtained result we present
another one, probably the simplest, interpretation of it. If the
quantity $aH=\dot{a}$ growth, then $\ddot{a}>0,$ which corresponds
to accelerated expansion of Universe. According to the first
Friedmann equation,
\begin{equation}\nonumber
    \frac{aH}{{{H}_{0}}}=\sqrt{\frac{{{a}^{3}}{{\Omega }_{\Lambda 0}}+{{\Omega }_{m0}}}{a}}\simeq \sqrt{\frac{0.73{{a}^{3}}+0.27}{a}}.
\end{equation}
It is easy to show that the function under radical starts to grow at
$a^*\simeq 0.573,$ which corresponds to $z^*=0.745.$ It is
interesting to note that the transition to accelerated expansion of
Universe $(z\simeq 0.75)$ occurred remarkably earlier than the dark
energy started to dominate $(z\simeq 0.4).$

In the SCM the dark energy exist in form of cosmological constant
with the state equation ${{p}_{\Lambda }}=-{{\rho }_{\Lambda }},$
i.e. the state parameter equals ${{w}_{\Lambda }}=-1.$ A natural
question rises: what is the limiting value of that parameter, which
still provides accelerated expansion of Universe in the present
time? As we have seen above, the condition for accelerated expansion
is the following $\sum\limits_{i}{\left( {{\rho }_{i}}+3{{p}_{i}}
\right)}<0.$ In SCM it is transformed to the following
\begin{equation}\nonumber
w_{_{DE}}<-\frac{1}{3}\Omega _{DE}^{-1};\quad w_{_{DE}}<-0.46.
\end{equation}
Of course, a substance with such state equation is not the
cosmological constant, and it can be realized, for instance, with
the help of scaler fields (see next section).

Now let us estimate absolute magnitude of the cosmological
acceleration using the SCM parameters. Taking time derivative of the
Hubble law, one gets
\begin{equation}\label{dot_V}
\dot{V}=\left( \dot{H}+{{H}^{2}} \right)R.
\end{equation}
The time derivative of the Hubble parameter is
\begin{equation}\nonumber
\dot{H}=\frac{\ddot{a}a-{{{\dot{a}}}^{2}}}{{{a}^{2}}}=\frac{{\ddot{a}}}{a}-{{H}^{2}}.
\end{equation}
Therefore
\begin{equation}\nonumber
    \dot{H}+{{H}^{2}}=\frac{{\ddot{a}}}{a}=-\frac{4\pi G}{3}\left( {{\rho }_{m}}-2{{\rho }_{\Lambda }} \right)=\frac{8\pi G}{3}\left( {{\rho }_{\Lambda }}-\frac{1}{2}{{\rho }_{m}} \right)=H^{2}\left( {{\Omega }_{\Lambda }}-\frac{1}{2}{{\Omega }_{m}} \right).
\end{equation}
And finally we get the analog of the Hubble law for the acceleration
$\dot{V}$
\begin{equation}\label{dot_VH}
\dot{V}=\tilde{H}R;\quad \tilde{H}=H^{2}\left( {{\Omega }_{\Lambda
}}-\frac{1}{2}{{\Omega }_{m}} \right).
\end{equation}
In the present time $({{\Omega }_{m}}={{\Omega }_{m0}},\ {{\Omega
}_{\Lambda }}={{\Omega }_{\Lambda 0}}),$ and on the distance, for
example, $R=1$ Mpc one gets
\begin{equation}\label{acc_num}
\dot{V}\simeq {{10}^{-11}}cm\:{\sec }^{-2}.
\end{equation}
A possible way to observe such effect is based on the fact that the
redshift for any cosmological object slowly changes due to the
acceleration (or deceleration) of the Universe expansion. Let us
estimate the magnitude of that effect. By definition
\[z=\frac{a\left( {{t}_{0}} \right)}{a(t)}-1,\]
where $t$ is the time of emission and ${{t}_{0}}$ is the detection
time, and it follows that
\begin{equation}\label{dzdt}
\frac{dz}{d{{t}_{0}}}=\frac{\dot{a}({{t}_{0}})a(t)-a({{t}_{0}})\dot{a}(t)\frac{dt}{d{{t}_{0}}}}{{{a}^{2}}(t)}.
\end{equation}
Taking into account that
$\frac{dt}{d{{t}_{0}}}=\frac{a(t)}{a({{t}_{0}})}=\frac{1}{(1+z)},$
one gets
\begin{equation}\label{dot_z}
\dot{z}=\frac{\dot{a}({{t}_{0}})}{a(t)}-\frac{a({{t}_{0}})\dot{a}(t)}{{{a}^{2}}(t)}\frac{1}{1+z}.
\end{equation}
Thus, the rate of redshift variation for the light emitted at time
$t$ and registered at the present time $t_0,$ is determined by the
relation
\[\dot{z}\equiv \frac{dz}{d{{t}_{0}}}={{H}_{0}}(1+z)-H(t).\]
In the SCM
\[H={{H}_{0}}{{\left( {{\Omega }_{m0}}(1+z \right)}^{3}}+{\Omega }_{\Lambda 0})^{1/2}.\]
Therefore for variation of the redshift on the time interval $\Delta
t$ one obtains
\begin{equation}\label{Delta_z}
\Delta z=\dot{z}\Delta t={{H}_{0}}\left[ 1+z-\left(\Omega_{m0}(1+z
\right)^3+\Omega_{\Lambda 0})^{1/2} \right].
\end{equation}
Note that for the two limiting cases ${{\Omega }_{\Lambda
0}}=1,\,{{\Omega }_{m0}}=0$ (accelerated expansion) and ${{\Omega
}_{\Lambda 0}}=0,\,{{\Omega }_{m0}}=1$ (decelerated expansion), as
was expected $\Delta z$ has opposite sign. For the SCM parameters
the redshift variation $\Delta z$ and the velocity increment $\Delta
V$ for a source with the redshift $z=4$ and duration of observation
$\Delta {{t}_{0}}=10$ equal respectively
\[ \Delta z\approx {{10}^{-9}}, ~~ \Delta V=c\frac{\Delta z}{1+z}\approx 6\,cm/\sec .\]
The result discourages by its smallness.  However, taking into
account fast progress in precision of observable cosmology, we
should not despair. Let us give an example. Today the list of
exoplanets -- planets out the Solar system -- counts more than 300
items. The most successful method to detect the exoplanets is to
measure radial velocity of the stars. A star accompanied by an
exoplanet experiences velocity oscillations ''to us and back from
us'', which can be measured observing the Doppler shift of the star
spectrum. For the first look it is impossible: affected by the
Earth, velocity of the sun annually oscillates with amplitude of
centimeters per second. Even effect of Jupiter produces variations
of only meters per second, while thermal broadening of the spectral
lines for the star corresponds to dispersion of velocity of order of
thousands kilometers per second. It means that even in the case of
Jupiter one have to measure shift of the spectral lines on a
thousandth of its width. It seems incredible, but the task was
successfully worked out.

\section{Dynamical forms of dark energy and evolution of Universe}
The cosmological constant represents one of many possible
realizations of the hypothetic substance called the dark energy,
introduced for explanation of the accelerated expansion of Universe.
As we have seen above, such substance have to have the parameter $w$
in the state equation $p=w\rho,$ which satisfies the condition
$w<-1/3$ (in absence of other components). Unfortunately the nature
of dark energy is absolutely unknown, which produces huge number of
hypotheses and candidates for the role of fundamental contributor to
the energy budget of the Universe. We told many times about
impressive progress of observational cosmology in the last decade.
However we still cannot answer the question about time evolution of
the state parameter $w.$ If it changes with time, we must seek an
alternative to the cosmological constant. For quite a short time a
plenty of such alternative (with respect to $w=-1$) possibilities
were investigated. The scalar fields, formed the post-inflation
Universe, are considered to be one of the principal candidates for
the role of dark energy. The most popularity acquired the version of
the scalar field $\varphi$ with appropriately chosen potential
$V\left( \varphi \right).$ In such models, unlike the cosmological
constant one, the scalar field is truly dynamical variable, and dark
energy density depends on time. The models differ by the choice of
the scalar field Lagrangian. Let us start from probably the simplest
dark energy model of such type, named the quintessence. We define
quintessence as the scalar field $\varphi$ in the potential $V\left(
\varphi \right),$ minimally coupled to gravity, i.e. experiencing
effect of the space-time curvature. Besides that we chose canonical
form for the kinetic energy term. The action for such field takes
the form
\begin{equation}\label{sec_6_1}
S=\int{{{d}^{4}}x\sqrt{-g}L=\int{{{d}^{4}}x\sqrt{-g}\left[
\frac{1}{2}{{g}^{\mu \nu }}\frac{\partial \varphi }{\partial
{{x}_{\mu }}}\frac{\partial \varphi }{\partial {{x}_{\nu }}}-V\left(
\varphi  \right) \right]}},
\end{equation}
where $g\equiv \det {{g}_{\mu \nu }}.$ Equation of motion for the
scalar field is obtained by variation of the action with respect to
the field,
\begin{equation}\label{sec_6_2}
\frac{1}{\sqrt{-g}}{{\partial }_{\mu }}\left( \sqrt{-g}{{g}^{\mu \nu
}}\frac{\partial \varphi }{\partial {{x}_{\nu }}}
\right)=-\frac{dV}{d\varphi }.
\end{equation}
In the case of flat Friedmannian Universe, namely in FRW-metric
\eqref{FRW} one obtain for homogeneous field $\varphi (t)$ the
following
\begin{equation}\label{sec_6_3}
\ddot{\varphi }+3H\dot{\varphi }+{V}'(\varphi )=0,
\end{equation}
where ${V}'(\varphi )\equiv \frac{dV}{d\varphi }.$ This relation is
sometimes called the Klein-Gordon equation.

The energy-momentum tensor for the scalar field can be found by
variation of \eqref{sec_6_1} with respect to metrics ${{g}^{\mu \nu
}},$ resulting in the following
\begin{equation}\label{sec_6_4}
{{T}_{\mu \nu }}=\frac{2}{\sqrt{-g}}\frac{\delta S}{\delta {{g}^{\mu
\nu }}}=\frac{\partial \varphi }{\partial {{x}_{\mu
}}}\frac{\partial \varphi }{\partial {{x}_{\nu }}}-{{g}_{\mu \nu }}L.
\end{equation}

In the case of homogeneous field$\varphi (t)$ in locally Lorentzian
reference frame, where metrics${{g}_{\mu \nu }}$ can be replaced by
the Minkowski one, we obtain the density and pressure of the scalar
field in the following form
\begin{equation}\label{sec_6_5}
{{\rho }_{\varphi }}={{T}_{00}}=\frac{1}{2}{{\dot{\varphi
}}^{2}}+V\left( \varphi  \right);\quad {{p}_{\varphi
}}={{T}_{ii}}=\frac{1}{2}{{\dot{\varphi }}^{2}}-V\left( \varphi
\right).
\end{equation}
The Friedmann equations for the flat Universe, filled by the scalar
field, take the form
\begin{subequations}
\begin{equation}\label{sec_6_6a}
{{H}^{2}}=\frac{8\pi G}{3}\left[ \frac{1}{2}{{{\dot{\varphi
}}}^{2}}+V(\varphi ) \right];
\end{equation}
\begin{equation}\label{sec_6_6b}
\dot{H}=-4\pi G\dot \varphi^2.
\end{equation}
\end{subequations}
The system (\ref{sec_6_6a} -- \ref{sec_6_6b}) should be complemented
by the Klein-Gordon equation (\ref{sec_6_3}) for the scalar field.
Note, that the latter can be obtained from the conservation equation
for the scalar field
\begin{equation}\label{sec_6_7}
{{\dot{\rho }}_{\varphi }}+3H\left( {{\rho }_{\varphi
}}+{{p}_{\varphi }} \right)=0.
\end{equation}
by substitution of the expressions (\ref{sec_6_5}) for energy
density and pressure.

Using (\ref{sec_6_5}), one get the state equation for the scalar
field
    \begin{equation}\label{sec_6_8}
   {{w}_{\varphi }}=\frac{{{p}_{\varphi }}}{{{\rho }_{\varphi }}}=\frac{{{{\dot{\varphi }}}^{2}}-2V}{{{{\dot{\varphi }}}^{2}}+2V}.
\end{equation}
As one can see, the parameter of the state equation for the scalar
field ${{w}_{\varphi }}$ varies in the range
\begin{equation}\label{sec_6_9}
-1\le {{w}_{\varphi }}\le 1.
\end{equation}
The state equation for the scalar field is conveniently transformed
into the following
\begin{equation}\label{sec_6_10}
w(x)=\frac{x-1}{x+1},\quad x\equiv \frac{\frac{1}{2}{{{\dot{\varphi
}}}^{2}}}{V(\varphi )}.
\end{equation}
The function $w(x)$ increases monotonously from the minimum value
${{w}_{\min }}=-1$ at $x=0$ up to the maximum asymptotic value
${{w}_{\max }}=1$ at $x\to \infty ,$ corresponding to $V=0.$ In the
slow-roll limit, $x\ll 1$ $({{\dot{\varphi }}^{2}}\ll V(\varphi ))$
the scalar field behaves as the cosmological constant,
${{w}_{\varphi }}=-1.$ It is easy to see that in such case ${{\rho
}_{\varphi }}=const.$ The opposite limit $x\gg 1$ $(\dot{\varphi
}\gg V\left( \varphi  \right))$ correspond to the hard matter
${{w}_{\varphi }}=1.$ In that case the scalar field energy density
evolves as ${{\rho }_{\varphi }}\propto {{a}^{-6}}.$ The
intermediate case $x\sim 1,\,p\sim 0$ corresponds to
non-relativistic matter.

Transforming (\ref{sec_6_7}) into the integral form
\begin{equation}\label{sec_6_11}
\rho ={{\rho }_{0}}\exp \left[ -3\int{\left( 1+{{w}_{\varphi }}
\right)\frac{da}{a}} \right].
\end{equation}
one finds that in the general case the scalar field energy density
behaves as
    \begin{equation}\label{sec_6_12}
\rho _\phi   \propto a^{ - m} ,\quad 0 < m < 6.
\end{equation}
The value ${{w}_{\varphi }}=-1/3$ represents boundary between the
regimes of decelerated and accelerated Universe expansions. Thus,
the case of accelerated expansion takes place for $0\le m<2.$ A
natural question arises: what are the potentials for scalar fields
that  can provide the accelerated expansion of Universe? The
question can be posed another way: what potentials can produce the
quintessence applicable for the role of dark energy? Consider
simplified version of the problem. Let us find the scalar field
potential resulting in power law for the scale factor growth:
\begin{equation}\label{sec_6_13}
a(t)\propto {{t}^{p}}.
\end{equation}
For accelerated expansion the condition $p>1$ must be satisfied.
Recall that $p=2/3$ for the case of non-relativistic matter
dominated Universe, and $p=1/2$ for radiation-dominated one, so in
both cases the expansion will be decelerated. Making use of the
Friedmann equations, one can express the potential $V(\varphi )$ and
$\dot{\varphi }$ in terms of $H$ and $\dot{H}.$ It allows to write
down the system of equations, which describe the parametric
dependence $V$ on $\varphi:$
\begin{subequations}
\begin{equation}\label{sec_6_14a}
V=\frac{3{{H}^{2}}}{8\pi G}\left( 1+\frac{{\dot{H}}}{3{{H}^{2}}}
\right);
\end{equation}
\begin{equation}\label{sec_6_14b}
\varphi ={{\int{dt\left( -\frac{{\dot{H}}}{4\pi G} \right)}}^{1/2}}.
\end{equation}
\end{subequations}
Excluding the time variable with the help of the relation
$\frac{\varphi }{{{m}_{Pl}}}=\sqrt{\frac{p}{4\pi }}\ln t,$ for the
case of the power law (\ref{sec_6_13}) one finds
\begin{equation}\label{sec_6_15}
V\left( \varphi  \right)={{V}_{0}}\exp \left( -\sqrt{\frac{16\pi
}{p}}\frac{\varphi }{{{m}_{Pl}}} \right).
\end{equation}
The obtained result implies that the scalar field in the potential
(\ref{sec_6_15}) under condition $p>1$ can be treated as the dark
energy, i.e. it provides accelerated expansion of Universe. As we
can see above, in the quintessence model the required dynamical
behavior is achieved by the choice of the scalar field potential.
Another dark energy model, realized by the scalar field with
modified kinetic term, is called the $k$-essence
\cite{k-Inflation,k-Inflation2}.

Let us define the quantity $X\equiv \frac{1}{2}{{g}^{\mu \nu
}}\frac{\partial \varphi }{\partial {{x}_{\mu }}}\frac{\partial
\varphi }{\partial {{x}_{\nu }}}$ and consider the action for the
scalar field of the form
\begin{equation}\label{sec_6_16}
    S=\int{{{d}^{4}}x\sqrt{-g}}L\left( \varphi ,X \right).
\end{equation}
where $L$ is generally speaking an arbitrary function of the
variables $\varphi$ and $X.$ The traditional scalar field action
corresponds to the case
\begin{equation}\label{sec_6_17}
    L\left( \varphi ,X \right)=X-V(\varphi ).
\end{equation}
We restrict our consideration to the following subset of
Lagrangians:
    \begin{equation}\label{sec_6_18}
L\left( \varphi ,X \right)=K(X)-V(\varphi ),
\end{equation}
where $K(X)$ is positive defined function of the kinetic energy $X.$
For description of homogeneous and uniform Universe we should choose
$X=\frac{1}{2}{{\dot{\varphi }}^{2}}.$ Using the standard definition
(\ref{sec_6_5}), one finds
\begin{subequations}
    \begin{equation}\label{sec_6_19a}
{{p}_{\varphi }}=L\left( \varphi ,X \right)=K(X)-V(\varphi );
\end{equation}
\begin{equation}\label{sec_6_19b}
{{\rho }_{\varphi }}=2X\frac{\partial K(X)}{\partial
X}-K(X)+V(\varphi ).
\end{equation}
\end{subequations}
Accordingly, the state equation for $k$-essence takes the form
    \begin{equation}\label{sec_6_20}
{{w}_{\varphi }}=\frac{K(X)-V(\varphi )}{2X\frac{\partial
K(X)}{\partial X}-K(X)+V(\varphi )}.
\end{equation}
We demonstrate the main features of $k$-essence in example of the
simplified model \cite{putter}, where the Lagrangian has the form
$L=F(X).$ Such model is called purely kinetic $k$-essence. In that
case one has
\begin{subequations}
   \begin{equation}\label{sec_6_21a}
    {{\rho }_{\varphi }}=2X{{F}_{X}}-F;\quad {{F}_{X}}\equiv \frac{\partial F}{\partial X};
\end{equation}
\begin{equation}\label{sec_6_21b}
     p=F;
\end{equation}
\begin{equation}\label{sec_6_21c}
    {{w}_{\varphi }}=\frac{F}{2X{{F}_{X}}-F}.
\end{equation}
\end{subequations}
The equations of motion for the field can be obtained either from
the Euler-Lagrange equation for the action (\ref{sec_6_16}), or by
substitution of density and pressure expressions (\ref{sec_6_21a}
and \ref{sec_6_21b}) respectively into the conservation equation for
the $k$-essence. It results in the following
\begin{equation}\label{sec_6_22}
{{F}_{X}}\ddot{\varphi }+{{F}_{XX}}{{\dot{\varphi
}}^{2}}\ddot{\varphi }+3H{{F}_{X}}\dot{\varphi }=0,
\end{equation}
or in terms of the kinetic energy $X$
\begin{equation}\label{sec_6_23}
\left( {{F}_{X}}+2{{F}_{XX}}X \right)\dot{X}+6H{{F}_{X}}X=0.
\end{equation}
The latter equation can be exactly solved
\begin{equation}\label{sec_6_24}
XF_{X}^{2}=k{{a}^{-6}}.
\end{equation}
where constant $k>0.$

The solution (\ref{sec_6_24}) $X(a)$ has one important property: all
principal characteristics of $k$-essence (${{\rho }_{\varphi
}},{{p}_{\varphi }},{{w}_{\varphi }};$ see (\ref{sec_6_21a} -
\ref{sec_6_21c})) as functions of the scale factor are completely
determined by the function $F(X)$ and do not depend on evolution of
other energy densities. All the dependence of $k$-essence on other
components appears only by means of $a(t).$ But this dependence is
trivially $a(t)\propto {{\rho }_{tot}},$ and it is determined by the
dominant component in the energy density. Solutions of such type are
called the tracker solutions and its occurrence allows to approach
the solution of the coincidence problem. It can be shown
\cite{Armendariz-Picon}, that such property is present not only in
purely kinetic $k$-essence, but also its general case.

Extensive set of available cosmological observations shows that the
parameter $w$ in the state equation for dark matter lies in narrow
vicinity of $w=-1.$ We considered above the region $-1\le w<-1/3.$
The lower bound of the region $w=-1$ corresponds to the cosmological
constant, and all remaining interval can be realized by scalar
fields with canonic Lagrangian. Recall that the upper bound $w=-1/3$
appears because of requirement to provide the observed accelerated
expansion of Universe. But can we go beyond the bounds of that
interval?

It is rather difficult question for the component of energy we know
so little about. General relativity usually impose certain
restrictions on possible values of the energy-momentum tensor
components, which are called the energy conditions(see section
\ref{SECIII}). One of the simplest restrictions of such type is the
condition $\rho +p\ge 0.$ Its physical motivation is to prevent the
instability of vacuum. Applied to dynamics of Universe this
condition requires that density of any allowed energy component
should not grow as Universe expands. The cosmological constant with
${{\dot{\rho }}_{\Lambda }}=0$ represents the limiting. Taking into
account our ignorance about the nature of dark energy, it is
reasonable to pose a question: may such a mysterious substance
differ so much from already known ''good'' energy sources and
violate the condition $\rho +p\ge 0$? Accounting the requirement of
positive defined energy density, valid even for dark one (it is
necessary to make the Universe flat), and negative pressure (in
order to explain the accelerated Universe expansion), the above
mentioned violation will result in $w<-1.$ Some time ago such
component, called the fantom energy, attracted attention of
physicists\cite{Caldwell12}. Action for the fantom field $\varphi,$
minimally coupled to gravity, differs only by sign of the kinetic
term from the canonic action for the scalar field. In that case the
energy density and pressure of the fantom field are determined by
the expressions
\begin{equation}\label{sec_6_25}
{{\rho }_{\varphi }}={{T}_{00}}=-\frac{1}{2}{{\dot{\varphi
}}^{2}}+V\left( \varphi  \right);\quad {{p}_{\varphi
}}={{T}_{ii}}=-\frac{1}{2}{{\dot{\varphi }}^{2}}-V\left( \varphi
\right),
\end{equation}
and state equation takes the form
\begin{equation}\label{sec_6_26}
{{w}_{\varphi }}=\frac{{{p}_{\varphi }}}{{{\rho }_{\varphi
}}}=\frac{{{{\dot{\varphi }}}^{2}}+2V(\varphi )}{{{{\dot{\varphi
}}}^{2}}-2V(\varphi )}.
\end{equation}

If ${{\dot{\varphi }}^{2}}<2V(\varphi ),$ then ${{w}_{\varphi
}}<-1.$ As an example let us consider the case of Universe,
containing only non-relativistic matter $(w=0)$ and phantom field
$({{w}_{\varphi }}<-1).$ Densities of the two components evolve
independently: ${{\rho }_{m}}\propto {{a}^{-3}}$ and ${{\rho }_{\varphi
}}\propto {{a}^{-3\left( 1+{{w}_{\varphi }} \right)}}.$ If matter
domination ends at the time moment ${{t}_{m}},$ then the solution
for the scale factor at $t>{{t}_{m}}$ reads
\begin{equation}\label{sec_6_27}
a(t)=a({{t}_{m}}){{\left[ -{{w}_{\varphi }}+(1+{{w}_{\varphi
}})\left( \frac{t}{{{t}_{m}}} \right)
\right]}^{\frac{2}{3(1+{{w}_{\varphi }})}}}.
\end{equation}

It immediately follows that for ${{w}_{\varphi }}<-1$ at the time
moment ${{t}_{BR}}=\frac{{{w}_{\varphi }}}{(1+{{w}_{\varphi
}})}{{t}_{m}}$ the scale factor and whole set of the cosmological
characteristics (such as scalar curvature and energy density of the
fantom field) for the Universe diverge. This catastrophe was called
the Big Rip, which is precede by a specific regime which is called
super-acceleration. Let us explain the origin of the
super-acceleration regime on a simple example. Consider differential
equation
    \begin{equation}\label{sec_6_28}
    \frac{dx}{dt}=A{{x}^{2}}.
\end{equation}
In the case $A>0$ the equation  (\ref{sec_6_24}) realizes the
non-linear positive feedback. Fast growth of the function $x(t)$
leads to the ``Big Rip'' (divergence of the function to infinity) on
finite time period. Indeed, the general solution of the equation
reads
\begin{equation}\label{sec_6_29}
x(t)=-\frac{1}{A(t+B)}.
\end{equation}
where $B$ is the integration constant. At $t=-B$ the Big Rip occurs.

It is easy to see that the model \eqref{sec_6_28} represents a
particular version of the Friedmann equation for ${{w}_{\varphi
}}<-1.$ Since ${{\rho }_{\varphi }}\propto {{a}^{-3(1+{{w}_{\varphi
}})}},$ then the first Friedmann equation can be presented in the
form
\begin{equation}\label{sec_6_30}
\dot{a}=A{{a}^{-\frac{3}{2}\left( 1+{{w}_{\varphi }} \right)+1}}.
\end{equation}
For instance, with ${{w}_{\varphi }}=-\frac{5}{3}$ the equation
(\ref{sec_6_30}) exactly coincides with (\ref{sec_6_29}).

\section{$f(R)$-Gravity, Braneworld Cosmology and MOND}
Although SCM can explain the present accelerated expansion of
Universe and well agrees with current observational data,
theoretical motivation for this model can be regarded as quite poor.
Consequently, there have been several attempts to propose dynamical
alternatives to dark energy. We considered them in the previous
section. Unfortunately, none of these attempts are problem-free. An
alternative and more radical approach is based on assumption that
there is no dark energy, and the acceleration is generated because
of gravity weakened on very large scales due to modification of
general relativity. In frames of such broad approach three main
directions can be selected: $f(R)$ gravity, braneworld cosmology and
modified Newtonian dynamics (MOND). Let us briefly consider those
alternatives to SCM from the point of view under our interest:
slowdown or speedup?
\subsection{$f(R)-$ gravity}
Theory of $f(R)$ gravity is created by direct generalization of the
Einstein-Hilbert action with the replacement $R\to f(R).$ The
transformed action reads
    \begin{equation}\label{f_r_1}
    S=\frac{1}{16\pi G }\int{{{d}^{4}}x\sqrt{-g}f(R)}.
\end{equation}
For the generalization we take the function $f(R)$ depending only on Ricci scalar $R,$ not including other invariants, such as ${{R}_{\mu
\nu }}{{R}^{\mu \nu }}.$ Motivation for such choice is the
following: the action with $f(R)$ is sufficiently generic in order
to reflect main features of the gravity, and with all that it is
simple enough to avoid technical difficulties in calculations. We
remark that the function $f(R)$ must satisfy the stability
conditions
\begin{equation}
{f}'(R)>0,\quad {f}''(R)>0,
\end{equation}
were the primes denote differentiation with respect to the scalar
Ricci curvature $R.$ Complete action for $f(R)$ gravity thus reads
\begin{equation}\label{f_r_2}
S=\frac{1}{2\kappa }\int{{{d}^{4}}x\sqrt{-g}f(R)+{{S}_{m}}\left(
{{g}_{\mu \nu }},\psi  \right)\quad }.
\end{equation}
Here $\psi $ is common symbol for all material fields. Variation
with respect to metrics after some manipulations gives
\begin{equation}\label{f_r_3}
{f}'(R){{R}_{\mu \nu }}-\frac{1}{2}f(R){{g}_{\mu \nu }}-\left(
{{\nabla }_{\mu }}{{\nabla }_{\nu }}-{{g}_{\mu \nu }}\square
\right){f}'(R)=8\pi G {{T}_{\mu \nu }},
\end{equation}
where
\begin{equation}\label{f_r_4}
{{T}_{\mu \nu }}=\frac{-2}{\sqrt{-g}}\frac{\delta {{S}_{m}}}{\delta
{{g}^{\mu \nu }}},
\end{equation}
and ${{\nabla }_{\mu }}$ is the covariant derivative, associated
with the Levi-Civita connection of the metric and $\square \,\equiv
{{\nabla }_{\mu }}{{\nabla }^{\mu }}.$

Leaving aside the complications connecting with variation procedure,
let us concentrate our attention on the field equations
(\ref{f_r_3}). They represent partial differential equation of
fourth order in metrics, as the Ricci scalar $R$ already contains
second order derivatives of the latter. For the action linear on
$R,$ the fourth-order derivatives (the two latter terms on the left
hand side of (\ref{f_r_3})) turn to zero and theory reduces to the
standard general relativity.

Note that the trace of (\ref{f_r_3}) gives
\begin{equation}\label{f_r_5}
{f}'(R)R-2f(R)+3\square {f}'(R)=8\pi G T,
\end{equation}
where $T={{g}^{\mu \nu }}{{T}_{\mu \nu }}$ links $R$ with $T$
differential relation, unlike algebraic one in general relativity,
where $R=-8\pi GT.$ This a direct hint on the fact that the field
equation of the $f(R)$-theory allow much wider set of solutions
compared to general relativity. To illustrate such statement we
remark that the Jebsen-Birkhoff's theorem, stating that the
Schwarzschild solution represents the unique spherically symmetric
vacuum solution, does not hold in the $f(R)$-theory. Leaving aside
the details, we note, that now $T=0$ does not require that $R=0$ or
even is constant.

The equation (\ref{f_r_5}) appears very useful for investigation of
different aspects of $f(R)$-gravity, especially for stability of
solutions and weak field limit. In particular it is conveniently
used for analysis of the so-called maximally symmetric solutions. It
is the solutions with $R=const.$ For $R=const$ and ${{T}_{\mu \nu
}}=0$ the equation (\ref{f_r_5}) reduces to
\begin{equation}\label{f_r_6}
    {f}'(R)R-2f(R)=0.
\end{equation}

For a given function $f(R)$ this equation is algebraic for $R.$ If
$R=0$ is a root of the equation, then (\ref{f_r_3}) reduces in that
case to ${{R}_{\mu \nu }}=0$ and maximally symmetric solution
represents the Minkovski space-time. If otherwise the root of
equation (\ref{f_r_6}) equals $R=const=C,$ then (\ref{f_r_3})
reduces to ${{R}_{\mu \nu }}=\frac{C}{4}{{g}_{\mu \nu }}$ and
maximally symmetric solution corresponds to the space of de Sitter
or anti-de Sitter (the cosmological constant in general relativity)
depending on sign of $C.$

Now let us consider how dynamics of Universe are immediately
described in $f(R)$ cosmology. Inserting FRW-metric into
(\ref{f_r_3}) and using the stress-energy tensor in the form
\begin{equation}\label{f_r_7}
{{T}_{\mu \nu }}=\left( \rho +p \right){{u}_{\mu }}{{u}_{\nu
}}-p{{g}_{\mu \nu }},
\end{equation}
one obtains
\begin{equation}\label{f_r_8}
{{H}^{2}}=\frac{8\pi G}{3{f}'}\left[ \rho +\frac{1}{2}\left( R{f}'-f
\right)-3H\dot{R}{f}'' \right];
\end{equation}
\begin{equation}\label{f_r_9}
2\dot{H}+3{{H}^{2}}=-\frac{8\pi G }{{{f}'}}\left[
p+{{{\dot{R}}}^{2}}{f}'''+2H\dot{R}{f}''+\ddot{R}{f}''+\frac{1}{2}\left(
f-R{f}' \right) \right].
\end{equation}
As was mentioned above, the main motivation for $f(R)$-gravity is in
the fact that it leads to accelerated expansion of Universe without
any dark energy. The easiest way to see it is to introduce the
effective energy density and effective pressure
    \begin{equation}\label{f_r_10}
    {{\rho }_{eff}}=\frac{R{f}'-f}{2{f}'}-\frac{3H\dot{R}{f}''}{{{f}'}};
\end{equation}
    \begin{equation}\label{f_r_11}
    {{p}_{eff}}=\frac{{{{\dot{R}}}^{2}}{f}'''+2H\dot{R}{f}''+\ddot{R}{f}''+\frac{1}{2}\left( f-R{f}' \right)}{{{f}'}}.
\end{equation}
In the spatially flat Universe the density ${{\rho }_{eff}}$ must be
non-negative, as it follows (\ref{f_r_8}) in the limit $\rho \to 0.$
Then the equations (\ref{f_r_8}), (\ref{f_r_9}) take the form of
standard Friedmann equations
    \begin{equation}\label{f_r_12}
    {{H}^{2}}=\frac{8\pi G }{3}{{\rho }_{eff}};
\end{equation}
\begin{equation}\label{f_r_12}
    \frac{{\ddot{a}}}{a}=-\frac{4\pi G }{3}\left( {{\rho }_{eff}}+3{{p}_{eff}} \right).
\end{equation}
The effective parameter ${{w}_{eff}}$ in the state equation in that
case equals
\begin{equation}\label{f_r_14}
{{w}_{eff}}\equiv \frac{{{p}_{eff}}}{{{\rho
}_{eff}}}=\frac{{{{\dot{R}}}^{2}}{f}'''+2H\dot{R}{f}''+\ddot{R}{f}''+\frac{1}{2}\left(
f-R{f}' \right)}{\frac{R{f}'-f}{2}-3H\dot{R}{f}''}.
\end{equation}
The denominator in (\ref{f_r_14}) is definitely positive, so sign of
${{w}_{eff}}$ is determined by the numerator. In general case of
metric $f(R)$-model in order to reproduce (mimic) the equation of
state for de Sitter (cosmological constant) with ${{w}_{eff}}=-1,$
the following condition must be satisfied
\begin{equation}\label{f_r_15}
    \frac{{{f}'''}}{{{f}''}}=\frac{\dot{R}H-\ddot{R}}{{{{\dot{R}}}^{2}}}.
\end{equation}
Let us consider two examples (regardless its realizability). The
firs is the function of the form $f(R)\propto {{R}^{n}}.$ We can
easily calculate ${{w}_{eff}}$ as function of $n,$ if we assume the
power law for time dependence of the scale factor
$a(t)={{a}_{0}}{{\left( t/{{t}_{0}} \right)}^{\alpha }}$ (arbitrary
dependence $a(t)$ results in time dependence for ${{w}_{eff}}$). The
result for $n\ne 1$ is
\begin{equation}\label{f_r_16}
    {{w}_{eff}}=-\frac{6{{n}^{2}}-7n-1}{6{{n}^{2}}-9n+3},
\end{equation}
and $\alpha $ in terms of $n$ reads
\begin{equation}\label{f_r_17}
    \alpha =\frac{-2{{n}^{2}}+3n-1}{n-2}.
\end{equation}
Appropriate choice of $n$ leads to desired value for ${{w}_{eff}}.$
For instance, $n=2$ leads to
\begin{equation}
    {{w}_{eff}}=-1,\quad \alpha =\infty.
\end{equation}
This result is expected from consideration of quadratic corrections
to Einstein-Hilbert action, which were used in the inflation scenery
by Starobinski \cite{starobin12}.

The second example is
\begin{equation}
f(R)=R-\frac{{{\mu }^{2\left( n+1 \right)}}}{{{R}^{n}}}.
\end{equation}
In that case, assuming again the power law for time dependence of
the scale factor, one gets
\begin{equation}\label{f_r_18}
    {{w}_{eff}}=-1+\frac{2(n+2)}{3(2n+1)(n+1)}.
\end{equation}
The condition for accelerated expansion $w_{eff}<-1/3$ for the two
above considered models is transformed into the form
\[n^2\mp n -1
>0,\] where the signs $\mp$ respond to the first and the second examples respectively.
In particular, in the second case, for $n=1$ we find that
${{w}_{eff}}=-2/3.$ We remark that in such case the positive values
of $n$ imply the presence of terms inversely proportional to $R,$
unlike the above considered case.
\subsection{Braneworld Cosmology}
All models explaining the accelerated expansion of Universe share
one common property: they involve one or another way to weaken the
gravity. It is the negative pressure in SCM and in the scalar field
models, transformation of the universal gravity law in the modified
gravitation model or voids in the models with inhomogeneities. An
original approach to suppress the gravity is realized in the
Braneworld scenario \cite{Randall}. According to latter we live on a
three-dimensional brane, which is embedded in the volume of higher
dimension (in the simplest case it is four-dimensional). All
material fields are restricted to the brane, while the gravitation
penetrates both the brane and whole embedding volume (see Lecture
notes by R.Maartens \cite{Maartens}). Higher dimension of the space
where the gravity acts, results in its suppression ($F\propto
{{R}^{-(D-1)}},$ where $D$ is the dimension of the space).

According such scenario the equation of motion for the scalar field
on the brane reads
\begin{equation}\label{bran_1}
    \ddot{\varphi }+3H\dot{\varphi }+{V}'(\varphi )=0.
\end{equation}
where
\begin{subequations}
\begin{equation}\label{bran_2a}
    {{H}^{2}}=\frac{8\pi }{3{{M_{Pl}}^{2}}}\rho \left( 1+\frac{\rho }{2\sigma } \right)+\frac{{{\Lambda }_{4}}}{3}+\frac{\varepsilon }{{{a}^{4}}};
\end{equation}
\begin{equation}\label{bran_2b}
     \rho =\frac{1}{2}{{{\dot{\varphi }}}^{2}}+V\left( \varphi  \right).
\end{equation}
\end{subequations}
Here $\varepsilon $ is the integration constant, which transforms
the volume gravity to the brane restricted one. The brane tension
$\sigma $ provides connection between the for-dimensional Planck
mass $M_{Pl}$ and five-dimensional one $M_{Pl}^{(5)}:$
\begin{equation}\label{bran_3}
   M_{Pl}=\sqrt{\frac{3}{4\pi }}\left( \frac{{{M_{Pl}^{(5)}}^{3}}}{\sqrt{\sigma }} \right).
\end{equation}
The tension $\sigma$ also links the four-dimensional cosmological
constant ${{\Lambda }_{4}}$ on the brane with the five-dimensional
volume one ${{\Lambda }_{b}}$ by:
\begin{equation}\label{bran_4}
    {{\Lambda }_{4}}=\frac{4\pi }{{{M_{Pl}^{(5)}}^{3}}}\left( {{\Lambda }_{b}}+\frac{4\pi }{3{{M_{Pl}^{(5)}}^{3}}}{{\sigma }^{2}} \right).
\end{equation}
Remark, that the equations (\ref{bran_2a} - \ref{bran_2b}) contain
an additional term $\propto {{\rho }^{2}}/\sigma ,$ whose presence
is connected with the conditions on the brane-volume boundary. The
latter term is responsible for dynamical enhancement of the decay
experienced by the scalar field while it rolls down the potential.
In such a case inflation regime can be realized even for those
potentials which are too steep to satisfy the slow-roll inflation
conditions in the standard approach. Indeed, the slow-roll
parameters in the braneworld models in the limit $V/\sigma \gg 1$
read:
\begin{equation}\label{bran_5}
    \varepsilon \simeq 4{{\varepsilon }_{FRW}}{{\left( \frac{V}{\sigma } \right)}^{-1}};\quad \eta \simeq 2{{\eta }_{FRW}}{{\left( \frac{V}{\sigma } \right)}^{-1}}.
\end{equation}
It follows now that the slow-roll inflation condition $\left(
\varepsilon ,\eta \ll 1 \right)$ is easier satisfied than in the FRW
cosmology under the condition $V/\sigma \gg 1.$ The inflation thus
can take place for very steep quintessence potentials, such as
$V\propto {{e}^{-\lambda \varphi }},\ V\propto {{\varphi }^{-\alpha
}}$ and others. This in turn let us hope that inflation and
quintessence can be generated by the same scalar field.

A radically different approach to provide the accelerated expansion
of Universe was proposed in papers \cite{Deffayet},
\cite{Deffayet2}, \cite{sahni21}.

The sharp distinction of the DDG model from the RS one is in the
fact that both the volume cosmological constant and brane tension
are set to zero, while a curvature term is introduced into the brane
action. The theory is based on the following action
\begin{equation}\label{bran_DDG}
    S={{M_{Pl}^{(5)}}^{3}}\int_{bulk}{R+{{M_{Pl}}^{2}}\int_{brane}{R+}}\int_{brane}{{{L}_{matter}}}.
\end{equation}
The physical sense of the term $\int_{brane}{R}$ is unclear.
Probably, it accounts for quantum effects caused by material fields,
which lead to the same term in the Einstein action, as was first
noted by A.D. Sakharov in its induced gravity theory\cite{Sakharov}.

The Hubble parameter in the DDG braneworld takes the form
\begin{equation}\label{bran_7}
H=\sqrt{\frac{8\pi G{{\rho
}_{m}}}{3}+\frac{1}{l_{c}^{2}}}+\frac{1}{{{l}_{c}}},
\end{equation}
where ${{l}_{c}}={{M_{Pl}}^{2}}/{{M_{Pl}^{(5)}}^{3}}$ is a new length scale, defined
by the 4D Planck mass $M_{Pl}$ and 5D one $M_{Pl}^{(5)}.$ An important property of
such a model is that the accelerated expansion of Universe does
require presence of dark energy. Instead, since gravity becomes five
dimensional on length scales $R>{{l}_{c}}=2{{H}^{-1}}{{\left(
1-{{\Omega }_{m}} \right)}^{-1}},$ one finds that the expansion of
the Universe is modified during late times instead of early times as
in RS model.

More general class of braneworld models, which includes both RS and
DDG models as particular cases, is described by the action
    \begin{equation}\label{bran_gen}
    S={{M_{Pl}^{(5)}}^{3}}\int_{bulk}{(R-2{{\Lambda }_{b}})+\int_{brane}{({{M_{Pl}}^{2}}R-2\sigma )+}}\int_{brane}{{{L}_{matter}}}.
\end{equation}
For $\sigma ={{\Lambda }_{b}}=0$ the action \eqref{bran_gen}
transforms into the DDG model action \eqref{bran_DDG}, and for $m=0$
it reduces to RS-model.

As was shown in \cite{sahni21}, the action (\ref{bran_gen})
describes the Universe, where the accelerated expansion era occurs
on later evolution stages with the Hubble parameter
\begin{equation}\label{bran_9}
    \frac{{{H}^{2}}(z)}{H_{0}^{2}}={{\Omega }_{m}}{{(1+z)}^{3}}+{{\Omega }_{\sigma }}+2{{\Omega }_{l}}\mp 2\sqrt{{{\Omega }_{l}}}\sqrt{{{\Omega }_{m}}{{(1+z)}^{3}}+{{\Omega }_{\sigma }}+{{\Omega }_{l}}+{{\Omega }_{{{\Lambda }_{b}}}}},
\end{equation}
where
\begin{equation}\label{bran_10}
    {{\Omega }_{l}}=\frac{1}{l_{c}^{2}H_{0}^{2}},\ {{\Omega }_{m}}=\frac{{{\rho }_{0m}}}{3{{M_{Pl}}^{2}}H_{0}^{2}},\ {{\Omega }_{\sigma }}=\frac{\sigma }{3{{M_{Pl}}^{2}}H_{0}^{2}},\ {{\Omega }_{{{\Lambda }_{b}}}}=-\frac{{{\Lambda }_{b}}}{6H_{0}^{2}}.
\end{equation}
Signs $\mp $ correspond to two possible ways to embed a brane in to
the volume. As it was in DDG model, ${{l}_{c}}\sim H_{0}^{-1}$ if
$M_{Pl}^{(5)}\sim 100 MeV.$ On shorter length scales $r\ll {{l}_{c}}$ and on
early times we recover the general relativity, while on large length
scales $r\gg {{l}_{c}}$ and long time periods the brane effects
start to be important. Indeed, setting $M_{Pl}^{(5)}=0\ ({{\Omega }_{l}}=0)$
the equation (\ref{bran_9}) reduces to the $\Lambda CDM$ model:
\begin{equation}\label{bran_11}
    \frac{{{H}^{2}}(z)}{H_{0}^{2}}={{\Omega }_{m}}{{(1+z)}^{3}}+{{\Omega }_{\sigma }}.
\end{equation}
while for the case $\sigma ={{\Lambda }_{b}}=0$ the relation
(\ref{bran_9}) reproduces the DDG model. An important feature of the
action (\ref{bran_gen}) is the fact that it generates the effective
state equation ${{w}_{eff}}\le -1.$ It can be easily seen
\cite{sahni21} from the expression for current value of the
effective parameter in the state equation
\begin{equation}\label{bran_12}
    {{w}_{0}}=\frac{2{{q}_{0}}-1}{3\left( 1-{{\Omega }_{m}} \right)}=-1\pm \frac{{{\Omega }_{m}}}{1-{{\Omega }_{m}}}\sqrt{\frac{{{\Omega }_{l}}}{{{\Omega }_{m}}+{{\Omega }_{\sigma }}+{{\Omega }_{l}}+{{\Omega }_{{{\Lambda }_{b}}}}}}.
\end{equation}
Taking the lower sign, one gets ${{w}_{0}}<-1.$

It is worth noting that in the latter model the accelerated Universe
expansion phase is the transient phenomenon, which comes to end when
the Universe returns to the matter dominated phase.

\subsection{MOND}
The so-called modified Newtonian dynamics (MOND) is sometimes
considered as an alternative to the DM (dark matter) concept. MOND
represents such modification of Newton physics, which allows to
explain the flat rotational curves for galaxies without attributing
to any assumptions on DM. MOND assumes that the second Newton law
$F=ma$ must be modified for sufficiently low accelerations
$(a\ll{{a}_{0}})$ so that
\begin{equation}\label{mond_1}
    \vec{F}=m\vec{a}\mu (a/{{a}_{0}}).
\end{equation}
where $\mu (x)=x$ if $x\ll1,$ and $\mu (x)=1$ if $x\gg1.$ It is easy
to see that such law leads to the modification of the traditional
formula for the gravitational acceleration
$\vec{F}=m{{\vec{g}}_{N}}\ ({{g}_{N}}=GM/{{r}^{2}}).$ Relation
between the ''correct'' acceleration and traditional Newtonian one
reads
\begin{equation}\label{mond_2}
    a=\sqrt{{{a}_{0}}{{g}_{N}}}.
\end{equation}
For a rotating point mass one gets $a={{v}^{2}}/r$ (this purely
kinematic relation is independent on choice of the dynamic model).
But here $a$ stands already for ''correct'' acceleration. It follows
that
\begin{equation}
    {{v}^{4}}=GM{{a}_{0}}.
\end{equation}
i.e. for sufficiently small accelerations the rotational curves for
isolated body of mass $M$ do not depend on radial distance $r,$
where the velocity is measured.  In other words, the considered
theory predicts not only flat rotational curves but also the fact
that individual halo, associated with a galaxy, has infinite
extension. This prediction may cause a serious problem for MOND, as
recent observations of galaxy-galaxy lensing support the result that
the maximum halo extension is about 0.5 Mpc. The value ${{a}_{0}},$
needed to explain the observations
\begin{equation}
    {{a}_{0}}\sim {{10}^{-8}}\,cm/{{s}^{2}}.
\end{equation}
is of the same order as $c{{H}_{0}}.$ It supports the hypothesis
that MOND ``can reflect the influence of cosmology on local particle
dynamics''.

Although the results of MOND agree very well with the observations
of individual galaxies, it is unclear whether they will be as well
successful for description of cluster structure, where strong
gravitational lensing points out considerably denser mass
concentration in the cluster center than was predicted by MOND.
Another difficulty connected with MOND, is the fact that it is quite
problematic to incorporate it in more general relativistic gravity
theory. Presently it is not clear what predictions are given by the
MOND-like theories for complicated gravity effects like the
gravitational lensing.

\section{Dynamics of Universe with interaction in the dark sector}
SCM considers dark matter and dark energy as independent components
of energy budget of Universe. The dark energy postulated in form of
cosmological constant, introduced as early as by Albert Einstein in
order to make possible the creation of stationary Universe model.
The assumed absence of interaction between the two components means
that the energy densities for each component obey independent
conservation equations
\[{{\dot{\rho }}_{i}}+3H\left( {{\rho }_{i}}+{{p}_{i}} \right)=0.\]

Coupling between the two components leads to modified evolution of
Universe. In particular, the energy density for non-relativistic
component will not evolve according the law ${{a}^{-3}},$ and the
dark energy density (in form of the cosmological constant) will not
remain constant any more. From one hand, such modification of the
theory opens for us new possibilities for solution of principal
problems in cosmology. So, for example, solution of the coincidence
problem\footnote{the problem is that during all the Universe history
the two densities decay by different laws, so it is necessary to
impose very strict limitations on their values in early Universe in
order to make them be of comparable order nowadays} reduces to
appropriate choice of interaction parameters, which can provide
satisfaction of the condition \[\frac{{{\Omega }_{de}}}{{{\Omega
}_{dm}}}={\mathrm O}(1),\] in the present time consistently with the
condition $\ddot{a}>0.$ From the other hand, introduction of the
interaction will modify the relations between the observable
parameters. In particular, the modification will affect the
fundamental relation between the photometric distance and the
redshift, which the evidence of accelerated expansion of Universe is
mainly based on. It imposes strict limitations both on the form of
interaction and its parameters.

Taking into account that dark matter and dark energy represent
dominant components in the Universe, then from point of view of
field theory it is naturally to consider interaction between them
\cite{InteractionQ1,InteractionQ2,InteractionQ3,InteractionQ4,InteractionQ5,InteractionQ6,InteractionQ7,0801.1565,ZKGuo}.
Appropriate interaction between the dark matter and dark energy can
facilitate solution of the coincidence problem
\cite{InteractionQ10,InteractionQ8,InteractionQ9}. Non-minimal
coupling in the dark sector can considerably affect the history of
cosmological expansion of Universe and evolution of density
fluctuations, thus modifying the rate of the cosmological structure
growth. The possibility of dark matter and dark energy to interact
with each other is widely discussed in recent literature
\cite{InteractionQ1,InteractionQ2,InteractionQ3,InteractionQ4,InteractionQ5,InteractionQ6,InteractionQ7,
InteractionQ10,InteractionQ8,InteractionQ9,0801.1565,ZKGuo}.
Different and independent data of many observations, such as
Wilkinson Microwave Anisotropy Probe, SNe Ia and BAO, were specially
analyzed in order to investigate the limitations on the intensity
and form of the interaction in the dark sector. Some researchers
also suggest \cite{BaldiClusters}, that dynamical equilibrium of
collapsing structures, such as clusters, will essentially depend on
form and sign of the interaction between dark matter and dark
energy.
\subsection{Model of Universe with time dependent cosmological constant}
The simplest example of the model with interacting DM and DE is the
cosmological model with decaying vacuum. Actually the
$\Lambda$(t)CDM cosmology represents one of the cases where
parameter $w$ for dark energy equals to $-1.$

This model is based on the assumption that the dark energy is
nothing that physical vacuum, and energy density of the latter is
calculated on the curved space background with subtraction of
renormalized energy density of physical vacuum in the flat space
\cite{0711.2686}. The resulting effective energy density of physical
vacuum depends on space-time curvature and decays from high initial
values in early Universe (at strong curvature) to almost zero
magnitude in the present time.

Due to the Bianchi identity, the decay of vacuum must be accompanied
by creation or mass increase of dark energy particles, which is
common property of the decaying vacuum models, or in more general
case, for the models with interacting dark matter and dark energy.

Now let us give the formulation of the $\Lambda$(t)CDM. In the
present section we will use the system of units where the reduced
Planck mass equals to unity: ${{M}_{Pl}}={{\left( 8\pi G
\right)}^{-1/2}}=1.$

For the case of flat Universe described by FRW-metric, the first
Friedmann \eqref{Frid_1} and the conservation equation can be
presented in the form
\[ \rho _{tot}  = 3H^2,~~\dot \rho _{tot}  + 3H(\rho _{tot}  + p_{tot} ) = 0,\]
where $\rho _{tot}  = \rho _m  + \rho _\Lambda,$  and taking into
account that $p_m  = 0$ and $p_\Lambda   = - \rho_\Lambda,$ the
conservation equation takes the form:
\begin{equation}\label{3.1.1}
    \dot \rho _m  + 3H\rho _m  =  - \dot \rho _\Lambda.
\end{equation}
The right hand side of the latter contains an additional term which
plays role of a source --- it is the decaying cosmological constant.
The problem with such approach is in fact that we end up with the
same number of equations, but acquire additional unknown function
$\Lambda(t).$ The above described approach to define the physical
vacuum density, though intuitively clear, faces certain principal
difficulties \cite{BertolamiLdec}. That is why a phenomenological
approach prevails in the literature. Below we present a simple but
exactly solvable model. Consider the case when $\Lambda$ depends on
time as \cite{Carneiro}:
\begin{equation}\label{3.1.2}
    \Lambda  = \sigma H.
\end{equation}
It is interesting to note that with the choice $\sigma \approx
m_{\pi }^{3}$ (${{m}_{\pi }}$ is the energy scale of the QCD vacuum
condensation) the relation (\ref{3.1.2}) gives the value of
$\Lambda$ which is very close to the observed one.

For the considered components the first Friedmann equation takes the
form
\begin{equation}\label{lamd_4}
{{\rho }_{\gamma }}+\Lambda =3{{H}^{2}}.
\end{equation}
The equations (\ref{3.1.1}) and (\ref{lamd_4}) together with the
conservation equation
\[{{p}_{\gamma }}=w{{\rho }_{\gamma }}\equiv \left( \gamma -1 \right){{\rho }_{\gamma }},\]
and the decay law for the cosmological constant (\ref{3.1.1})
completely determine the scale factor time evolution. Combining
those equation, one obtains the evolution equation in the following
form
\[2\dot{H}+3\gamma {{H}^{2}}-\sigma \gamma H=0.\]

Under condition $H>0$ the solution for the latter equation reads
\[a(t)=C{{\left[ \exp \left( \sigma \gamma t/2 \right)-1 \right]}^{\frac{2}{3\gamma }}},\]
where $C$ is one of the two integration constants (recall that the
equation for the scale factor is of second order). The second
integration constant is determined from the condition $a(t=0)=0.$
Using such solution, one can find the densities of matter
(radiation) and time-dependent cosmological constant
\begin{equation}\label{lamd_8}
{{\rho }_{\gamma }}=\frac{{{\sigma }^{2}}}{3}{{\left( \frac{C}{a}
\right)}^{\frac{3}{2}\gamma }}\left[ 1+{{\left( \frac{C}{a}
\right)}^{\frac{3}{2}\gamma }} \right];
\end{equation}
\begin{equation}\label{lamd_9}
\Lambda =\frac{{{\sigma }^{2}}}{3}\left[ 1+{{\left( \frac{C}{a}
\right)}^{\frac{3}{2}\gamma }} \right].
\end{equation}
Let us now analyze the history of Universe expansion in the scenario
under consideration. In the radiation-dominated epoch $\gamma
=w+1=4/3$ and, therefore,
    \[a(t)=C{{\left[ {{e}^{\frac{2}{3}\sigma t}}-1 \right]}^{\frac{1}{2}}}.\]
For small values of time $\left( \sigma t\ll 1 \right)$ and we
recover the well-known dependence $a\propto {{t}^{1/2}},$
    \[a(t)=C\sqrt{\frac{2}{3}}\sigma {{t}^{\frac{1}{2}}}.\]
The densities ${{\rho }_{\gamma }}$(\ref{lamd_8}) and $\Lambda $
(\ref{lamd_9}) in the radiation dominated epoch transform into
\[{{\rho }_{\gamma }}\to {{\rho }_{r}}=\frac{{{\sigma }^{2}}{{C}^{4}}}{3}\frac{1}{{{a}^{4}}}+\frac{{{\sigma }^{2}}{{C}^{2}}}{3}\frac{1}{{{a}^{2}}};\]
\[\Lambda =\frac{{{\sigma }^{2}}}{3}+\frac{{{\sigma
}^{2}}{{C}^{2}}}{3}\frac{1}{{{a}^{2}}}.\] In the linit $a\to 0\
\left( t\to 0 \right)$ one has
    \[{{\rho }_{r}}=\frac{{{\sigma }^{2}}{{C}^{4}}}{3}\frac{1}{{{a}^{4}}}=\frac{3}{4{{t}^{2}}};\]
    \[\Lambda =\frac{{{\sigma }^{2}}{{C}^{2}}}{3}\frac{1}{{{a}^{2}}}=\frac{\sigma }{2t}.\]

Now consider the matter dominated era. In that case $\gamma =w+1=1$
and
    \[a(t)=C{{\left[ {{e}^{\frac{1}{2}\sigma t}}-1 \right]}^{\frac{2}{3}}}.\]
For $\sigma t\ll 1$ we reproduce the standard law for matter
evolution
    \[a(t)=C{{\left( \frac{\sigma }{2} \right)}^{\frac{2}{3}}}{{t}^{\frac{2}{3}}}.\]
For the densities ${{\rho }_{\gamma }}$ and $\Lambda $ in the matter
dominated era one finds
\[{{\rho }_{\gamma }}\to {{\rho }_{m}}=\frac{{{\sigma }^{2}}{{C}^{3}}}{3}\frac{1}{{{a}^{3}}}+\frac{{{\sigma }^{2}}{{C}^{3/2}}}{3}\frac{1}{{{a}^{3/2}}};\]
\[\Lambda =\frac{{{\sigma }^{2}}}{3}+\frac{{{\sigma }^{2}}{{C}^{3/2}}}{3}\frac{1}{{{a}^{3/2}}}.\]

Note that in the limit of long times $(\sigma t\gg 1)$ the scale
factor growth exponentially as
    \[a(t)=C{{e}^{\frac{\sigma }{3}t}}.\]

In the same limit the matter density tends to zero, and the
time-dependent function $\Lambda (t)$ transforms into the ''real''
cosmological constant.

In the matter dominated epoch the Friedmann equation (\ref{lamd_4})
can be presented in the form
    \[H(z)={{H}_{0}}\left[ 1-{{\Omega }_{m0}}+{{\Omega }_{m0}}{{(1+z)}^{3/2}} \right].\]
This expression can be used for calculation of the deceleration
parameter $\left( q=\frac{1+z}{H}\frac{dH}{dz}-1 \right),$
    \begin{equation}\label{lamd_22}
    q(z)=\frac{\frac{3}{2}{{\Omega }_{m0}}{{(1+z)}^{3/2}}}{1-{{\Omega }_{m0}}+{{\Omega }_{m0}}{{(1+z)}^{3/2}}}-1.
\end{equation}

One thus finds for the current value of the deceleration parameter
    \[q(z=0)=\frac{3}{2}{{\Omega }_{m0}}-1.\]
Therefore the accelerated expansion of Universe occurs under the
following condition
\[{{\Omega }_{m0}}<\frac{2}{3},\]
This condition is in fact satisfied for the observed value ${{\Omega
}_{m0}}\approx 0.23.$ From (\ref{lamd_22})  it follows that the
transition from the decelerated expansion to the accelerated one
took place at
    \[z^*={{\left[ 2\frac{1-{{\Omega }_{m0}}}{{{\Omega }_{m0}}} \right]}^{2/3}}-1.\]
This value $\left( z^*\approx 1.2 \right)$ exceeds (being however of
the same order ${\mathrm O}\,(1)$) the corresponding value for SCM
$\left( z^*\approx 0.75 \right),$ which is the result of the matter
production in the vacuum decay process.

Therefore the model based on the time-dependent cosmological
constant, where vacuum density linearly depends on Hubble parameter,
appears to be quite competitive. It sufficiently accurately
reproduces the ''canonic'' results, relative both to the radiation
dominated and matter dominated eras. The present Universe expansion
is accelerated according to the model under consideration.
Verification of the model with the latest observational data
obtained for SN1a leads to the results (e.g., $0.27\le {{\Omega
}_{m0}}\le 0.37 $), that agree very well with the currently accepted
estimates for the accelerated Universe expansion parameters.

\subsection{Interacting dark matter and dark energy}
One of the most interesting properties of dark matter is its
possible interaction with dark energy. Although the most realistic
models (SCM in particular) postulate that the dark matter and dark
energy are uncoupled, there is no serious evidence to consider such
assumption as a principle. In the present time many active
researches are provided to investigate the possible consequences of
such interaction
\cite{InteractionQ1,InteractionQ2,InteractionQ3,InteractionQ4,InteractionQ5,InteractionQ6,InteractionQ7,
InteractionQ10,InteractionQ8,InteractionQ9, 0801.1565,ZKGuo,
BaldiClusters, Chimento, Lima}. As we have mentioned above, the
interaction of dark components can at least soften some sharp
cosmological problems, such as the coincidence problem for example.
The dark energy density is approximately three times as high as the
dark matter one. Such coincidence could be explained if dark matter
was somehow sensitive to dark energy.

Remark that the possibility of interaction between the dark energy
in form of scalar field and dark matter lies in the basis of warm
inflation model. Unlike the cold inflation scenario, the former does
not assume that the scalar field in the inflation period is isolated
(uncoupled) from other fields. That is why instead of overcooled
Universe in the inflation period, in the Universe model under
discussion a certain quantity of radiation is always conserved,
which is sufficiently remarkable to manifest in post-inflation
dynamics.

Interaction of the scalar field quantum, responsible for the
inflation start -- the inflaton -- with other fields imply that its
master equation contains the terms describing the process of energy
dissipation from inflaton system to other particles. Barera and Fang
\cite{Berera} initially assumed that consistent description of the
inflaton field with energy dissipation requires the master equation
in form of the Langeven equation, where the fluctuation-dissipation
relation is present, which uniquely links the field fluctuations and
dissipated energy. Such equation lies in the the basis of
description of the warm inflation process.

Interaction between the components in the Universe must be
introduced in such a way that preserves the covariance of the
energy-momentum tensor $T^{\,\,\mu\nu}_{(tot)\,\,;\nu}=0,$ therefore
$T^{\,\,\mu\nu}_{_{DM}\,\,;\nu}=-T^{\,\,\mu\nu}_{_{DE}\,\,;\nu}\neq
0,$ where $u_\nu$ is the 4-velocity. The conservation equations in
that case take the form:
\begin{equation}\label{ref1}
 u_\nu T^{\,\,\mu\nu}_{_{DM}\,;\mu}
 =-u_\nu T^{\,\,\mu\nu}_{_{DE}\,;\mu}=
 -Q.
\end{equation}
For FRW-metric the equations \eqref{ref1} transform to:
\begin{eqnarray}
  &&\dot{\rho}_m+3H\rho_m=Q,\label{eq55}\\
 &&\dot{\rho}_{_{DE}}+3H\rho_{_{DE}}(1+w_{_{DE}})=-Q,\label{eq56}
\end{eqnarray}
where $\rho_m$ and $\rho_{_{DE}}$ are densities of dark matter and
dark energy respectively, $w_{_{DE}}$ is the parameter of state
equation for for dark energy, $H\equiv\dot{a}/a$ is the Hubble
parameter.
$$
Q~\left\{ \begin{array}{l} >0\\ <0 \end{array} \right. ~~
\Rightarrow ~~ \mbox{energy transits}~\left\{ \begin{array}{l} \mbox{DE $\to$ DM};\\
\mbox{DM $\to$ DE}\end{array} \right.
$$
If $Q < 0,$ then dark matter continuously decays into dark energy,
if $Q> 0$ then vice versa.
Note that the equations \eqref{eq55} and \eqref{eq56} obey the
conservation equation:
\begin{equation}\label{eq57}
     \dot{\rho}_{tot}+3H\rho_{tot}(\rho_{tot}+p_{tot})=0,
\end{equation}
where $\rho_{tot}=\rho_{_{DE}}+\rho_m$ is the total energy density.

The interaction between the dark matter and dark energy is
effectively equivalent to modification of the state equation for the
interacting components. Indeed, the equations \eqref{eq55} can be
transformed to the standard form of the conservation law written for
uncoupled components:
\[\begin{array}{l} {\dot{\rho }_{_{DE}} +3H\rho _{_{DE}} (1+w_{_{DE,eff}} )=0} \\ {\dot{\rho }_{_{DM}} +3H\rho _{_{DM,eff}} =0}, \end{array},\]
where
\begin{equation}\label{w_eff}w _{_{DE,eff}} =w_{_{DE}} -\frac{Q}{3H\rho_{_{DE}} } ;\quad w_{_{DM,eff}} =\frac{Q}{3H\rho _{_{DM}} }, \quad \end{equation}
play role of effective state equations for dark energy and dark
matter respectively.

As we do not know the nature of both the dark energy and dark
matter, we cannot derive the coupling $Q$ from the first principles
\cite{secondref}. However it is clear from dimension analysis that
this quantity must depend on the energy density of one of the dark
component, or a combination of both components with dimension of
energy density, times the quantity of inverse time dimension. For
the latter it is natural to take the Hubble parameter $H.$

Different forms of $Q$ the most often used in the literature, are
the following:
 \begin{equation}\label{Q}
    Q=3H\gamma \rho_{m},$~~ $Q=3H\gamma \rho_{_{DE}},$~~ $Q=3H\gamma (\rho_{m}+\rho_{_{DE}}).
 \end{equation}
For example we consider the simplest case for such model
\[
\begin{array}{l}
 \dot \rho _{m}  + 3H\rho _{m}  = \delta H\rho _{m},  \\
 \dot \rho _{_{DE}}  + 3H\left( {\rho _{_{DE}}  + p_{_{DE}} } \right) =  - \delta H\rho
 _{m},
 \end{array}
\]
where $\delta $ is the dimensionless coupling constant.

Integration of the latter equations yields
  \begin{equation}\label{Hrho_delta}
   \begin{array}{l}
     \rho _{_{DE}}  = \rho _{_{DE0}} a^{ - \left[ {3(1 + w_{_{DE}} ) + \delta } \right]} ; \hfill  \\
      \rho _{m}  = \frac{{ - \delta \rho _{_{DE0}} }}
{{3w_{_{DE}} + \delta }}a^{ - \left[ {3(1 + w_{_{DE}} ) + \delta }
\right]}  + \left( {\rho _{m0}  + \frac{{\delta \rho _{_{DE0}}
}} {{3w_{_{DE}} + \delta }}} \right)a^{ - 3}.
   \end{array}
  \end{equation}
Inserting the obtained densities into the first Friedmann equation
\eqref{Frid_1} and transforming from scale factor to redshift, one
gets for  $H^2(z)$ the following
\begin{equation}\label{H2_delta}
H^2 = \frac{(1+z)^3H_0^2}{3(3w_{_{DE}}+\delta)}\left[
3w_{_{DE}}\Omega_{DE}(1+z)^{(3w_{_{DE}}+\delta)}+\Omega_{m}(3w_{_{DE}}+\delta)\right].
\end{equation}
Note that in the considered case the deceleration parameter $q(z)$
can be easily determined by the following
\[ q(z)=\frac{1+z}{2 H^2}\frac{d H^2}{dz}-1,\] and direct substitution of \eqref{H2_delta} results in
explicit dependence of deceleration parameter on the redshift:
\begin{equation}
   q(z) =-1 +\frac{3}{2}\frac{ w_{_{DE}}\Omega_{DE}(3(1+w_{_{DE}})+\delta)(1+z)^{(3w_{_{DE}}+\delta)}+\Omega_{m}+\delta/w_{_{DE}}}
   {3w_{_{DE}}\Omega_{DE}(1+z)^{(3w_{_{DE}}+\delta)}+\Omega_{m}+\delta/w_{_{DE}}}.
  \end{equation}
With $w_{_{DE}}=-1$ and $\delta = 0,$ this expression coincides with
the deceleration parameter value obtained in SCM. Note that in the
considered model appropriate choice of the coupling parameter value
$\delta$ can essentially soften or even get rid of the compatibility
problem for the densities of dark matter and dark energy. Indeed,
using the densities \eqref{Hrho_delta}, one finds the following
relation $R=\rho_{m}/\rho _{_{DE}}:$
\[ R =-\frac{ \delta }{{3w_{_{DE}} + \delta }} + \left( R_0  + \frac{\delta}
{3w_{_{DE}} + \delta} \right)a^{3w_{_{DE}} + \delta}, \] where $R_0
=\rho_{m0}/\rho _{_{DE0}} $ is the ratio of the densities in the
present time. In SCM it is known to be $R \sim a^{-3},$ which
differs from the considered model on the $\delta$ in exponent.

Let us now consider the inverse problem, and instead of interaction
we postulate the ratio
\begin{equation}\label{eq76}
     \frac{\rho_{m}}{\rho_{_{DE}}}=f(a),
\end{equation}
where $f(a)$ is arbitrary differentiable function of scale factor.
Thus one obtains:
\begin{eqnarray}
&&\rho_m=\rho_{_{DE}}f(a),\label{eq66}\\
&&\rho_{_{DE}}=\frac{\rho_m}{f(a)},\label{eq67}\\
&&\dot{\rho}_m=\dot{\rho}_{_{DE}}f+\rho_{_{DE}}f^\prime \dot{a},\label{eq68}\\
&&\dot{\rho}_{_{DE}}=\frac{\dot{\rho}_m}{f}-
\frac{\rho_m\dot{a}f^\prime}{f^2}.\label{eq69}
\end{eqnarray}
Inserting the expressions \eqref{eq68} and \eqref{eq66} into the
equation \eqref{eq55} we obtain:
\begin{equation}\label{eq700}
    \dot{\rho}_{_{DE}}f+\rho_{_{DE}}f^\prime\dot{a}+3H\rho_{_{DE}}f=Q,
\end{equation}
where prime denotes the differentiation with respect to the scale
factor. Using here the expression for $\dot{\rho}_{_{DE}},$ obtained
from \eqref{eq56}, one gets:
\begin{equation}\label{eq70}
    Q=\frac{H\rho_{_{DE}}f}{1+f}\left(\frac{f^\prime a}{f}-3w_{_{DE}}\right).
\end{equation}
The first Friedmann equation takes on the form:
\begin{equation}\label{eq701}
3H^2=\rho_{_{DE}}+\rho_m=\rho_{cr},
\end{equation}
where $\rho_{cr}$ is the critical density. Accordingly one can write
down that
\begin{eqnarray} &&\Omega_{_{DE}}=\frac{\rho_{_{DE}}}{\rho_{cr}}=\frac{1}{1+f};\label{eq72}\\
 &&\Omega_m=\frac{\rho_m}{\rho_{cr}}=\frac{f}{1+f}.\label{eq730}
\end{eqnarray}
Now, substituting \eqref{eq730} into \eqref{eq70}, we finally obtain
the expression for $Q$:
\begin{equation}\label{eq71}
Q=H\rho_{_{DE}}\Omega_m\left(\frac{f^\prime a}{f}-3w_{_{DE}}\right)
=H\rho_m\Omega_{_{DE}}\left(\frac{f^\prime a}{f}-3w_{_{DE}}\right).
\end{equation}
It is worth noting that in the case when $f(a)\propto a^\xi$ we
obtain the expression
\begin{equation}\label{eq60}
Q=H\rho_m\Omega_{_{DE}}\left(\xi-3w_{_{DE}}\right)
 =H\rho_{_{DE}}\Omega_m\left(\xi-3w_{_{DE}}\right).
\end{equation}
From the equation \eqref{eq57} it follows that:
\begin{equation}\label{eq702}
    \frac{d \ln\rho_{tot}}{d\ln a}=-3(1+w_{eff}),
\end{equation}
where
\begin{equation}\label{eq73}
w_{eff}=\frac{p_{tot}}{\rho_{tot}}
=\frac{\rho_{_{DE}}w_{_{DE}}}{\rho_{_{DE}}+\rho_m}
=\frac{w_{_{DE}}}{1+f}=\Omega_{_{DE}}w_{_{DE}},
\end{equation}
and therefore
\begin{equation}\label{eq74}
      \frac{d \ln\rho_{tot}}{d\ln a}=-3(1+\Omega_{_{DE}}w_{_{DE}}).
\end{equation}
From the latter equation we obtain:
\begin{equation}\label{eq75}
\rho_{tot}=Ca^{-3}\exp\left(-\int 3\Omega_{_{DE}}w_{_{DE}}\,d\ln a
 \right)
\end{equation}
where $C$ is the integration constant determined from the
requirement that $\rho_{tot}(a=1)=\rho_{tot,0}=3H_0^2/(8\pi G).$
Using the expression for $\rho_{tot},$ one can easily obtain from
the Friedmann equations the relation for $E\equiv H/H_0,$ which is
used to verify cosmological models and establish restrictions on
cosmological parameters. Expression for
 $\rho_{_{DE}}=\Omega_{_{DE}}\rho_{tot}$ and $\rho_m=\Omega_m\rho_{tot}$ can also easily obtained from the equations
 \eqref{eq72} and \eqref{eq75}.

At last it is worth noting that one can obtain from the equation
\eqref{eq76},
\begin{equation}\label{eq77}
     f_0=f(a=1)=\frac{\rho_{m0}}{\rho_{_{DE0}}}
 =\frac{\Omega_{m0}}{1-\Omega_{m0}}=\frac{1}{\Omega_{_{DE0}}}-1.
\end{equation}
Let it be now
\begin{equation}\label{eq79}
     f(a)=f_0\,a^\xi,
\end{equation}
where $\xi$ is a constant and $f_0$ is defined above. Substituting
the equation (\ref{eq79}) into the equation (\ref{eq75}) and
requiring that $\rho_{tot}(a=1)=\rho_{tot,0},$ we can determine the
integration constant and find that
\begin{equation}\label{eq80}    \rho_{tot}=\rho_{tot,0}\,a^{-3}\left[\,\Omega_{m0}+
\left(1-\Omega_{m0}\right)a^\xi\,\right]^{-3w_{eff}/\,\xi}.
\end{equation}
Inserting this expression into the Friedmann equation, one finds:
\begin{equation}\label{eq81}
\begin{gathered}
E^2=\frac{H^2}{H_0^2}=a^{-3}\left[\,\Omega_{m0}+\left(1-\Omega_{m0}\right)a^\xi\,
 \right]^{-3w_{_X}/\,\xi}\\
 =(1+z)^3 \left[\,\Omega_{m0}+\left(1-\Omega_{m0}\right)
 (1+z)^{-\xi}\,\right]^{-3w_{_X}/\,\xi}.
\end{gathered}
\end{equation}
Using the above obtained formulae, it is easy to find the following
expression for the deceleration parameter in the considered model
\begin{equation}\label{q_f(a)}
q = 1+\frac{3}{2}\left(w_{_{DE}}\Omega_{_{DE}} +
w_m\Omega_{m}\right),
\end{equation}
where relative densities $\Omega_{_{DE}},~\Omega_{_{DM}}$ has the
following form
\begin{equation}\label{q_f(a)}
\Omega_{_{DE}}=
\frac{\Omega_{_{DE0}}}{\Omega_{_{DE0}}+(1-\Omega_{_{DE0}})a^{\xi}},~~~\Omega_{_{DM}}=
\frac{(1-\Omega_{_{DE0}})a^{\xi}}{\Omega_{_{DE0}}+(1-\Omega_{_{DE0}})a^{\xi}},
\end{equation}
and we finally obtain
\begin{equation}\label{q_f(a)_fin}
q =
\frac{1}{2}+\frac{3}{2}\frac{w_{_{DE}}\Omega_{_{DE0}}(1+z)^{\xi}}{1-\Omega_{_{DE0}}+\Omega_{_{DE0}}(1+z)^{\xi}},
\end{equation}

As the dark energy density is not constant in the considered model,
unlike the SCM, it facilitates solution of the coincidence problem
under condition  $\xi<3.$ In the model under consideration the
accelerated expansion phase starts at the redshift $z^*,$ which is
defined by the relation $q(z^*)=0,$ and it equals to
\begin{equation}\label{z_f(a)_fin}
  z^*=\left(\frac{1-\Omega_{_{DE0}}}{{(1+3w_{_{DE}}})\Omega_{_{DE0}}}\right)^{\frac{1}{\xi}} -1.
\end{equation}
This relation naturally generalizes the expression \eqref{z_trSCM},
obtained for SCM. The difference  of the value \eqref{z_f(a)_fin}
from the corresponding one for SCM depends on magnitude of the
distinction of the parameters $\xi$ and $w_{_{DE}}$ from 3 and -1
respectively. In the case of zero difference the quantities
\eqref{z_trSCM} and \eqref{z_f(a)_fin} coincide. Note that at $z\to
\infty$ Universe expanded with deceleration $q\to \frac12,$ the case
$z\to -1$ responds to $q\to \frac12-\frac23 w_{_{DE}}.$ Therefore,
as it was expected, dynamics of the considered model is
asymptotically (at $z\to \infty$ and $z\to -1$) identical to the
case of two-component Universe, filled by non-relativistic matter
and dark energy with the state equation $p = w_{_{DE}}\rho.$

\subsection{Cosmological models with the new type of interaction}
In the present subsection we consider one more type of interaction
$Q,$ \cite{HaoWei_Q(q)} which change its sign, i.e. direction of
energy transfer, when the expansion changes from decelerated regime
to accelerated one and vice versa.

Some recently appeared papers \cite{r14} make attempt to determine
both the very possibility of interaction in the dark sector and
specify its form and sign, basing solely on analysis o observational
data. The analysis splits all available set of redshift data $z$
into some parts, within which the function $\delta(z)=Q/(3H)$ is
considered to be constant. This analysis allowed to establish that
the function $\delta(z)$ the most likely takes zero values
$(\delta=0)$ in the interval of redshifts $0.45\lesssim z\lesssim
0.9.$ It turns out that this remarkable result produces new
problems. Indeed, as we have already mentioned, the researchers
mostly consider the interaction of the form \eqref{Q}: $Q=3H\gamma
\rho_{m},~ Q=3H\gamma \rho_{_{DE}},~ Q=3H\gamma
(\rho_{m}+\rho_{_{DE}})$ and for the model under consideration it is
always either positive or negative defined, and therefore it cannot
change its sign. Sign changes are possible only in the case when
$\gamma=f(t),$ which can change the sign of $Q$ or  $Q=3H (\alpha
\rho_{m}+\beta\rho_{_{DE}}),$ where $\alpha$ and $\beta$ have
opposite signs.

As authors of \cite{r14} mention, solution of such problem requires
to introduce a new type of interaction, which can change its sign
during the evolution of Universe.

An interaction $Q$ of such type was proposed in the paper
\cite{HaoWei_Q(q)}, where its cosmological consequences were
considered. As it was noted in the above cited paper, the redshift
interval, where the function $\delta(z)$ must change its sign,
includes the transition point when the Universe expansion stopped to
decelerate and started the acceleration (see \eqref{z_trSCM}).
Therefore the simplest type of the interaction which can explain the
above mentioned property is the case when the source $Q$ is
proportional to the deceleration parameter $q:$
\begin{equation}
\label{Q(q)}
  Q=q(\alpha\dot{\rho}+3\beta H\rho)\,
\end{equation}
where $\alpha$ and $\beta$ are dimensionless constants, and the sign
of $Q$ will change with the transition of universe from the
decelerated expansion stage $(q>0)$ to the accelerated one $(q<0).$
The authors also consider the case
\begin{eqnarray}
&&Q=q(\alpha\dot{\rho}_m+3\beta H\rho_m),\label{q_eq5}\\
 &&Q=q(\alpha\dot{\rho}_{tot}+3\beta H\rho_{tot}),\label{q_eq6}\\
 &&Q=q(\alpha\dot{\rho}_{_{DE}}+3\beta H\rho_{_{DE}})\label{q_eq7}.
\end{eqnarray}
The paper \cite{HaoWei_Q(q)1} considers a model of Universe with
decaying cosmological constant
$$
 \dot{\rho}_\Lambda=-Q\,.
$$
The Friedmann and Raychaudhuri equation take thus the form
\begin{eqnarray}
   H^2 &=& \frac{\kappa^2}{3}\rho_{tot}=  \frac{\kappa^2}{3}(\rho_\Lambda+\rho_m)\,,\label{q-eq6}\\
   \dot{H} &=&-\frac{\kappa^2}{2}(\rho_{tot}+p_{tot})=
 -\frac{\kappa^2}{2}\rho_m,\label{q-eq7}
 \end{eqnarray}
where $\kappa^2\equiv 8\pi G.$ Following the paper
\cite{HaoWei_Q(q)1}, in the succeeding subsections we consider
cosmological models with interaction of the type
\eqref{q_eq5}-\eqref{q_eq7}.
\subsubsection{Case  $Q=q(\alpha\dot{\rho}_m+3\beta H\rho_m)$}
For the beginning we consider the case when the interaction takes
the form \eqref{q_eq5} and insert such expression into the
conservation equation \eqref{eq55}, resulting in the following
 \begin{equation}\label{q-eq19}
 \dot{\rho}_m=\frac{\beta q-1}{1-\alpha q}\cdot 3H\rho_m\,.
\end{equation}
Substituting the obtained expression into the equation
\eqref{q_eq5}, we finally get
\begin{equation}\label{q-eq20}
 Q=\frac{\beta-\alpha}{1-\alpha q}\cdot 3qH\rho_m.
\end{equation}
From the equation \eqref{q-eq7} one obtains
 \begin{equation}\label{q-eq21}
\rho_m=-\frac{2}{\kappa^2}\dot{H}.
\end{equation}
Inserting it into the equation \eqref{q-eq19} one finds that
 \begin{equation}\label{q-eq22}
\ddot{H}=\frac{\beta q-1}{1-\alpha q}\cdot 3H\dot{H}\,,
\end{equation}
Thus we obtained the second order differential equation for the
function $H(t).$ Now transform from the time derivative to
differentiation with respect to the scale factor (denoted by the
prime ${}^\prime$), then the equation \eqref{q-eq22} takes on the
form
 \begin{equation}\label{q-eq23}
 aH^{\prime\prime}+\frac{a}{H}H^{\prime\,2}+H^\prime=
 \frac{\beta q-1}{1-\alpha q}\cdot 3H^\prime\, .
\end{equation}
This expression represents a second order differential expression
for the function $H(a).$ Note that the deceleration parameter
\[
 q=-1-\frac{\dot{H}}{H^2}=-1-\frac{a}{H}H^\prime\,,
\]
is also function of $H$ and $H^\prime$ and, except the case
 $\alpha\not=0,$ the equation has no exact solution and it
 represents a transcendental differential equation of second order.
 That is why we consider solely the case $\alpha=0.$ Thus the interaction \eqref{q_eq5} takes the form
 \[
 Q=3\beta qH\rho_m.
\]
With $\alpha=0$ the solution \eqref{q-eq23} can be presented in the
form
  \begin{equation}\label{q-eq26}
 H(a)=C_{12}\left[\,3C_{11}(1+\beta)-(2+3\beta)
 \,a^{-3(1+\beta)}\,\right]^{1/(2+3\beta)},
 \end{equation}
where $C_{11}$ and $C_{12}$ are the integration constants determined
below. We find the relative density of dark matter as the following
 \begin{equation}\label{q-eq27}
 \Omega_m\equiv \frac{\kappa^2 \rho_m}{3H^2}
 =-\frac{2\dot{H}}{3H^2}=-\frac{2aH^\prime}{3H}\,.
  \end{equation}
Inserting the equation \eqref{q-eq26} into \eqref{q-eq27}, one gets
\begin{equation}\label{q-eq28}
 \Omega_m=\frac{2\left(1+\beta\right)}{2+3\beta
 -3C_{11}\left(1+\beta\right)\,a^{3\left(1+\beta\right)}}\,.
 \end{equation}
With the requirements $\Omega_m(a=1)=\Omega_{m0}$ and $H(a=1)=H_0,$
the integration constants take the form
 \begin{eqnarray}
   C_{11} &=& \frac{\Omega_{m0}(2+3\beta)
 -2(1+\beta)}{3\Omega_{m0}(1+\beta)}, \label{q-eq29}\\
   C_{12} &=& H_0\left[3C_{11}(1+\beta) -(2+3\beta)\right]^{-1/(2+3\beta)}. \label{q-eq30}
 \end{eqnarray}

Substitution of the expressions \eqref{q-eq29} and \eqref{q-eq30}
into the equation \eqref{q-eq26} finally gives the result
 \begin{equation}\label{q-eq31}
 E\equiv\frac{H}{H_0}=\left\{1-
 \frac{2+3\beta}{2(1+\beta)}\,\Omega_{m0}\left[1-
 (1+z)^{3(1+\beta)}\right]\right\}^{1/(2+3\beta)}.
 \end{equation}
The model contains two free parameters $\Omega_{m0}$ and
 $\beta.$ We note that if $\beta=0,$ then the equation \eqref{q-eq31} reduces to
 $E(z)=\left[\Omega_{m0}(1+z)^3+\left(1-\Omega_{m0}\right)
 \right]^{1/2},$ which is equivalent to $\Lambda$CDM model.
Using the relation \[q(z)=
-\frac{(1+z)}{E(z)}\frac{d}{dz}\left(\frac{1}{E(z)}\right)-1,\] one
finds the dependence of deceleration parameter on the redshift in
the considered model
\begin{equation}\label{Q(q)_1}
   q(z)= -1+\frac{3}{2}\Omega_{m0}\frac{(1+z)^{3(1+\beta)}}{E^{(2+3\beta)}}.
\end{equation}
The effective parameter of the state equation is known to equal
\[w_{\rm eff}\equiv \frac{p_{tot}}{\rho_{tot}}=\frac{(2q-1)}{3}.\]
 \begin{center}
 \begin{figure}[tbhp]
 \centering
 \includegraphics[width=1.0\textwidth]{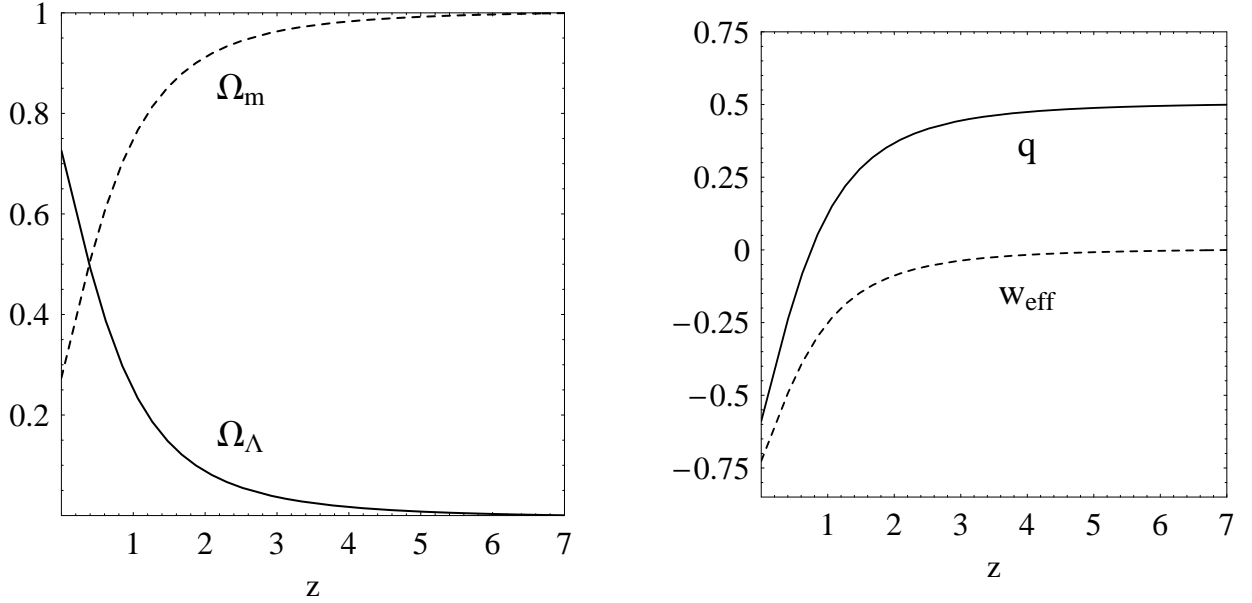}
 \caption{\label{q-fig2} $\Omega_m,$ $\Omega_\Lambda,$ $q$ and
 $w_{\rm eff}$ as function of the redshift $z$ at $\Omega_{m0}=0.2738$ and $\beta=-0.010$ in the case $Q=3\beta qH\rho_m.$}
 \end{figure}
 \end{center}
 The figure \ref{q-fig2} presents the plots for dependence of some
 cosmological parameters on the redshift $z.$ The free parameters $\Omega_{m0}$
 and $\beta$ were chosen to provide the best agreement with observations.
One can find that in the considered model the transition from the
decelerated expansion $(q>0)$ to the accelerated one $(q<0)$ took
place at $z_t=0.7489,$ the parameter $\beta$ is negative and
therefore dark matter decays into dark energy when $z>z_t,$ and vice
versa at $z<z_t.$ The Universe is interaction-free in dark sector at
$z_t.$
\subsubsection{Case
 $Q=q(\alpha\dot{\rho}_{tot}+3\beta H\rho_{tot})$}
Now we consider the case \eqref{q_eq6}, and proceeding completely
analogous the above considered case one obtains
\begin{equation}\label{q-eq32}
    Q=\frac{3qH^3}{\kappa^2}\left(2\alpha\frac{\dot{H}}{H^2}
 +3\beta\right).
\end{equation}
Inserting the equations \eqref{q-eq21} and \eqref{q-eq32} into
\eqref{eq55}, and transforming to differentiation with respect to
the scale factor, one finds
 \begin{equation}\label{q-eq34}
 aH^{\prime\prime}+\frac{a}{H}H^{\prime\,2}+
 \left(4+3\alpha q\right)H^\prime+\frac{9\beta qH}{2a}=0\,.
\end{equation}
As in the previous case, we obtained again the differential equation
of second order for the function $H(a).$ Exact solution exists only
in the case $\alpha=0:$
 \begin{equation}\label{q-eq36}
 H(a)=C_{22}\cdot a^{-3(2-3\beta+r_1)/8}\cdot\left(a^{3r_1/2}+C_{21}\right)^{1/2},
 \end{equation}
where $C_{21},$ $C_{22}$ are integration constants and  $
r_1\equiv\sqrt{4+\beta\left(4+9\beta\right)}.$ Inserting
\eqref{q-eq36} into \eqref{q-eq27}, we get
 \begin{equation}\label{q-eq38}
 \Omega_m=\frac{1}{4}\left[2-3\beta
 +\left(\frac{2C_{21}}{a^{3r_1/2}+C_{21}}-1\right)r_1\right].
 \end{equation}
The integration constants are determined as usual from the condition
$\Omega_m(a=1)=\Omega_{m0},~H(a=1)=H_0:$
 \begin{equation}\label{q-eq39}
 C_{21}=-1+\frac{2\,r_1}{2-3\beta-4\Omega_{m0}+r_1},~~C_{22}=H_0\left(1+C_{21}\right)^{-1/2}.
 \end{equation}
We finally get
 \begin{equation}\label{q-eq41}
 E\equiv\frac{H}{H_0}=(1+z)^{3(2-3\beta+r_1)/8}\cdot
 \left[\frac{(1+z)^{-3r_1/2}+C_{21}}{1+C_{21}}\right]^{1/2}.
 \end{equation}
In the considered model there are also two free parameters
$\Omega_{m0}$ and $\beta.$ Using the condition $0\leq\Omega_m\leq 1$
with $a\to 0,$ from the equation \eqref{q-eq38} one gets $\beta\geq
0.$ The best agreement of the model under consideration with
observational data occurs with $\Omega_{m0}=0.2701$ and $\beta=0.0.$
It means that the considered interaction model gives worse agreement
with observations than $\Lambda CDM.$ More detailed discussion an
interested reader finds in the paper \cite{HaoWei_Q(q)1} by the
author of the considered model. The transition form the decelerated
expansion phase $(q>0)$ to the accelerated one $(q<0)$ occurs at
$z_t=0.7549.$
 \begin{center}
 \begin{figure}[tbhp]
 \centering
 \includegraphics[width=1.0\textwidth]{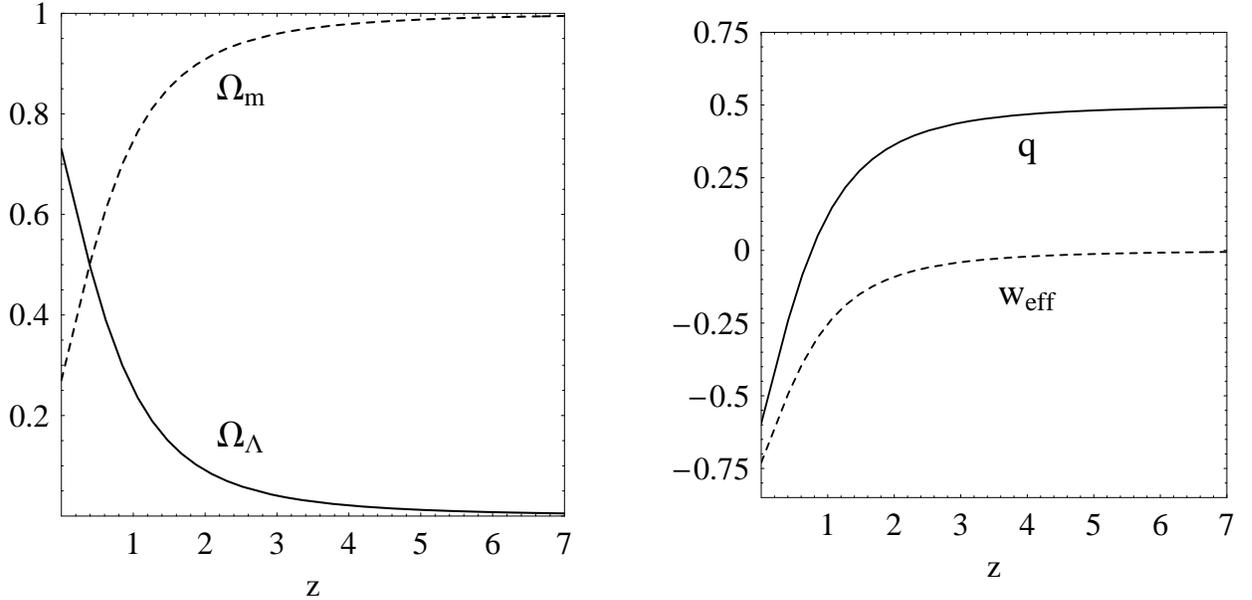}
 \caption{\label{q-fig4} The same as on Fig.~\ref{q-fig2}, but in the case of interaction of the form
 $Q=3\beta qH\rho_{tot}$ under condition $\beta\geq 0.$}
 \end{figure}
 \end{center}
\subsubsection{The case of
 $Q=q(\alpha\dot{\rho}_\Lambda+3\beta H\rho_\Lambda)$}
For the conclusion we consider the case \eqref{q_eq7}. Following the
same procedure as in the two preceding case, one obtains
 \begin{equation}\label{q-eq43}
 Q=\frac{3\beta qH\rho_\Lambda}{1+\alpha q}\,.
 \end{equation}
With the equation \eqref{q-eq21} one has
 \begin{equation}\label{q-eq44}
 \rho_\Lambda=\frac{3}{\kappa^2}H^2-\rho_m=
 \frac{1}{\kappa^2}\left(3H^2+2\dot{H}\right).
\end{equation}
Therefore the equation for the Hubble parameter in terms of the
scale factor takes the form:
 \begin{equation}\label{q-eq46}
 aH^{\prime\prime}+\frac{a}{H}H^{\prime\,2}
 +\left(4+\frac{3\beta q}{1+\alpha q}\right)H^\prime+
 \frac{9\beta qH}{2a(1+\alpha q)}=0.
\end{equation}
Exact solution can be obtained in the case, $ Q=3\beta
qH\rho_\Lambda,$ namely
\begin{equation}\label{q-eq48}
 H(a)=C_{32}\cdot a^{-3(2-5\beta+r_2)/[4(2-3\beta)]}\cdot
 \left(a^{3r_2/2}+C_{31}\right)^{1/(2-3\beta)},
\end{equation}
where $C_{31},$ $C_{32}$ are the integration constants, and
$r_2\equiv\sqrt{\left(2-\beta\right)^2}=\left|\,2-\beta\,\right|\,.$
Inserting \eqref{q-eq48} into \eqref{q-eq27}, we get
\begin{equation}\label{q-eq50}
 \Omega_m=\frac{1}{2\left(2-3\beta\right)}\left[2-5\beta+
 \left(\frac{2C_{31}}{a^{3r_2/2}+C_{31}}-1\right)r_2\right]\,.
 \end{equation}
Assuming that $\Omega_m(a=1)=\Omega_{m0}$ and $H(a=1)=H_0,$ we can
write
 \begin{equation}\label{q-eq51}
 C_{31}=-1+\frac{2\,r_2}{2-5\beta+
 r_2+2\Omega_{m0}\left(3\beta-2\right)},~~C_{32}=H_0\left(1+C_{31}\right)^{1/(3\beta-2)},
  \end{equation}
and finally get
 \begin{equation}\label{q-eq53}
 E\equiv\frac{H}{H_0}=(1+z)^{3\left(2-5\beta+r_2\right)/\left[
 4\left(2-3\beta\right)\right]}\cdot\left[\frac{(1+z)^{-3r_2/2}
 +C_{31}}{1+C_{31}}\right]^{1/\left(2-3\beta\right)}.
\end{equation}
As before, the model has two free parameters $\Omega_{m0},\beta.$
The maximum plausibility method for the free parameters of the
considered model gives the result \cite{HaoWei_Q(q)1}
$\Omega_{m0}=0.2717,~\beta=0.0136.$ Unlike the two preceding models,
the observational data analyzed in \cite{HaoWei_Q(q)1}, give
evidence in favor of $\beta>0.$ More detailed discussion the
interested reader can find in the paper \cite{HaoWei_Q(q)1} of the
author of the model.
 \begin{center}
 \begin{figure}[tbp]
 \centering
 \includegraphics[width=1.0\textwidth]{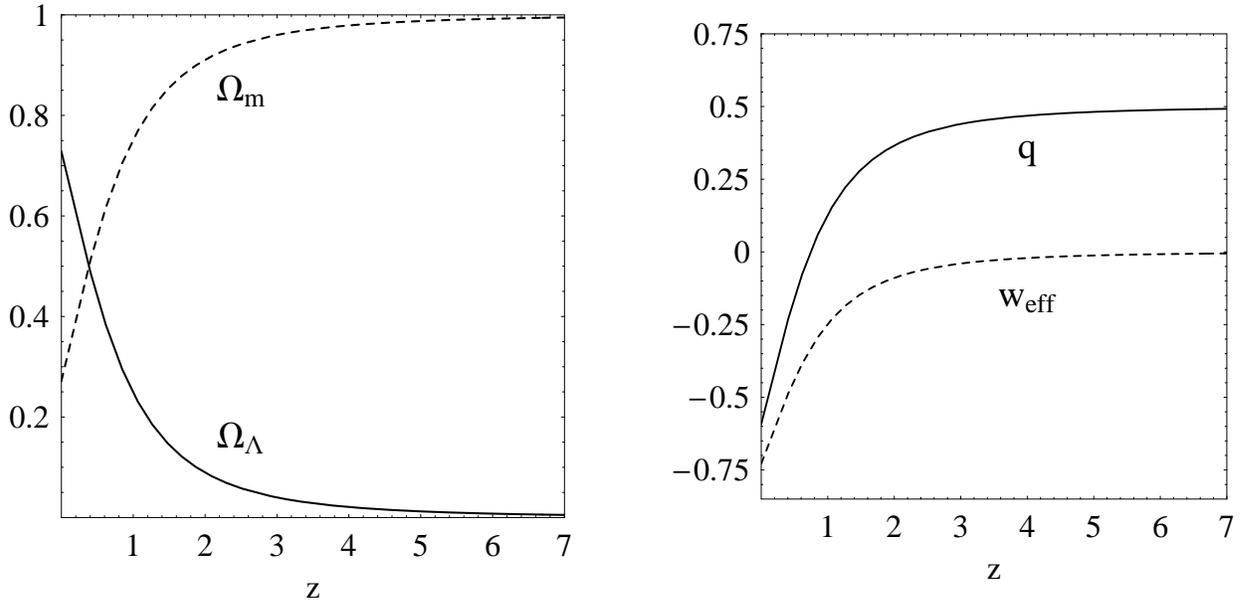}
 \caption{\label{fig8} The same as Fig.~\ref{q-fig2}, but for the case $Q=3\beta qH\rho_\Lambda.$}
 \end{figure}
 \end{center}
The plot \ref{fig8} displays the dependence for deceleration
parameter and effective parameter of the state equation $w_{\rm
eff}\equiv p_{tot}/\rho_{tot}=(2q-1)/3$ as functions of the redshift
$z$ with the parameters obtained by the maximum plausibility method.
It is easy to show that the transition from the decelerated
expansion phase $(q>0)$ to the accelerated one $(q<0)$ takes place
at $z_t=0.7398.$ As the parameter $\beta$ obtained from observations
satisfies $\beta>0,$ then dark energy decays into dark matter
($Q>0$) for $z>z_t,$ and vice versa $(Q<0)$ for $z<z_t.$
\subsection{Statefinders for the interacting dark energy}
In this part of the review, we discuss recently introduced ``statefinder parameters'' \cite{sahni}, that include the third derivative of the cosmic scale factor, are useful tools to characterize
interacting quitessence models \cite{pavon}.

Accelerated expansion of Universe is normally described by the
deceleration parameter $q = - \ddot{a}/(aH^{2}).$ As was mentioned
above, current value of the Universe deceleration parameter $q$ is
negative, however its amplitude is difficult to determine from the
observations. Since the accelerated dynamics of Universe is predicted by many models, they can be used to obtain an additional
information on $q.$

In particular, an example of such models is given by the cosmological models where the dominant components -- dark matter and
dark energy -- interact with each other. The models, where the dominating components do not evolve independently and interact with
each other instead, are of special interest because, as was mentioned above, they can facilitate or even completely solve the
``coincidence problem''.

Let us consider the Universe filled by two components of
non-relativistic matter (indexed by $m$) with negligibly small
pressure $p_m \ll \rho_m$ and dark energy (indexed by  $x$) with the
state equation $p_{x} = w \rho_{x},$ where $ w < 0.$ The dark energy
decays into dark matter according to
\begin{eqnarray}
\dot{\rho}_m & + & 3H \rho_m = Q \, , \nonumber\\
\dot{\rho}_x & + & 3H(1 + w) \rho_x = -Q \, , \label{conserv}
\end{eqnarray}
where $Q \geq 0$ measures the strength of the interaction. For further convenience we
will write it as $Q = - 3 \Pi H,$ where $\Pi$ is a new variable with dimension of pressure.

The Einstein equations for spatially flat Universe have the form
\begin{eqnarray}
H^{2} &=& \frac{8\, \pi G}{3} \rho \ ,  \\
\dot{H} & = & - \frac{8\, \pi G}{2} (\rho + p_x) \, , \label{efe}
\end{eqnarray}
where $\rho = \rho_m + \rho_x$ is the total energy density. The
quantity $\dot H$ is connected by the deceleration parameter $q$ by
the relation $q =- 1-(\dot H/H^2) = (1 + 3 w \Omega_{x})/2 ,$ where
$\Omega_{x} \equiv 8\, \pi G \rho_x/(3H^{2}).$ Evidently the
deceleration parameter does not depend on interaction between the
components. Nevertheless, differentiation of $\dot H$ yields
\begin{equation}
\frac{\ddot{H}}{H^{3}} = \frac{9}{2}\left(1 + \frac{p_x}{\rho}
\right) + \frac{9}{2}\left[w\left(1 + w \right)\frac{\rho_x}{\rho} -
w \frac{\Pi}{\rho} - \frac{\dot w}{3H}\frac{\rho_x}{\rho} \right]\ .
\label{doubledotH}
\end{equation}
In contrast with $H$ and $\dot H,$  the second order derivative
$\ddot H$ in fact depends on the interaction of the components.
Therefore, in order to distinguish models with different types of
interaction, or between interacting and free models, we need the
cosmology dynamical description involving the parameters explicitly
dependent on $\ddot {H}.$ In the papers \cite{alam} and \cite{sahni}
Sahni and Alam introduced the pair of parameters (the so-called
''statefinders''), which seem to be promising candidates for
investigation of distinction between the models with interacting
components.
\begin{equation}
r = \frac{\dddot{a}}{aH^{3}}, \qquad  s = \frac{r-1} {3(q
-\textstyle{1\over{2}})} \, . \label{statef}
\end{equation}
In the considered context of interacting components they take on the
form
\begin{equation} r = 1+ \frac{9}{2}\frac{w}{1+ \kappa} \left[1 + w
-\frac{\Pi}{\rho _x} - \frac{\dot{w}}{3wH} \right] \ , \label{r2}
\end{equation}
where $\kappa \equiv \rho_m /\rho_x,$ and
\begin{equation}
s = 1+ w -\frac{\Pi}{\rho_{x}} - \frac{\dot{w}}{3wH}\, . \label{s2}
\end{equation}
For the interaction-free models with $\Pi =0,$ the statefinders reduce to the expressions that were studied in \cite{sahni} and \cite{alam}. Let us show how the statefinders can be useful for analysis of cosmological models, where the dominant components interact with each other. Note that the third order derivative of
the scale factor is necessary for description of any variations in the general state equation for the cosmic medium
\cite{Visser_Jerk,Dabrowski}. It is obvious from the general relation
\cite{alam}
\begin{equation}
r - 1 = \frac{9}{2}\frac{\rho + P}{P} \frac{\dot{P}}{\dot \rho} \ ,
\label{}
\end{equation}
where $P$ is the total density of cosmic medium which in our case
leads to the relation $P \approx p_x.$ Since
\begin{equation}
\left(\frac{d}{dt}\right)\left(\frac{P}{\rho}\right) = \frac{\dot
\rho}{\rho} \left[\frac{\dot{P}}{\dot{\rho}} -
\frac{P}{\rho}\right]\ , \label{}
\end{equation}
it is evident that the interaction term in the relation
$\dot{P}\approx \dot{p}_x = \dot{w}\rho_x + w \dot{\rho}_x$ in
accordance with (\ref{conserv}) will additionally affect the time
dependence of the common state parameter $P/\rho.$

Models with interacting components suggest a dynamical approach to
the coincidence problem. The principal quantity in that context is
the density ratio $\kappa,$ which was defined above in expression
(\ref{r2}). Strict solution of the coincidence problem requires that
the density ratio should be of order of unity on late stages of
Universe evolution. At least it has to vary sufficiently slowly for
time interval of order $H^{-1}.$ The ratio $\kappa$ is governed by
the equation (\ref{conserv})
\\
\begin{equation}
\dot{\kappa} = - 3H \left[\left( \frac{\rho_{x} + \rho_{m}}
{\rho_{m} \, \rho_{x}}\right) \Pi - w \right] \kappa \, .
\label{evolr}
\end{equation}
Now let us consider the statefinders for different solutions of that
equation.
\subsection{Scaling solutions}
\begin{figure}[tbp]
\includegraphics*[scale=0.5]{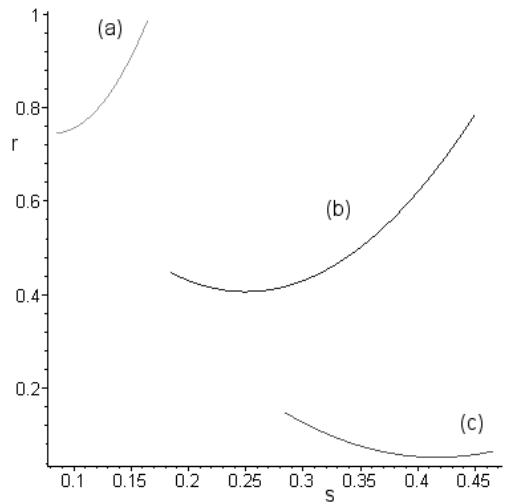}
\caption{The curves $r(s)$ are drawn for the redshift range $[0,6]$
(from left to right) with $w=-0.95$ and $\kappa_{0} = 3/7$ for 3
different values of the $\xi$ parameter:  (a) $2.5$; (b) $1.5$; (c)
$0.5.$} \label{fig1}
\end{figure}

\begin{figure}[tbh!]
\centerline{ \psfig{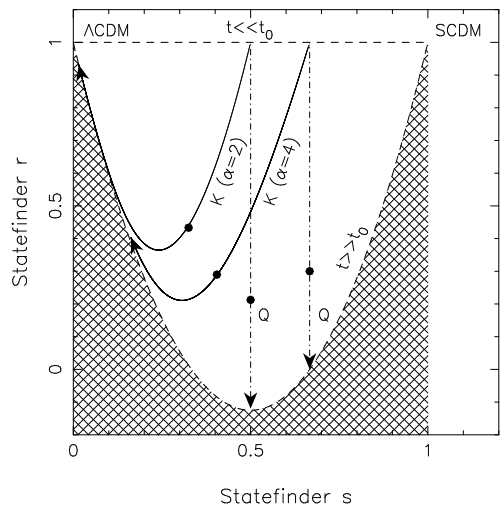} }
\caption{\small The Statefinder pair ($r,s$) is shown for different
forms of dark energy. In quintessence (Q) models ($w = $constant
$\neq -1$) the value of $s$ remains fixed at $s = 1+w$ while the
value of $r$ asymptotically declines to $r(t \gg t_0) \simeq 1 +
\frac{9w}{2}(1+w).$ Two models of quintessence corresponding to $w_Q
= -0.25, -0.5$ are shown. Quintessence (K) models are presented by a
scalar field (quintessence) rolling down the potential $V(\phi)
\propto \phi^{-\alpha}$ with $\alpha = 2,4.$ $\Lambda$CDM
($r=1,s=0$) and SCDM in the absence of $\l$ ($r=1,s=1$) are the
fixed points of the system. The hatched region is disallowed in
quintessence models and in the quintessence model which we consider.
The filled circles show the {\em current values} of the statefinder pair ($r,s$) for the Q and K models with $\Omega_{m0} = 0.3.$\label{fig:state}}
\end{figure}

In the paper \cite{scaling} it was shown by the authors that the
so-called scaling solutions, i.e. the solutions of the form
$\rho_m/\rho_x \propto a^{-\xi},$ where $\xi$ is constant parameter,
lying in the range $[0,3],$ can be calculated in the case when when
dark energy decays into dark matter (see eq.(\ref{conserv})). The
solutions of such type make interest because they facilitate
solution of the coincidence problem \cite{dalal}. Indeed, the model
with $\xi = 3$ reduces to $\Lambda$CDM one for $w = -1$ and $\Pi
=0.$ For $\xi = 0$ the Universe dynamics becomes stable with $\kappa
= const ,$ and the coincidence problem does not appear \cite{ZPC}.
Thus for the parameter values $\xi<3$ the coincidence problem
weakens. In this scheme with $w = const,$ one can easily see that
the interaction which generates the scaling solutions, can be
calculated as the following
\begin{equation}
\frac{\Pi}{\rho_{x}} = \left(w + \frac{\xi}{3}\right) \,
\frac{\kappa_{0} (1+z)^{\xi}}{1 + \kappa_{0} (1+z)^{\xi}}\  ,
\end{equation}
where $\kappa_0$ denotes the current ratio of the energy densities
and $z = (a_{0}/a) - 1$ is the redshift. Substituting this
expression into the equations (\ref{r2}) and (\ref{s2}), one finds
the statefinder parameters:
\begin{equation}
r =  1 + \frac{9}{2} \frac{w}{1 + \kappa_{0} (1+z)^{\xi}} \left[1 +
w - \left( w + \frac{\xi}{3} \right) \frac{\kappa_{0} (1+z)^{\xi}}
{1 + \kappa_{0} (1+z)^{\xi}}\right] \, , \label{r3}
\end{equation}
and
\begin{equation}
s = 1 + w -  \left( w + \frac{\xi}{3} \right)
 \frac{\kappa_{0} (1+z)^{\xi}}{1 + \kappa_{0} (1+z)^{\xi}}.
\label{s3}
\end{equation}
Figure \ref{fig1} presents the dependence $r(s)$ for different
values of $\xi.$ The less is the value $\xi,$ the lower is the
corresponding curve on $s$--$r$ diagram and the less actual is the
coincidence problem. Remark that those curves are qualitatively
identical to those drawn for the models with interaction-free
components (see Fig.\ref{fig:state} \cite{sahni}). Compare also that
 $\Lambda$CDM model ($\Pi = 0,$ $w
= -1$) corresponds to the point $s = 0,$ $r = 1$ (not shown on the
figure).
\noindent
The present values $r_0$ and $s_0$ of the statefinder parameters are
important both from the observational point of view and for the
purpose of distinction between different models. For the scaling
models one has
\begin{equation}
r_0 = 1 + \frac{9}{2} \frac{w}{1 + \kappa_0}s_0 \, , \qquad {\rm and
}\qquad s_0 = 1+ w -  \left(w + \frac{\xi}{3}\right)
\frac{\kappa_0}{1 + \kappa_0} \, . \label{rsscaling}
\end{equation}
Having in mind that
\begin{equation}
q_0  = \frac{1}{2} \frac{1 + \kappa_0 + 3w}{1 + \kappa_0}  \ ,
\label{}
\end{equation}
and introducing
\begin{equation}
q_{0\Lambda}\equiv q_0\left(w = -1\right)  = -\frac{1}{2} \frac{2 -
\kappa_0}{1 + \kappa_0} \Leftrightarrow \frac{3}{2}\frac{\kappa_0}{1
+ \kappa_0} = 1 + q_{0\Lambda}\ , \label{}
\end{equation}
we can classify different models by its dependence $s_0(q_0),$ which
is
\begin{equation}
s_0 = \frac{2}{3}\left[ q_0 - q_{0\Lambda} + \left( \frac{\xi}{3} -
1 \right)\left( 1 + q_{0\Lambda} \right)\right]\ . \label{reads}
\end{equation}
The first part of the bracket in the righthand side describes
deviation from $w = - 1,$ and the second responds for deviation from
$\Lambda$CDM value $\xi = 3.$ For models with $w = - 1,$ which have
the same deceleration parameter $q_0 = q_{0\Lambda},$ we obtain $s_0
= \frac{2}{3}\left(\frac{\xi}{3} - 1 \right) \left(1 +
q_{0\Lambda}\right).$ Of course, $\xi = 3$ corresponds to
$\Lambda$CDM model with $s_0 = 0.$ If the conditions $\kappa_0 =
3/7$ and $\xi = 1$ hold, then $s_0\left(\xi = 1\right) = - 0.2,$
while for $\xi = 0$ one has $s_0\left(\xi = 0\right) = - 0.3.$

Analogous considerations are valid also for other values of $w.$
Therefore the parameter $s_0$ enable us to distinguish between
different scaling models, which have the same deceleration
parameter.
\subsection{Concluding remarks}
The statefinder parameters introduced in \cite{sahni} and
\cite{alam}, as was assumed, became useful tools for verification of
models with interacting components, which in turn solve or weaken
the coincidence problem --- a stone of obstacle for many models with
late acceleration. While the deceleration parameter does not depend
on interaction between the dark energy and dark matter, the state
finders $(r, s)$ contain the dependence explicitly.
\section{Holographic dynamics: entropic acceleration}
In the previous section we presented the description of Universe
dynamics based on general relativity with some generalizations. The
present section is intended to show that there exists a principally
independent approach, which enables us both to reproduce all
achievements of the traditional description and resolve a number of
problems the latter faced to.

The traditional point of view assumed that space filling fields
constitute the dominant part of degrees of freedom in our World.
However it became clear soon that such estimate much hardened the
development of the quantum gravity theory: for the latter to make
sense one had to cutoff on small distances all the integrals
appeared in the theory. Consequently our World was described on a
three-dimensional discrete lattice with cell size of order of Planck
length.

Recently some physicists came up with even more radical point of
view: complete description of Nature required only two-dimensional
lattice, situated on space edge of our World, instead of the
three-dimensional one. Such approach is based on so-called
''holographic principle''
\cite{Hooft,Susskind,Witten,Maldacena_Gravity,Susskind_Hologram,Gubser_Klebanov_Polyakov,TASI_AdS_CFT}.
The term comes from the optical holography, which represent nothing
but two-dimensional recording of three-dimensional objects. The
holographic principle is composed of the two main statements:
\begin{enumerate}
\item  all information contained in some region of space can be
''recorded'' (presented) on boundary of that region;
\item the theory contains at most one degree of freedom per Planck
area on boundaries of the considered space region
\begin{equation}\label{N}
N\le \frac{A{{c}^{3}}}{G\hbar }.
\end{equation}
\end{enumerate}

Therefore central place in the holographic principle is occupied by
the assumption that all information about the Universe can be coded
on some two-dimensional surface --- the holographic screen. Such
approach leads to possibly new interpretation of cosmological
acceleration and to completely novel concept of gravity. The
gravitation is now defined as entropic force generated by variation
of information connected to positions of material bodies.

Let us consider a small piece of holographic screen and approaching
particle of mass $m.$ According to the holographic principle, the
particle affects the information amount (and thus on entropy), which
is stored on the screen. It is naturally to assume that the entropy
increment near the screen is linear on the displacement $\Delta x,$
\begin{equation}\label{DeltaS}
    \Delta S=2\pi {{k}_{B}}\frac{mc}{\hbar }\Delta x.
\end{equation}
The factor $2\pi$ is introduced for convenience reasons, to be clear
below. To make clear why that quantity is proportional to mass, let
us imagine that the particle splits on two or more particles of
lower mass. Each of them then carries its own entropy increment with
displacement on $\Delta x.$ As both entropy and mass are additive
function, then they are proportional to each other. According to the
first principle of thermodynamics, the entropic force, caused by
information variation, reads
\begin{equation}\label{FDelta}
    F\Delta x=T\Delta S.
\end{equation}
If the entropy gradient is given, or calculated using the relation
(\ref{DeltaS}), and the screen temperature is fixed, then one can
find the entropy force. As is known, an observer, moving with
acceleration $a,$ effectively feels the so-called Unruh temperature
\cite{Unruh}
\begin{equation}\label{k_bT}
    {{k}_{B}}T=\frac{1}{2\pi }\frac{\hbar }{c}a.
\end{equation}
Let us assume that the total energy of the system equals $E.$ A
simple assumption follows, that it is uniformly distributed on all
$N$ bits of information on the holographic screen. The temperature
is then defined as average energy per bit
\begin{equation}\label{E}
    E=\frac{1}{2}N{{k}_{B}}T.
\end{equation}
The formulae (\ref{N})-(\ref{E}) enable us to formulate the
holographic dynamics, and in particular dynamics of Universe,
without any notion of gravity.

The above presented ideology is essentially based on successful
description of black holes physics
\cite{Bekenstein,Bekenstein1,Bekenstein2,Bekenstein3}. On the first
face, there is nothing common between the extremely diluted Universe
with density $\rho \sim {{10}^{-29}}g/c{{m}^{3}}$ and ''typical''
black hole of stellar mass with critical density $\rho \sim
{{10}^{14}}g/c{{m}^{3}}\,(M=10{{M}_{\odot }}).$ But the situation
sharply changes if we transit to more massive black holes. As we
have seen above, the black hole radius scales ${{r}_{g}}\propto M,$
therefore for the black hole critical density one has $\rho \propto
{{M}^{-2}}.$ Super-massive black holes situated in the galactic
nuclei reach masses of order ${{10}^{10}}{{M}_{\odot }},$ and their
gravitational radius (${{r}_{g}}\sim 3\times {{10}^{15}}cm$) five
times exceeds the dimensions of Solar system. The critical density
in the latter case equals $\rho \sim {{10}^{-4}}g/c{{m}^{3}},$ which
is one order of magnitude less than the room air density ($1.3\times
{{10}^{-3}}g/c{{m}^{3}}$). Let us consider, how close are the
parameters of our observed Universe to those of black hole. We
estimate mass of the observed Universe taking the Hubble radius for
its radial dimension ${{H}^{-1}}$ The mass contained inside the
Hubble sphere thus reads
\begin{equation} \label{M_univ}
    {{M}_{univ}}=\frac{4\pi }{3}R_{H}^{3}\rho.
\end{equation}
In the case of flat space \footnote{as we have repeatedly mentioned
above, deviations from the flatness are very small}, the density
$\rho$ of the observable part of Universe enclosed by Hubble sphere
can be reasonably replaced by the critical one \eqref{Omega} $\rho
={3{{H}^{2}}}/{(8\pi G)},$ and therefore the formula \eqref{M_univ}
takes the form
\[ {{M}_{univ}}=\frac{{{R}_{H}}}{2G}.\]
Inserting the obtained mass value into the expression for
gravitational radius of the Universe, ${{r}_{g}}=2G{{M}_{univ}},$
one finds
\begin{equation}
\label{r_{g,univ}}
    r_g=R_H.
\end{equation}

The obtained result for many reasons is nothing but a rough
estimate, however it obviously favors the above cited black hole
argumentation in application to the observable Universe. Remark that
in the case of the Sun the ratio of the physical radius ${{R}_{\odot
}}=695\,500 km$ to the gravitational one ${{r}_{g\odot }}=3km$
counts more than five orders of magnitude.

Now let us identify the holographic screen with the Hubble sphere of
radius $R={{H}^{-1}}$ (which is valid for flat Universe) and we try,
making use of the holographic principle, to reproduce the Friedmann
equations, using neither Einstein equations nor Newtonian mechanics.
The holographic screen spans the area $A=4\pi {{R}^{2}}$ and carries
the (maximum) information $N=4\pi {{R}^{2}}/L_{Pl}^{2}$ bits. The
change of information quantity $dN$ for the time period $dt,$ caused
by the Universe expansion $R\to R+dR,$ equals
\[dN=\frac{dA}{L_{Pl}^{2}}=\frac{8\pi R}{L_{Pl}^{2}}dR.\]
Here we use the system of units $c={{k}_{B}}=1.$ Variation of Hubble
radius leads to change in Hawking temperature
 $\left( T=\frac{\hbar }{2\pi R}
\right)$
\[dT=-\frac{\hbar }{2\pi {{R}^{2}}}dR.\]
From the equipartition law it follows that
\begin{equation}\label{dE}
    dE=\frac{1}{2}NdT+\frac{1}{2}TdN=\frac{\hbar }{L_{Pl}^{2}}dR=\frac{dR}{G},
\end{equation}
$\left( L_{Pl}^{2}=\frac{\hbar G}{{{c}^{3}}} \right).$ The quantity
$dR$ can be presented in the form
\begin{equation}\label{dR}
    dR=-H\dot{H}{{R}^{3}}dt.
\end{equation}
From the other hand, the energy flow through the Hubble sphere can
be calculated if the energy-momentum tensor is given for the
substance filling the Universe. Considering the latter substance as
an ideal liquid, (see chapter 2) and using the relation ${{T}_{\mu
\nu }}=(\rho +p){{u}_{\mu }}{{u}_{\nu }}+p{{g}_{\mu \nu }},$ we
obtain
\begin{equation}\label{dE1}
dE=A\left( \rho +p \right)dt.
\end{equation}
Equating (\ref{dE}) to (\ref{dE1}) accounting (\ref{dR}), one finds
\[\dot{H}=-4\pi G\left( \rho +p \right).\]
It is well known that the system of equations
\begin{equation}\label{dR}
  \dot{H}=-4\pi G\left( \rho +p \right),~ \dot{\rho }+3H\left( \rho +p \right)=0.
\end{equation}
is equivalent to the system  \eqref{Frid_1}-\eqref{Frid_2}. Thus the
declared aim is achieved. We remark that already in the year 1995 a
more general problem was solved \cite{Jacobson}: the Einstein
equations were obtained from the thermodynamic considerations. This
important result followed from the fact that entropy was
proportional to horizon area and from the assumption that the
relation \eqref{k_bT}  holds for any accelerated moving observer situated
inside her own casual horizon, and $T$ stands for Unruh temperature.

Derivation of the Friedmann equations from the holographic principle
is of course an important result, but by itself it represents
nothing but reproduction of something already well-known. A natural
question is whether one can develop a novel approach to description
of Universe dynamic basing on the holographic principle? If yes,
than is it possible to overcome the unsolvable difficulties of the
traditional approach in frames of the new one? Let us start from the
logic scheme which the holographic dynamics of Universe is based on.

The standard procedure to develop any field theory involves the
formulation of equations of motion obtained from some effective
action corresponding to the energy scale characteristic to the
phenomenon to be described by the theory. General relativity theory
is not an exception from the rule: equations of motion in general
relativity are obtained by variation of effective action with
respect to dynamical variables present in it. The standard approach
to evaluate the integrals that arise in variations of the action is
the integration by parts, which produces the terms corresponding to
boundaries of the region under consideration. According to such
approach a quite reasonable assumption is made that on boundary of
the considered region, far from the field sources, all the dynamical
variables, be it either fields, or gravitational potentials or
connectivity components (as in general relativity), turn to zero. In
holographic physics instead the surface is the place where the main
action takes goes on. The natural conclusion follows that the
contribution of surface terms now should be necessarily accounted.

Let us show that consideration of the boundary term in the
Einstein-Hilbert action is equivalent to inclusion of non-zero
energy-momentum tensor in the standard Einstein equations. The
action for gravitational field takes the form \fnote{\dag}{We use
the signature $(+, -, -, -),$ and definition
$R^{\rho}_{~\sigma\mu\nu} =
\partial_{\mu}\Gamma^{\rho}_{\nu\sigma}-
 \partial_{\nu}\Gamma^{\rho}_{\mu\sigma}+\Gamma^{\rho}_{\mu\lambda}\Gamma^{\lambda}_{\nu\sigma} - \Gamma^{\rho}_{\nu\lambda}\Gamma^{\lambda}_{\mu\sigma} ,$ ${ R}_{\nu \mu} = R_{\nu \beta
\mu}^ {~~~~~\beta},$ $R = g^{\mu \nu}R_{\mu \nu}.$}
\begin{equation}
    S_{EH} \,=\,- \frac{c^{3}}{16 \pi G} \int d^{4} x \,\sqrt{-g} \,R\, .
\label{S_action}
\end{equation}
Variation of that action with respect to metrics  $g_{\mu \nu}$ in a
compact region $\Omega,$ one obtains
\begin{eqnarray}
   \lefteqn{
\delta \int_{\Omega} d^{4}x \sqrt{-g}\, R \, =}  \nonumber \\
& &
 \int_{\Omega}d^{4} x
\sqrt{-g} \Biggl[ \left( g^{\mu \nu}\nabla^2 \delta g_{\mu \nu}\,-
\nabla^{\mu} \nabla^{\nu} \delta g_{\mu \nu} \right)
 -\left( R^{\mu \nu} - \frac{1}{2}\, g^{\mu \nu} R
\right) \delta g_{\mu \nu} \Biggr] \, , \label{2.2}
\end{eqnarray}
where $\nabla^2 \,=\, g^{\alpha \beta} \nabla_{\alpha} \,
\nabla_{\beta}.$ Full derivative in (\ref{S_action}) can be
presented as the contribution of the region boundary  $\partial \,
\Omega,$ where those contribution do not cancel. In order to obtain
the Einstein equations from the least action principle without
boundary terms, the Einstein-Hilbert action should be complemented
by a functional intended to compensate the contribution of full
derivative in (\ref{S_action}). We denote it as  $S_{Boundary}[g]..$
Then the full action takes on the form

\begin{equation}
    S \,=\, - \frac{1}{16 \pi G} \int_{\Omega} d^{4} x \sqrt{- g} \,R +\,
 S_{Boundary}[g] \, + \, S_{Source}\, , \label{S_action_full}
\end{equation}
where we included possible sources  $S_{Source}$ of gravitational
field, coupled with the material fields. The functional
$S_{Boundary}[g],$ thus takes the form
\begin{equation}
    \delta S_{Boundary}[g]\,=\,-\, \frac{1}{16 \pi G} \int_{\Omega}d^{4}x \sqrt{-g}
\Bigl( g^{\mu \nu} \nabla^2 \delta g_{\mu \nu}\,-\, \nabla^{\mu}
\nabla^{\nu} \delta g_{\mu \nu} \Bigr) \, , \label{S_Boundary}
\end{equation}
In the context of holographic physics where the boundaries play
crucial role, the functional $S_{Boundary}[g]$ can be treated as
action for holographic dark energy. The Einstein equations including
the contributions of material fields and boundary terms have the
following form
\begin{eqnarray}
    R^{\mu \nu} - \frac{1}{2}\, g^{\mu \nu} R & = & 8 \pi G\,\left( T^{\mu \nu}_{Source}+ T^{\mu \nu}_{Boundary}\right)\, , \nonumber \\
 T^{\mu \nu}_{Source} & =  &\frac{2}{\sqrt{- g}} \, \frac{\delta S_{Source}}
{\delta g_{\mu \nu}}, ~ T^{\mu \nu}_{Boundary}  = \frac{2}{\sqrt{-
g}} \, \frac{\delta S_{Boundary}} {\delta g_{\mu \nu}}\,.
\label{Eq_b_M}
\end{eqnarray}
Remark that in the flat space case, when $g_{\mu \nu}=0,$ the
boundary action $S_{Boundary}[g] = 0,$ and Einstein equations take
the standard form. We also note that such modification of Einstein
equations preserves the general structure of the equations, which,
as we will see below, guarantees the conservation of the Friedmann
equation structure.

Inclusion of the additional terms in the Einstein equations requires
the so-called ''holographic correction'' of the Friedmann equations.
The easiest way to perform the correction is to insert the entropic
force into the second Friedmann equation \cite{EntropicCosmology}.
The entropic term structure can be guessed from the dimensional
considerations
\begin{equation}
    \frac{{\ddot{a}}}{a}=-\frac{4\pi G}{3}\left( \rho +3p
    \right)+\frac{{{a}_{e}}}{{{L}_{b}}}.
\end{equation}
Here ${{a}_{e}}$=${{F}_{e}}/m$ is the acceleration caused by the
entropic force and ${{L}_{b}}$ is the scale length, determined by
the holographic screen position. If one takes the Hubble sphere as
the holographic screen, then the scale length coincides with the
Hubble radius: ${{L}_{b}}={{R}_{H}}={{H}^{-1}}.$ The entropic
acceleration magnitude can be expressed through the holographic
screen temperature ${{T}_{b}}$
\begin{equation}\label{a_e}
    {{a}_{e}}=\frac{{{F}_{e}}}{m}={{T}_{b}}\frac{\Delta S}{\Delta x}\frac{1}{m}=2\pi {{T}_{b}}.
\end{equation}
Does the considered scheme agree with the main observational data at
least? We have shown above in frames of SCM that absolute value of
the cosmological acceleration of the Hubble sphere equals
$\dot{V}\simeq 4\times {{10}^{-10}}m{{\sec }^{-2}}.$ Let us estimate
that magnitude in the holographic approach.

It is naturally to assign the Unruh temperature to the holographic
screen ${{T}_{b}}={{T}_{U}}.$ Let us link it with the Hawking
radiation temperature ${{T}_{H}},$ which is equal to
\begin{equation}\label{T_H}
{{T}_{H}}=\frac{\hbar }{8\pi {{k}_{B}}GM}=\frac{\hbar g}{2\pi
{{k}_{B}}},
\end{equation}
where $g$ is the free fall acceleration on the Hubble sphere.
Comparing the latter expression with Unruh temperature
\begin{equation}\label{T_U}
{{T}_{U}}=\frac{\hbar a}{2\pi {{k}_{B}}},
\end{equation}
one gets
\begin{equation}
{{T}_{H}}={{T}_{U}}(a=g).
\end{equation}
Thus the Unruh temperature coincides with that for the Hawking
radiation, but depends on the reference frame acceleration instead
of surface gravitation (free fall acceleration). According to the
equivalence principle, the free fall acceleration on the Hubble
surface is equivalent to acceleration of a reference frame and we
can relate it to each other. Treating the Hubble sphere as an
analogue of events horizon for a black hole, we can present the
formula (\ref{T_H}) in the following form
\begin{equation}\label{T_H_aprox}
    {{T}_{H}}=\frac{\hbar }{{{k}_{B}}}\frac{H}{2\pi }\sim 3\times
    {{10}^{-30}}\,K.
\end{equation}
Comparing (\ref{T_H_aprox}) and (\ref{T_U}), one finds that the
Hubble sphere acceleration ${{a}_{H}}$ equals
\begin{equation}\label{a_H}
    {{a}_{H}}=H\simeq {{10}^{-9}}m\,{{\sec }^{-2}}.
\end{equation}
The latter result is in convincing qualitative agreement with the
one obtained in SCM.

Certainly, quality of a novel model is not defined by its ability to
reproduce the results established by its predecessors --- it is
nothing but minimum task. Its main purpose and virtue is to solve
the problems untouchable by former models. The holographic dynamics
is very promising from that point of view as it gives hope to solve
both main challenges of the SCM --- they are cosmological constant
magnitude and coincidence problem.

Let us start from the former. The cosmological constant problem
stands for the huge disagreement \footnote{about $120$ orders of
magnitude !!!} between the observed density of dark energy in form
of the cosmological constant and its ''expected'' value. The
expectations are based on rather natural assumptions on the cutoff
parameter for the integral, representing the zero fluctuations of
vacuum. The holographic principle enables us to replace the
''natural assumptions'' by more thorough quantitative estimates.
\subsection{Universe with holographic dark energy}
In any effective quantum field theory defined in space region with
typical length scale $L$ and using the ultraviolet cutoff $\Lambda,$
the system entropy takes the form $S\propto \Lambda^3L^3.$ For
instance, fermions, placed in nodes of space lattice with
characteristic size $L$ and period $\Lambda^{-1},$ occupy one of the
number $2^{{(L\Lambda)}^3}$ states. Therefore entropy of such system
is $S\propto \Lambda^3L^3.$ According to the holographic principle,
this quantity must obey the inequality \cite{CohenKaplanNelson}
\begin{equation}
\label{conditionQFT}
    L^3\Lambda^3\le S_{BH}\equiv \frac{1}{4}\frac{A_{BH}}{l_{Pl}^2}=\pi
    L^2M_{Pl}^2,
\end{equation}
where $S_{BH}$ is the black hole entropy and $A_{BH}$ stands for its
event horizon area, which in the simplest case coincides with the
surface of sphere with radius $L.$ This reasoning shows that
magnitude of the infrared cutoff cannot be chosen independently of
the ultraviolet one.

Thus we formulate the important result \cite{CohenKaplanNelson}: in
frames of holographic dynamics the infrared cutoff magnitude is
strictly linked to the ultraviolet one. In other words, physical
properties on small UV-scales depends on physical parameters on
large IR-scales. In particular, if the inequality
\eqref{conditionQFT} holds, then one gets
\begin{equation}
\label{LsimLambda{-3}}
    L\sim {{\Lambda }^{-3}}M_{Pl}^{2}.
\end{equation}

Effective field theories with UV-cutoff \eqref{LsimLambda{-3}}
obviously include many states with the gravitational radius
exceeding the region where the theory was initially defined. In
other words, for arbitrary cutoff parameter one can find
sufficiently large volume where entropy in an  effective field theory exceeds the Bekenstein limit.
 In order check it let us remark that usually an
effective quantum field theory must be able to describe the system
under temperature $T\le \Lambda .$ While $T\gg 1/L,$ the system has
thermal energy $M\sim L^3T^{4}$ and entropy $M\sim L^3T^3.$ The
condition (\ref{conditionQFT}) is satisfied at $T\le \left(
M_{Pl}^2/L \right)^{1/3},$ which responds to gravitational radius
$r_g\sim L(LM_{Pl})\gg L.$ In order to get rid of that difficulty an
even more strict limitation \cite{CohenKaplanNelson} is imposed on
the infrared cutoff $L\sim \Lambda^{-1},$ which excludes all the
states localized within limits of their gravitational radii. Taking
into account all above mentioned and using the expression
(\ref{rho_{vac}})
\begin{equation}
    \label{rho_{vac}}
        \rho_{vac}\approx \frac{\Lambda^4}{16\pi^2},
\end{equation} the condition (\ref{conditionQFT}) can be presented in the form
\begin{equation}
   L^3\rho_\Lambda\le L M_{Pl}^2\equiv 2M_{BH},
\end{equation}
where $M_{BH}$ is the black hole mass with gravitational radius $L.$
Thus total energy contained in the region of dimensions $L$ cannot
exceed the mass of a black hole of roughly the same size. This
result agrees surprisingly well with the expression
\eqref{r_{g,univ}} provided the IR-cutoff scale is identified with
the Hubble radius $H^{-1},$ besides that such choice is well
motivated in cosmology context. The quantity $\rho_{\Lambda }$ is
commonly called the holographic dark energy.

In the cosmological aspect under interest the link between the small
and large scales can be obtained from rather natural condition, that
total energy contained within a volume of dimension $L$ cannot
exceed mass of a black hole of the same size:
\begin{equation}\label{rho-L}
L^3\rho_\Lambda\le M_{BH}\sim LM_{Pl}^2.
\end{equation}
Here $\rho_\Lambda$ is the energy density. Would this inequality be
violated, the Universe was composed of black holes solely. Applying
this relation to the Universe as whole it is naturally to identify
the IR-scale with the Hubble radius $H^{-1}.$ Then for the upper
bound of the energy density one finds
 \begin{equation}\label{LsimLambda{-2}}
 \rho_\Lambda\sim L^{-2}M_{Pl}^{2}\sim H^2M_{Pl}^{2}.
 \end{equation}
The quantity $\rho_\Lambda$ is usually called the holographic dark
energy. We will below denote its density as $\rho_{DE}.$ Accounting
that
\[
  M_{Pl} \simeq  1.2\times {{10}^{19}} GeV; ~~~  H_0 \simeq  1.6\times 10^{-42} GeV,
\]
one finds
\begin{equation}\label{rho_{DE}}
\rho_{DE}\sim 10^{-46}\,GeV^4.
\end{equation}
This expression will agrees with the observed value of the dark
energy density $\rho_{DE}\simeq 3\times 10^{-47}GeV^4.$ Therefore
the holographic dynamics is free from the cosmological constant
problem.

Though the obtained dark energy density corresponds to the correct
one, the choice of the IR-cutoff scale equal to the Hubble radius a
problem arises connected with the state equation problem: the
holographic dark energy cannot provide the accelerated expansion in
that case \cite{Hsu}. It can be easily seen from the following
simple arguments. Consider the Universe composed of holographic dark
energy with density given by the relation \eqref{LsimLambda{-2}} and
of normal matter component. In that case $\rho ={{\rho }_{\Lambda
}}+{{\rho }_{M}}.$ From the first Friedmann equation it follows that
$\rho \propto {{H}^{2}}.$ If ${{\rho }_{\Lambda }}\propto
{{H}^{2}},$ then dynamical behavior of holographic dark energy
coincides with that of normal matter, thus the accelerated expansion
is impossible. If the Universe was initially dominated by matter,
then the dark energy defined by \eqref{LsimLambda{-2}} is in fact a
tracker solution, as it reproduces dynamics of the dominating
component. The presence of tracker solutions in frames of the
holographic models gives a hope to solve the coincidence problem,
but it contradicts to the observed dynamics of Universe.

In order to produce the accelerated expansion of Universe in frames
of holographic dark energy model we will try to use the IR-cutoff
spatial scale different from the Hubble radius. First which comes in
mind is to replace the Hubble radius by the particle horizon $R_p$:
\begin{equation}\label{R_p}
R_p=a\int_{0}^{t}\frac{dt}{a}=a\int_{0}^{a}\frac{da}{Ha^2}.
\end{equation}
Unfortunately such substitution cannot give the desired result. It
can be made clear again using the initially matter-dominated
Universe model. Due to the causality principle the gravity influence
cannot extend to the regions situated on distances exceeding the
causality horizon size. Therefore the vacuum energy density present
in the Friedmann equations cannot be an arbitrary function of $L.$
In the matter dominated era the causality (particle) horizon scales
as $a^{3/2},$ therefore $\rho_{_{DE}}(L)\sim a^{-3}.$ As $\rho
(t)\sim a(t)^{-3(1+w)},$ then it immediately follows that for the
case of holographic energy with IR-cutoff on the particle horizon
one gets $w=0,$ which again clearly contradicts the observations.

Let us convince ourself now that the above discovered problem
preserves also in the case of Universe dominated by holographic dark
energy. Present the density of the latter in the following form
\cite{Hsu}
\begin{equation}\label{rho_{DE}c}
    \rho_{DE}=3c^2M_{Pl}^2L^{-2}.
\end{equation}
The coefficient $3{{c}^{2}}\,(c>0)$ is introduced for convenience,
and $M_{Pl}$ further stands for reduced Planck mass:
$M_{Pl}^{-2}=8\pi G.$ Replacing $L\to {{R}_{h}}$ in
\eqref{rho_{DE}c}, one obtains the first Friedmann equation in the
form
\begin{equation}\label{R_pc}
R_pH=c,
\end{equation}
and it immediately follows that
\begin{equation}\label{Ha2}
\frac{1}{Ha^2}=c\frac{d}{da}\left(\frac{1}{Ha}\right).
\end{equation}
One now easily finds that
\begin{equation}
H^{-1}=\alpha a^{1+\frac{1}{c}},
\end{equation}
where $\alpha $ is constant. As $R_h=cH^{-1},$ and
$\rho_{\Lambda}=3{{c}^{2}}M_{Pl}^{2}R_{h}^{-2},$ then
\begin{equation}
\rho_{DE}=3\alpha^2M_{Pl}^2a^{-2\left(1+\frac{1}{c}\right)}.
\end{equation}
How can we find the parameter $w$ in the state equation if the
dependence $\rho =f(a)$ is given? One possible approach is the
following. Let us use the conservation equation
    $$\dot{\rho }+3H\rho (1+w)=0.$$
In the case $\rho =f(a)$ it takes the form
\begin{equation}\nonumber
    {f}'(a)\dot{a}+3Hf(a)(1+w)=0,
\end{equation}
and it follows that
\begin{equation}\label{w(f(a))}
w=-\frac{1}{3}\frac{{f}'(a)}{f(a)}a-1=-\frac{1}{3}\frac{d\ln
f(a)}{da}a-1.
\end{equation}
It is easy to see that for well-known cases of matter $\rho \propto
{{a}^{-3}},$ radiation $\rho \propto {{a}^{-4}}$ and cosmological
constant $\rho =const$ one obtains the correct values $w=0,1/3,-1.$
From the expression \eqref{w(f(a))} it follows that
\begin{equation}
w = -\frac 13 + \frac 2{3c}>-\frac 13.
\end{equation}
Thus one can see that the above described component does not
deserves the name of ``dark energy'' in proper sense, as it cannot
serve its main purpose --- to provide the accelerated expansion of
universe. The difficulties originate from the fact that, as it
follows from \eqref{Ha2}, the derivative obeys
$\frac{d}{da}({{H}^{-1}}/a)>0.$ In order to obtain the accelerated
expansion of Universe, one should ''slow down'' the growth of the
IR-cutoff scale. It turns out \cite{Holographic_Dark_Energy}, that
it can be done if one replaces the particle horizon by the events
one. Recall (see Chapter 2), that the size of spatial region such
that a signal emitted at time $t$ from any point of it will reach
the immobile observer at infinitely far future, equals
\begin{equation}\label{R_e}
R_e=a(t)\int_{t}^{\infty }{\frac{d{t}'}{a\left( t' \right)}}.
\end{equation}
Assuming again that the dark energy dominates and finding solution
of the first Friedmann equation in the form
\begin{equation}
\int_{a}^{\infty }{\frac{da}{H{{a}^{2}}}}=\frac{c}{Ha},
\end{equation}
one finds
\begin{equation}
    \rho_{DE}=3{{c}^{2}}M_{Pl}^{2}R_{e}^{-2}=3{{\alpha }^{2}}M_{Pl}^{2}{{a}^{-2\left( 1-\frac{1}{c}
    \right)}},
\end{equation}
or
    $w=-\frac{1}{3}-\frac{2}{3c}<-\frac{1}{3}.$
We obtained a component which behaves as the dark energy, i.e. it
provides the accelerated expansion of Universe. If $c=1$ then it
behaves as the cosmological constant. If $c<1,$ then $w<-1.$ Such
value of $w$ responds to phantom model in the traditional approach.

In the above considered cases we investigated properties of
holographic dark energy in two limiting cases: the matter dominated
and dark energy dominated one. Let us now consider the common case
situation, i.e. the Universe dynamics at arbitrary relation between
the both components densities \cite{Q-GHuang}. For simplicity we
restrict to the case of flat Universe and set the IR-cutoff scale to
the event horizon ${{R}_{e}}$ (35). Introducing the relative density
of the holographic dark energy ${{\Omega }_{DE}}={{\rho
}_{DE}}/{{\rho }_{cr}}$ $({{\rho }_{cr}}=3M_{Pl}^{2}{{H}^{2}}),$ we
represent \eqref{rho_{DE}c} for $L={{R}_{e}}$ in the form
\begin{equation}\label{HR_e}
HR_e=\frac{c}{\sqrt{\Omega_{DE}}}.
\end{equation}
Of course, with $\Omega_{DE}=1$ and replacement $R_e\to R_h$ the
equation \eqref{HR_e} transits into \eqref{R_pc}. Taking derivative
with respect to time from both sides of \eqref{R_e}, one gets
\begin{equation}
    \dot{R}_e=H{{R}_{e}}-1=\frac{c}{\sqrt{\Omega_DE}}-1
\end{equation}
From the definition \eqref{rho_{DE}c} it follows that
\begin{equation}
\frac{d{{\rho }_{{DE}
}}}{dt}=-6{{c}^{2}}M_{Pl}^{2}R_{e}^{-3}{{\dot{R}}_{e}}=-2H\left(
1-\frac{\sqrt{{{\Omega }_{{DE}}}}}{c} \right){{\rho }_{{DE}}}.
\end{equation}
Due to the energy conservation law
\begin{equation}
\frac{d}{da}\left( {{a}^{3}}{{\rho }_{{DE}}}
\right)=-3{{a}^{2}}{{p}_{{DE} }}.
\end{equation}
and it follows that
\begin{equation}
{{p}_{{DE}}}=-\frac{1}{3}\frac{d{{\rho }_{{DE}}}}{d\ln a}-{{\rho
}_{DE}}.
\end{equation}
Thus the state equation reads
\begin{equation}
{{w}_{{DE}}}=\frac{{{p}_{{DE} }}}{{{\rho
}_{{DE}}}}=-\frac{1}{3}\frac{d\ln {{\rho }_{{DE}}}}{d\ln
a}-1=-\frac{1}{3}\left( 1+\frac{2}{c}\sqrt{{{\Omega }_{{DE}}}}
\right).
\end{equation}

We used the fact that $d\ln a=Hdt.$ This result can be obtained
without calculation of the pressure ${{p}_{\Lambda }}$ by the
relation \eqref{w(f(a))}. The obtained expression for $w_{_{DE}}$ is a
consequence of the holographic dark energy definition in form of
\eqref{rho_{DE}c}, therefore it is independent on other energy
components. It follows from the obtained result that $w_{_{DE}}\simeq
-1/3$ in the case of domination of other energy components and
$w_{_{DE}}=-\frac{1}{3}\left( 1+\frac{2}{c} \right)$ in the dark energy
dominated case. The latter result coincides with the expression
\eqref{w(f(a))}, obtained above for the Universe filled by solely
holographic dark energy.

For the first impression the declared task is completed. The
holographic dark energy with density \eqref{rho_{DE}c} from the one
hand provides correspondence between the observed density and the
theoretical estimate, and from the other it leads to the state
equation which is able to generate the accelerated expansion of
Universe. However the holographic dark energy with IR-cutoff on the
event horizon still leaves unsolved problems connected with the
causality principle: according to the definition of the event
horizon the holographic dark energy dynamics depends on future
evolution of the scale factor. Such dependence is hard to agree with
the causality principle.

In order to seek the way out from that dead end let us address once
more the cosmological constant problem, which is the huge gap
between the theoretical estimates and observed values of the dark
energy density. The simplest type of the dark energy --- the
cosmological constant --- is attribute to vacuum average of
quantized fields and can be measured in gravitational experiments.
Therefore the cosmological constant problem is the quantum gravity
one. Although the complete quantum gravity theory is absent, a
combination of quantum mechanics with general relativity can shed
light on the question under discussion.

From the first days of quantum mechanics the concept of
measurements, either real or gedanken, played fundamental role for
understanding of physical reality. General relativity states that
classical physical laws can be verified with arbitrary unlimited
precision. The above revealed connection between the macroscopic
(infrared) and microscopic scales dictates the necessity of deeper
analysis of the measurement process. The uncertainty relations
combined with general relativity generates a fundamental space-time
scale --- the planck length ${{L}_{Pl}}\sim {{10}^{-33}}\,cm.$

The existence of a fundamental length scale critically influence the
measurement process \cite{PCBGFL}. Assume that the fundamental
length scale is $L_f.$ As the space-time reference frame must make
sense, it must be linked to physical bodies. Therefore the
fundamental length postulate is equivalent to limitations for
possibilities to realize the precise reference frames. In terms of
light signal experiments this means, for instance, that the time
interval required for the light signal to propagate from A to b and
back, measured by clock in A, is subject to uncontrolled
fluctuations. The latter should be treated as evidence for the
fluctuations of metric, i.e. of the gravitational field. Thus the
fundamental length postulate is equivalent to gravity field
fluctuations.

The existence of quantum fluctuations in the metric
\cite{Sasakura,Maziashvili_1,Maziashvili_2,0707.4049} directly leads
to the following conclusion, related to the problem of distance
measurements in the Minkowski space: the distance $t$
\footnote{recall that we use the system of units where the light
speed equals $c=\hbar =1,$ so that
${{L}_{Pl}}={{t}_{Pl}}=M_{Pl}^{-1}$} cannot be measured with
precision exceeding \cite{Karolyhazy} the following
\begin{equation}
\label{Karolyhazy_rel} \delta t=\beta t_{Pl}^{2/3}{{t}^{1/3}},
\end{equation}
where $\beta $ is a factor of order of unity. Following
\cite{CohenKaplanNelson} we can consider the result
\eqref{Karolyhazy_rel} as the relation between the UV and IR scales
in frames of effective quantum field theory, which correctly
describes the entropy features of black holes. Indeed, presenting
the relation \eqref{LsimLambda{-2}} in terms of length and replacing
$\Lambda \to \delta t,$ we now recover \eqref{Karolyhazy_rel} in
holographic context.

The relation \eqref{Karolyhazy_rel} together with quantum mechanical
energy-time uncertainty relation enables us to estimate the quantum
fluctuations energy density in the Minkowski space-time. According
to \eqref{conditionQFT} we can consider the region of volume $t^3$ as composed of
cells $\delta {{t}^{3}}\sim t_{Pl}^{2}t.$ Therefore such a cell
represents the minimally detectable unit of space-time for the scale
$t.$ If the age of the chosen region equals $t,$ than it follows
from the energy-time uncertainty relations that its existence cannot
be realized with energy less than $\sim {{t}^{-1}}.$ Thus we come to
conclusion: if lifetime (age) of some spatial region of linear size
$t$ equals $t,$ then there exists a minimal cell $\delta {{t}^{3}},$
with energy that cannot be less than
 \begin{equation}
  \label{E_delta_t^3}
  E_{\delta t^3} \sim t^{-1}.
  \end{equation}
From \eqref{Karolyhazy_rel} and \eqref{E_delta_t^3} it immediately
follows that due to the energy-time uncertainty principle the energy
density of quantum fluctuating metric in the Minkowski space equals
\cite{Sasakura,Maziashvili_2,0707.4049}
\begin{equation}\label{rho_q}
{{\rho }_{q}}\sim \frac{{{E}_{\delta {{t}^{3}}}}}{\delta
{{t}^{3}}}\sim \frac{1}{t_{Pl}^{2}{{t}^{2}}}.
\end{equation}
A fact of principal importance is that the dynamical behavior of the
metric fluctuations density \eqref{rho_q} coincides with the above
defined holographic dark energy \eqref{LsimLambda{-2}},
\eqref{rho_{DE}c}, though the expression were derived based on
completely different physical principles. The holographic dark
energy density was obtained on the basis of entropy restrictions ---
the holographic principle, while the metric fluctuations energy
density of the Minkowski space is caused only by their quantum
nature, namely with the uncertainty principle.

The relation \eqref{rho_q} allows to introduce an alternative model
for holographic dark energy \cite{0707.4049}, which uses the age of
Universe $T$ for IR-cutoff scale. In such a model
\begin{equation}\label{rho_qn}
\rho_q=\frac{3{{n}^{2}}M_{Pl}^{2}}{{{T}^{2}}},
\end{equation}
where $n$ is a free parameter of model, and the number coefficient
$3$ is introduced for convenience. So defined energy density
\eqref{rho_qn} with $T\sim H_{0}^{-1},$ where ${{H}_{0}}$ is the
current value of the Hubble parameter, leads to the observed value
of the dark energy density with the coefficient $n$ value of order
of unity. Thus in SCM, where ${{H}_{0}}\simeq 72km\,{{\sec
}^{-1}}Mp{{c}^{-1}},\ {{\Omega }_{DE}}\simeq 0.73,\ T\simeq
13.7\,Gyr,$ one finds that $n\simeq 1.15.$

Now let us address the crucial question, whether the holographic
energy density in form \eqref{rho_qn} result in accelerated
expansion of Universe. For simplicity consider the Universe free of
other energy components. In such case the first Friedmann equation
reads
\begin{equation}\label{H2rho_qn}
H^2=\frac{1}{3M_{Pl}^{2}}{{\rho }_{q}}.
\end{equation}
The Universe age $T$ in \eqref{rho_qn} is linked with scale factor
by the relation
\begin{equation}\label{AGE_U}
T=\int_{0}^{a}{\frac{d{a}'}{H{a}'}}.
\end{equation}
Solution of the equation \eqref{H2rho_qn} with energy density
\eqref{rho_qn} reads
\begin{equation}\label{aAGE_U}
    a={{\left[ n\left( {{H}_{0}}t+\alpha  \right) \right]}^{n}}.
\end{equation}
The integration constant can be determined from the conditions
$a_0=1.$ Evaluating the second order derivative from the scale
factor, it can be shown that the accelerated expansion of Universe
takes place under the condition $n>1.$ We note that the above
obtained value $n\simeq 1.15,$ corresponding to the observed value
of the dark energy density, satisfies the above mentioned condition,
which can be obtained from the conservation equation for dark
energy, easily transformed to the following form
\begin{equation}\label{w_q}
{{w}_{q}}=-1-\frac{{\dot{\rho }}}{3H\rho }.
\end{equation}
Using \eqref{rho_qn}, \eqref{rho_qn} and \eqref{H2rho_qn} we present
the expression \eqref{w_q} in the form
\begin{equation}\label{w_qn}
{{w}_{q}}=-1+\frac{2}{3n}.
\end{equation}
As was repeatedly mentioned above, the accelerated expansion of
Universe requires $w<-1/3,$ which is equivalent to the above
obtained condition $n>1.$ Like in the previous model of the
holographic dark energy, we transit to more general case: the
Universe where dark energy coexists with matter with density ${{\rho
}_{m}}.$ Such Universe is described by the Friedmann equation
\begin{equation}\label{H2rho_qnpho_m}
{{H}^{2}}=\frac{1}{3M_{Pl}^{2}}\left( {{\rho }_{q}}+{{\rho }_{m}}
\right).
\end{equation}

Transforming to relative densities ${{\Omega }_{m}}=\frac{{{\rho
}_{m}}}{3{{H}^{2}}M_{pl}^{2}}$ and ${{\Omega }_{q}}=\frac{{{\rho
}_{q}}}{3{{H}^{2}}M_{Pl}^{2}}=\frac{{{n}^{2}}}{{{T}^{2}}{{H}^{2}}},$
we present the Friedmann equation \eqref{H2rho_qnpho_m} in the form
 \begin{equation}
 \label{Omega_q}
 \frac{d\Omega_q}{d\ln a}=
 (3-\frac{2}{n}\sqrt{\Omega_q})(1-\Omega_q)\Omega_q.
 \end{equation}
The equation \eqref{Omega_q} can be solved exactly to give
 \begin{eqnarray}
 \label{eq13}
 \frac{1}{n}\ln a + c_0  &=& -\frac{1}{3n-2}\ln (1-\sqrt{\Omega_q})
-\frac{1}{3n+2}\ln (1+\sqrt{\Omega_q}) \nonumber \\
   && + \frac{1}{3n} \ln \Omega_q +\frac{8}{3n(9n^2-4)}\ln
   (\frac{3n}{2}-\sqrt{\Omega_q}).
 \end{eqnarray}
The integration constant can be determined from the condition
${{\Omega}_{q}}\simeq 0.73$ at $a=1.$ Let us analyze the dynamics of
relative dark energy density in two limiting cases: the
matter-dominated and dark energy-dominated ones. In the former case
it follows from \eqref{Omega_q} that
\begin{equation}
{\Omega _q} \approx {c_1}{a^3}.
\end{equation}
Fast growth of relative contribution of the dark energy,
irrespective of $n$ value, results in the dark energy dominated era,
when
\begin{equation}
{{\Omega }_{q}}\approx 1-{{c}_{2}}{{a}^{-(3n-2)/n}}.
\end{equation}
The state equation for the dark energy can be obtained from the
relation \eqref{w_q}.
\begin{equation}\label{w_qOmega_q}
{{w}_{q}}=-1+\frac{2}{3n}\sqrt{{{\Omega}_{q}}}.
\end{equation}
In the early Universe during matter-dominated era one gets ${{\Omega
}_{q}}\to 0$ and ${{w}_{q}}\to -1,$ i.e. during that period the
holographic dark energy in the considered model behaves like
cosmological constant. In later epoch of dark energy domination,
when ${{\Omega }_{q}}\to 1,$ the state equation \eqref{w_qOmega_q}
naturally transforms to the above obtained relation \eqref{w_qn}.
Remark that fate of the Universe filled by matter and holographic
dark energy with density \eqref{rho_qn} is constant accelerated
expansion with power law time dependence \eqref{aAGE_U} of the scale
factor. Thus the holographic model for dark energy with IR-cutoff
scale set to the Universe age, allows the following:
\begin{enumerate}
  \item to obtain the observed value of the dark energy density;
  \item provide the accelerated expansion regime on later stages of
  the Universe evolution;
  \item resolve contradictions with the causality principle.
\end{enumerate}

However the reader should not hurry with the ultimate conclusions.
The first successes of holographic principle application from one
hand awoke hopes to create on that basis an adequate description of
the Universe dynamics, free of a number of problems which the
traditional approach suffers from. From the other hand, it was those
successes which provoke we would say unreasonable optimism. We
believe that papers entitled somewhat like ``Solution of dark energy
problem'' \cite{1004.1285} represent manifestation of specific
``holographic extremism''. We would like to remind the saying of
W.Churchill: ``Success means motion from failure to failure, not
loosing the enthusiasm''. The holographic dynamics is one of the
youngest directions of theoretical physics. On that path the
physicists experienced yet too few failures in order to pretend for
complete success.
\section{Transient acceleration}
Unlike the fundamental theories, the physical models only reflect
our current understanding of a process or a phenomenon, to describe
which they were created. The model efficiency significantly depends
on its flexibility, i.e. its ability to be modernized using the
new-coming information. That is why evolution of any actively living
model includes multiple generalizations, directed both to resolve
the conceptual problems and to describe ever growing set of
observational data. In the case of SCM those generalizations can be
divided on two types. The first includes replacements of
cosmological constant by more complicated dynamical forms of dark
energy, taking into account the possibility of interaction of the
latter with dark matter. Generalizations of the second type are of
more radical nature and pretend one consequent change of
cosmological paradigm. The ultimate (explicit or hidden) aim of them
is to reject the dark components due to modifications of Einstein
equations, and consequently the Newtonian laws two. The
generalizations of both types can be well viewed on example of the
phenomenon called the ``transient acceleration''.

As we have seen above in frames of SCM, the dependence of the
deceleration parameter $q$ on the redshift $z$ has the
characteristic monotonous trend to the limiting value $q(z)=-1$ for
$z\to -1.$ It physically means that after the beginning of dark
energy-dominated era (at $z\sim 1$), the Universe in SCM is
condemned to eternal acceleration.

Below we consider some cosmological models with dynamical forms of
dark energy, which lead to transient acceleration, and we also
discuss the information on current Universe expansion rate extracted
from observational data.
\subsection{Theoretical background}
J. Barrow \cite{Barrow} was one of the first to consider principal
possibility of the transient acceleration. He showed that many well
established scenarios, consistent with the current accelerated
expansion of Universe, do not exclude the possibility of recurrence
to non-relativistic matter-dominated era, and thus to the
decelerated expansion regime. Therefore the transition to the
accelerated expansion does not yet imply the eternal accelerated
expansion.

In order to show that we, following the Barrow work \cite{Barrow},
consider a homogeneous and isotropic flat Universe, filled by
non-relativistic matter and scalar field with the potential energy
$V(\varphi)$ and state equation $p=w\rho .$ We take the scalar field
potential in the form
\begin{equation}
V(\varphi) = V_p(\varphi) e^{-\lambda \varphi} \label{expVp}.
\end{equation}
In some versions of low-energy limits of string theory the potential
$V_p(\varphi)$ represents a polynomial. The exponential potential
with a shallow minimum was first suggested by Albrecht and Skordis
\cite{AS}. This minimum on the exponential potential background was
created due to the polynomial factor $V_p(\varphi)$ of the simplest
form
\begin{equation}
V_p(\varphi) = (\varphi -\varphi_0)^{2} + A\label{Vp}.
\end{equation}
In the considered case the potential takes the following form
\begin{equation}
V(\varphi )=e^{-\lambda \varphi }\left( A+(\varphi
-\varphi_0)^2\right). \label{V_Barrow}
\end{equation}
In order to relate the above considered potential to that of string
theory the constant parameters $A$ and $\varphi_0$ should be of
order of unity in Planck units. In such quintessence models the
accelerated expansion of Universe appears naturally on late stages
of evolution without any fitting of initial parameters, thus this
model is free of the precise tuning problem.

The accelerated expansion starts when the field rolls down the local
potential minimum at $\phi =\phi _0+(1\pm \sqrt{1-\lambda
^2A})/\lambda ,$ which is formed by the quadratic factor in
(\ref{V_Barrow}), where $1\geq\lambda ^2A.$ While the field stays in
the false vacuum state its kinetic energy is negligibly small $(\phi
\approx const ),$  and consequent domination $\rho_\phi $ is caused
by almost constant value of the potential energy, which runs the era
of the accelerated expansion of Universe which never ends.

It was discovered in the paper \cite{Barrow} that it is neither
solely possible nor most probable scenario.

The transient vacuum domination appears in the two following cases.
When $A\lambda ^2<1,$ the field $\varphi $ reaches the local minimum
with kinetic energy sufficient to overcome the potential barrier and
continues roll down the exponential part of the potential to the
region where $\varphi \gg\varphi _0.$ The kinetic energy is defined
by the scaling regime and thus by the parameters of potential rather
than by the initial conditions. The transition acceleration appears
also in the case when the condition $A\lambda ^2>1$ holds. Since $A$
grows proportional to $\lambda ^{-2},$ the potential looses its
local minimum and flattens near the inflection point. It is
sufficient to cause the temporal acceleration of the Universe
expansion, however the field never stops rolling down the potential
and once the Universe will be again dominated by matter with the
dependence $a(t)\propto t^{2/3}.$

Therefore the Universe quits the regime of eternal accelerated
expansion and returns to the decelerated expansion. Besides the the
well motivated family Albrecht-Skordis potentials the possibility of
transient acceleration regime appears to be more probable than the
eternal accelerated expansion.

\subsection{Different models with transient acceleration}
In order to show explicitly that the transient acceleration
represents a natural feature of different cosmological models, we
briefly consider some of them below.

Barrow considered in his work \cite{Barrow} a model of Universe
where dark energy is present in form of the scalar field with
potential \eqref{V_Barrow}. We consider some other examples of
cosmological models that provide alteration of the accelerated
expansion phase by the decelerated one.
\subsubsection{Scalar field, multidimensional cosmology and transient acceleration}
In the paper \cite{Russo} it was shown that the transition
acceleration phase can be realized in the exponentail potential as
well, and besides that a $d$-dimensional cosmological model was also
considered in the work. The action of the latter reads
\begin{equation}\label{Russo_act}
S=\int d^dx\sqrt{-g}\bigg(\frac{R}{ 2\kappa^2} + \frac{1}{2}
g^{\mu\nu} \partial_\mu \phi \partial_\nu\phi -V_0 \exp(-\lambda
\phi )\bigg),
\end{equation}
where  $\kappa_N^2=8\pi G_N=1/2$ and $V_0>0 ,$ $\lambda >0$  (the
case $\lambda <0 $ is connected to the case $\lambda >0$ by the
replacement $\phi\to -\phi $). Following the paper \cite{Russo} we
consider below the FRW-metric for flat Universe $(k=0)$
\begin{equation}
ds^2=dt^2-a^2(t ) dx^i dx^i \ ,\ \ \ \ i=1,...,d-1.
\end{equation}
In that case the action takes the form
\begin{equation}
S=\int d^{d}x  \bigg((d-1)(d-2) a^{d-3} \dot a^2+ a^{d-1} \big(
\frac{1}{2} \dot \phi ^2 -V(\phi ) \big)\bigg),
\end{equation}
and in variables $(u,v)$
\begin{equation}
\phi=\frac{1}{ \kappa}\sqrt{ d-2\over d-1 }(v-u)\ ,~~
a^{d-1}=e^{v+u}, \label{qqa}
\end{equation}
one obtains
\begin{eqnarray}
 S &=& \int d^{d-1}x dt \ e^{u+v}\bigg(  \frac{2 (d-2)}{\kappa^2 (d-1)}\dot
u \dot v -V_0e^{-2\alpha (v-u ) }\bigg),\nonumber
\\
\alpha &\equiv & \frac{1}{2\kappa }\sqrt{\frac{ d-2}{d-1 }} \lambda. \label{xxx}
\end{eqnarray}

Let us transform to new time variable $\tau $
\begin{equation}
\frac{d\tau}{dt}=\kappa  \sqrt{\frac{(d-1)V_0}{2(d-2)}} \ e^{\alpha
(u-v) }. \label{xxx1}
\end{equation}
and therefore
$$
S= \frac{1}{\kappa}\sqrt\frac{2(d-2)V_0}{d-1} \int d^{d-1}x d\tau \
e^{u+v} e^{\alpha (u-v) } \big( u'  v' -1 \big).
$$
Using \eqref{xxx}, \eqref{xxx1} and \eqref{qqa} it can be shown
\cite{Russo}, that the general solution for the case $\alpha<1$
takes the form:
\begin{eqnarray}
 d\mathfrak{s}^2 &=& \frac{2(d-2)}{\kappa^2(d-1)V_0}\ e^\frac{4\alpha ^2 \tau}{w}
\frac{(1+m e^{-2w\tau } )^\frac{2\alpha}{(1-\alpha)} }{ (1-m e^{-2w
\tau} )^\frac{2\alpha}{(1+\alpha)}}  d\tau ^2 -e^\frac{4 \tau}{sw}
(1+m e^{-2w\tau } )^\frac{2}{s(1-\alpha)}
 (1-m e^{-2w \tau} )^\frac{2}{s(1+\alpha)}dx^i dx^i ;\nonumber\\
\phi &=&\frac{1}{\kappa} \sqrt{\frac{d-2}{d-1}} \bigg( \frac{2\alpha
\tau }{w}- \frac{1}{1+\alpha}
 \log(1-m e^{-2w\tau } ) + \frac{1}{1-\alpha }\log(1+m e^{-2w \tau })\bigg),
\label{metid}
\end{eqnarray}
where $s=d-1,$ and $m$ is the integration
constant.

The solution (\ref{metid}) has the asymptotes:
\begin{eqnarray}
&&a \sim t^\frac{1}{d-1}\ ,~~~\phi=-\frac{1}{\kappa}\sqrt\frac{d-2}{
d-1}\log t\ ,\ \ \mbox{\rm for}\ \ t\cong 0\ ,\nonumber
\\
&&a \sim t^\frac{4\kappa^2}{(d-2)\lambda^2}\ ,\ \  \ \
\phi=\frac{2}{\lambda}\log t\ ,\ \ \ \ \ \mbox{\rm for}\ \ t\gg 1\ .
\label{dfg}
\end{eqnarray}
On early stages of evolution the state equation is extremely rigid
$p=\rho,$ and on later stages it is easy to show that
$$
p=\omega \rho , \ \ \ \
\omega=\frac{d-2}{2\kappa^2(d-1)}\lambda^2-1\ .
$$
In accordance with the with the equation (\ref{dfg}), the
accelerated expansion will continue forever under condition $\lambda
< \frac{2\kappa}{\sqrt{d-2} }.$ Let us show that under condition
$\frac{2\kappa}{\sqrt{d-2}} < \lambda < 2\kappa
\sqrt\frac{d-1}{d-2}$ the solution (\ref{metid}) with $m>0$ has a
transient acceleration stage. First let us find
$\frac{da}{dt}=\frac{da}{d\tau } \frac{d\tau}{dt}.$ Using
(\ref{metid}) it is easy to show that $\dot a$ is proportional to
positively defined quantities $m$ and $\tau .$ Setting $|m|=1,$
which can be easily done by shift along the $\tau$ axis, one finds
$\ddot a$ in the following form:
\begin{equation}
\ddot a =-({\rm positive}) \bigg( \big( (d-1)\alpha^2-1\big) Z^2
-2(d-2)\, {\rm sign}(m) \alpha Z+ d-1-\alpha^2\bigg), \label{asdx}
\end{equation}
where $Z\equiv \cosh(2w\tau ).$ If $(d-1)\alpha^2<1,$ which
corresponds to the case $\lambda< \frac{2\kappa}{\sqrt{d-2} } ,$
then we obtain the eternal accelerated expansion for arbitrary
values of $m,$ because only the first term in (\ref{asdx})dominates
on the later stages of evolution. Remark that such solution
represents an attractor. If $(d-1)\alpha^2>1$ holds for later times
then the solution always corresponds to to decelerated expansion of
Universe. For $(d-1)\alpha^2>1$ and $m<0$ the right hand side of the
equation (\ref{asdx}) is negatively defined, which corresponds to
decelerated expansion of Universe during whole evolution time. At
last, in the case $(d-1)\alpha^2>1$ and $m>0,$ the solution always
provides the transient acceleration phase, which is mentioned
in\cite{town}. Indeed, the equation (\ref{asdx}) has two roots
\begin{equation}
Z_\pm={(d-2)\alpha \pm \sqrt{d-1}(1-\alpha^2) \over
(d-1)\alpha^2-1}, \label{aqua}
\end{equation}
defined in the interval $\tau_-(\alpha)<\tau <\tau_+(\alpha ),$
corresponding to the accelerated expansion.

The root $\tau_{\pm }$ is real (and positive), because on the
interval $\alpha \in (1/\sqrt{d-1},1)$ we have $Z_{\pm }>1.$ In the
limit $\alpha\to 1$ the two roots coincide and duration of the
accelerated expansion phase tend to zero. In the opposite limit
$\alpha\to 1/\sqrt{d-1}$ one obtains $Z_-=d/(2\sqrt{d-1})$ and
$Z_+\to\infty ,$ which corresponds to infinite period of accelerated
expansion. For higher dimensions of space one obtains $Z_\pm
=1/\alpha \pm (1-\alpha^2)/(\alpha^2\sqrt{d})+O(1/d),$  and it means
that the transient acceleration duration is shorter in the Universe
with larger number of spatial dimensions.
\subsubsection{Transient acceleration in models with multiple scalar fields}

The existence of an event horizon in the case of a de Sitter phase is an obstacle to the implementation of string theory because the S-matrix formulation is no longer possible, hence eternal acceleration, leading to a de Sitter space in the asymptotic future is problematic \cite{HKS}.
The transient acceleration models allow to avoid that contradiction,
because they produce the accelerated expansion today and in recent
past, without any contradiction in future. In the paper
\cite{Polarski} the dynamics of homogeneous and isotropic Universe
was considered, where the dark energy is realized by two (generally
speaking, turn by turn) scalar fields $\Phi$ and $\psi.$ Let us
write down the equations of motion for that system
\begin{eqnarray}\nonumber
\dot{\rho_b} &=& -3 H \gamma_b \rho_b \\ \label{2Fields}
\ddot{\varphi} &=& -3H\dot{\varphi} - \partial_\varphi V \\
\nonumber \ddot{\psi} &=& -3H\dot{\psi} - \partial_\psi V, \nonumber
\end{eqnarray}
and also the first Friedmann equation:
\[
H^2 = \frac{8\pi G}{3}(\rho_b +\rho_Q)-\frac{k}{a^2}~.
\]
Here the dot denoted the derivative with respect to time $t,$ the
subscript $b$ marks the background components such as dark matter
($m$) or radiation ($r$), and $Q$ stands for dark energy in form of
the two scalar fields. It is convenient also to define
\[
p_{b,{}_Q}=(\gamma_{b,{}_Q}-1)\rho_{b,{}_Q}~,
\]
then $\gamma_m=1$ and $\gamma_r=\frac{4}{3},$  where, for any component $i,$ we have introduced for convenience the quantity  $\gamma_i\equiv 1 + w_i.$ The energy
density and pressure for the quintessence fields in the potential
$V(\varphi,\psi)$ have the form:
\[
\rho_Q =\frac{1}{2}\dot{\varphi}^2 + \frac{1}{2}\dot{\psi}^2 +
V(\varphi,\psi)~~ p_Q =\frac{1}{2}\dot{\varphi}^2 +
\frac{1}{2}\dot{\psi}^2 - V(\varphi,\psi).
\]
As always, the possibility of the transient acceleration crucially
depends on the potential form. Below we consider several
possibilities to obtain the transient acceleration. The paper
\cite{Polarski} considers a number of examples, including the case
when the two fields interact through the coupling potential
$V(\varphi,\psi),$ as well as the case when one of the fields is
free. Here we consider only the models with potentials depending on
both fields. Direct generalization of the potential from the paper
\eqref{V_Barrow} for the case of minimal coupling between the two
scalar fields enables us to obtain the transient acceleration.

Consider a potential of the form
\begin{equation}
    V(\varphi,\psi) = M^4 e^{-\lambda \varphi}(P_0+f(\psi)(\varphi-\varphi_c)^2+g(\psi)).
\label{Va}
\end{equation}
The above introduced additional scalar field $\psi$ will control
absence or presence o the potential minimum for $\varphi.$ The main
idea is that, from one hand, a potential minimum initially exists
for the scalar $\varphi,$ which provides accelerated expansion of
Universe, and from the other hand evolution of the field $\psi$
results in the extinction of the minimum and the Universe returns to
the decelerated expansion regime. The model of \cite{AS} is
recovered with $f \equiv 1$ and $g \equiv 0.$ For the potential
\eqref{Va} the position of the minimum is determined by the
expression
\begin{equation}
\varphi_{\pm}=\varphi_c+\frac{1}{\lambda}\left(1\pm\sqrt{1-\lambda^2\frac{P_0+g(\psi)}{f(\psi)}}\right).
\label{min}
\end{equation}
The function $g$ ($g > 0$) is responsible for the scalar field mass
and it usually takes the form $g \propto \psi^2,$ but as it does not
affect the system dynamics, we set for simplicity $g \equiv 0.$

The minimum \eqref{min} disappears under the condition
\begin{equation}
f(\psi) < \lambda^2 P_0 \equiv f(\psi_c). \label{cond}
\end{equation}
Note that the potential \eqref{Va} can be presented in the form
\[V(\varphi,\psi) = M^4/\lambda^2 e^{-\lambda \varphi}(f(\psi_c)+f(\psi)(\lambda\varphi-\lambda\varphi_c)^2).\]
We assume that $f$ is positive, continuous and monotonous function
for $\psi > 0$ and/or $\psi < 0.$ From the condition \eqref{cond} it
is easy to see that if $f$ decreases (increases), then for
$\psi<\psi_i$ ($\psi_i-$ is the initial value of the scalar field),
($\psi<\psi_c$) the accelerated expansion takes place, because the
minimum \eqref{min} exists. One is free to use different functions
$f$ to obtain the transient acceleration, for instance we present
the case $p,~\alpha \geq 0,~f = 1 + \alpha \psi^p,~f =
\tanh(\alpha\psi^p),$ or for arbitrary $p$ and $\alpha,~f =
\exp(\alpha\psi^p),~f = \cosh(\alpha\psi^p).$ During the evolution
of Universe the scalar field $\varphi$ rolls down the potential and
dominates in the exponential part so that $M_{Pl}^2 m^2_{\varphi} \sim
M_{Pl}^2 m^2_{\psi} \sim V \sim \dot{\varphi}^2 \sim M_{Pl}^2 H^2$ with
$\Omega_Q = 4/\lambda^2,$ $w_Q = 1/3$ during the radiation dominated
era, and with $\Omega_Q = 3/\lambda^2,$ $w_Q = 0$ for the
matter-dominated one, while $\varphi$ approach the minimum. When the
field $\varphi$ is in the minimum \eqref{min} and oscillates near
its minimum value, the universe expands with acceleration with $V
\gg \dot{\varphi}^2$ and therefore $w_Q \simeq -1$ until $\psi$
obeys $\psi \leq \psi_c,$ $(\psi\geq \psi_c),$ if the function $f$
grows (decays). In that moment the minimum \eqref{min} disappears
and the field $\varphi$ starts free roll acquiring large values
(speeding up) and the the Universe expansion becomes decelerated one
$q>0.$

When $\psi$ is initially larger (smaller) than $\psi_c$, provided $f$ is increasing (decreasing)
$\psi$ passes through $\psi_c$ because $\partial_\psi V = M^4e^{-\lambda \phi}(\phi-\phi_c)^2\frac{df}{d\psi}$ is positive (negative), acceleration occurs which is always transient; if on the contrary $\psi_i\leq \psi_c$ ($\psi_i\geq \psi_c$) quintessence domination is not possible. The critical value $\psi_c$ controls the presence or not of the minimum for $\phi$.

For a particular example let us consider the simplest case
\begin{equation}
f(\psi) = \psi^2~, \label{f}
\end{equation}
then the minimum for $\varphi$ disappears under condition $-\psi_c
\leq \psi \leq \psi_c,$ where $\psi_c \equiv \lambda\sqrt{P_0}.$ In
analogy with the model \eqref{V_Barrow}, one can obtain $\lambda
\gtrsim 9,$ which agrees with the limitations imposed by the
cosmological observations, provided the initial value $\varphi_i$ is
fixed and the critical value $\varphi_c$ must be finely tuned in
order to make the quintessence dominate in the present time.

\begin{figure}[h]
\begin{center}
\psfrag{z}[][][2.0]{$z$} \psfrag{w}[][][2.0]{$w_{_Q},q$}
\includegraphics[angle=-90,width=.75\textwidth]{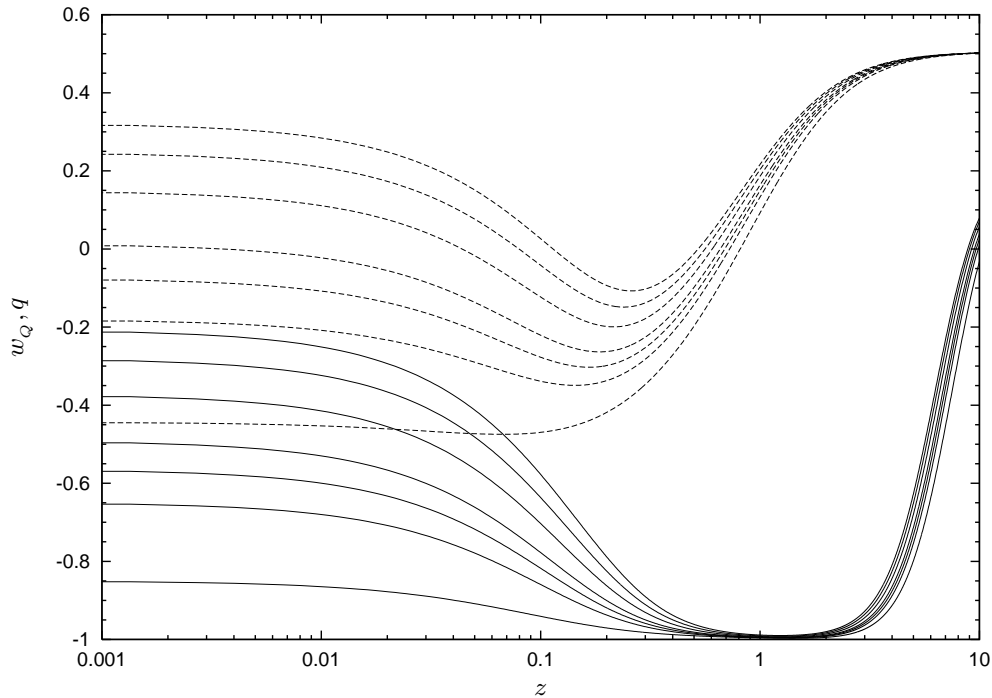}
\end{center}
\caption[]{The diagram plots the dependence of the state equation
parameter $w_{_Q}$ (solid line) and deceleration parameter $q$
(dashed line) as functions of the redshift $z$ for the potential
\eqref{Va} with the fixed parameter values $\lambda = 10,$ $\phi_c =
23.8$ and $P_0$ taking the following values values (upward from
below) $0.160, 0.162, 0.163, 0.164, 0.166, 0.168, 0.170$ $(\phi_i =
0$ and $\psi_i = 5).$  In the models with $p_0 = 0.164, 0.166,
0.168, 0.170$ the accelerated expansion is replaced by the
decelerated one in the modern epoch.} \label{eosAz}
\end{figure}

The diagram \ref{eosAz} shows the state equation parameter $w_{Q,0}
\simeq -0.491$ and $w_{eff,0} = -0.874.$ When the field $\varphi$
gets into the minimum, then $w_Q \simeq -1.$ The deceleration
parameter $q$ is also shown on the plot, and it points out on the
fact that in the present time the Universe already passed the
transient acceleration stage and currently it expands with
deceleration $q_0 \simeq 0.013.$ The accelerated expansion ($q < 0$)
starts at $z \simeq 0.658$ and stops at $z \simeq 0.0035,$ when the
Universe age equals $t_{end}/t_0 \simeq 0.996,$ while in the present
time one has $H_0t_0 \simeq 0.912.$

\subsubsection{Decaying dark energy as a scalar field}
As was already mentioned above, the cosmological models where the
dark energy decays have many attractive features, one of which being
the presence of transient acceleration --- alteration of the
accelerated Universe expansion period by the decelerated one, which
takes place when the dark energy density becomes sufficiently low.
Dynamics of such Universe clearly differs from its evolution
predicted by SCM. The considered model with the decaying dark energy
represents a pre-image of the commonly accepted inflation model,
where the field, generated the inflationary expansion of Universe,
experiences the decay.

\begin{figure}[ht]
\includegraphics[width=.7\textwidth]{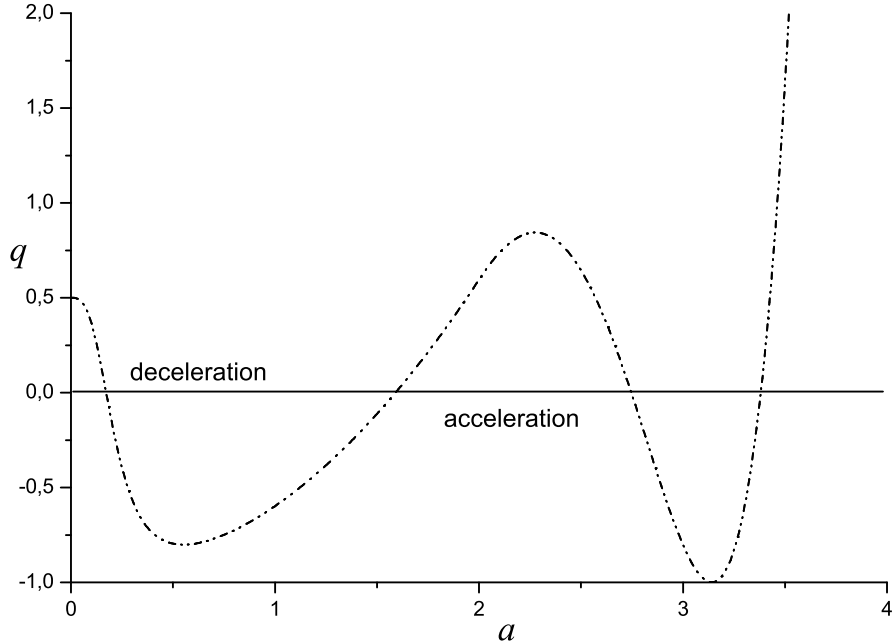}
\caption{Dependence of the deceleration parameter on the scale
factor for the Universe filled by matter and dark energy in form of
scalar field in the potential $V(\varphi ) = {V_0}\cos \left(
{\varphi m/\sqrt {{V_0}} } \right),$ with $m = 0.74;\,{V_0} =
150,\varphi (0) = 0.23,\varphi '(0) = 0.$ The present time
corresponds to $a = 1.$ The potential parameters are chosen so that
the deceleration parameter value appears to be the same as in SCM,
$q(1) \approx - 0.6.$} \label{cos}
\end{figure}

Consider a scalar field model with the potential that takes both
positive and negative values \cite{0302302}:
\begin{equation}\label{cos1}
V(\varphi ) = {V_0}\cos \frac{\varphi }{f},\,\,\,f = \frac{{\sqrt
{{V_0}}}}{m}.
\end{equation}
The assumption that the scalar field potential can take negative
values is very curious, but one should remember that except the
motivation to explain the observed transient acceleration this model
lacks any other observational support. Nevertheless such effective
potentials often appear in the super-gravity and M-theory. It is be
shown below how to obtain the transient acceleration in frames of
that model. Remark also that evolution of Universe in cosmological
models with negative potentials sharply differs from that of SCM.
So, for example, in the case when the Friedmann equation is
dominated by potential energy \eqref{cos1}, taking negative values
\begin{equation}\label{cos2}
{\left( {\frac{{\dot a}} {a}} \right)^2} = \frac{1} {3}\left( {\rho
+ \frac{1} {2}{{\dot \varphi }^2} + V(\varphi )} \right),
\end{equation}
even the spatially flat Universe can collapse, which is principally
impossible in SCM, but it can happen only on later stages of
Universe evolution, far before the transient acceleration ends. Up to
the moment when the Universe starts collapsing, many stages of
transient acceleration manage to occur (see Fig.\ref{cos}), which
become more often when approaching the collapse moment.
\subsubsection{Transient Acceleration In the Universe with the Interacting Components}
Consider the spatially flat Universe composed from three components
\cite{cosmolog}: dark energy, dark matter and baryons. The first friedmann
equation then takes the form:
\begin{equation}\label{GrindEQ__32_}
    3H^{2} =\rho _{_{DE}} +\rho _{m} +\rho _{b},
\end{equation}
where as usual $\rho _{_{DE}} $ is the dark energy density, $\rho
_{m} $ -- dark matter energy density, $\rho _{b} $ -- the density of
baryons, $H=\frac{\dot{a}}{a} $ -- Hubble parameter. The state
equation for the dark energy takes the form $p_{_{DE}} =w\rho
_{_{DE}} .$

The conservation equations in that case take the form:
\begin{equation}\label{GrindEQ__33_}
\begin{gathered}
{\dot{\rho }_{_{DE}} +3H(1+w)\rho _{_{DE}} =-Q;}\hfill \\
 {\dot{\rho }_{m} +3H\rho _{m} =Q,}
\end{gathered}
\end{equation}
where $Q$ characterizes the interaction. The conservation equation
for the baryons reads:
\begin{equation}\label{GrindEQ__34_}
\dot{\rho }_{b} +3H\rho _{b} =0\, \, \, \Rightarrow \, \, \, \rho
_{b} =\rho _{b0} \left(\frac{a_{0} }{a} \right)^{3}.
\end{equation}
Total density equals $\rho =\rho _{m} +\rho _{b} +\rho _{_{DE}} .$
Without lack of generality, we assume that the dark matter energy
density satisfies
\begin{equation}\label{GrindEQ__35_}
    \rho _{m} =\tilde{\rho }_{m0} \left(\frac{a_{0} }{a} \right)^{3} f\left(a\right),
\end{equation}
where $\tilde{\rho }_{m0} $ and $a_{0}$ are constants and $f(a)$ is
arbitrary time dependent function. From \eqref{GrindEQ__33_} and
\eqref{GrindEQ__35_} one obtains
\begin{equation}\label{GrindEQ__36_}
    Q=\rho _{m} \frac{\dot{f}}{f} =\tilde{\rho }_{m0} \left(\frac{a_{0} }{a} \right)^{3} \dot{f}.
\end{equation}
Let $f(a)$ takes the form
\begin{equation}\label{GrindEQ__37_}
    f(a)=1+g(a).
\end{equation}
Interaction free case corresponds to $f(a)=1,$ thus the function
$g(a)$ responds for the interaction. Taking into account that
\begin{equation}\label{GrindEQ__38_}
    \dot{f}=\dot{g}=\frac{dg}{da} \dot{a},
\end{equation}
one obtains
\begin{equation}\label{GrindEQ__39_}
    Q=\tilde{\rho }_{m0} \frac{dg}{da} \dot{a}\left(\frac{a_{0} }{a} \right)^{3} .
\end{equation}
For $\rho _{m} $ it means that
\begin{equation}\label{GrindEQ__40_}
    \rho _{m} =\tilde{\rho }_{m0} \left(1+g\right)\left(\frac{a_{0} }{a} \right)^{3},
\end{equation}
where $\rho _{m0} =\rho _{m} (a_{0} )$ in presence of interaction,
and analogously $\tilde{\rho }_{m0} $ in uncoupled case. The two
initial values of dark matter density are linked by the following
relation
\begin{equation}\label{GrindEQ__41_}
    \rho _{m0} =\tilde{\rho }_{m0} \left(1+g_{0} \right),
\end{equation}
where $g_{0} \equiv g(a_{0} ).$ As can be seen from
\eqref{GrindEQ__36_}, in the case when $Q$ is positive the dark
energy decays into dark matter $\frac{dg}{da}
>0$ and vice versa $\frac{dg}{da} <0$ otherwise. It follows from the equations \eqref{GrindEQ__33_}
and \eqref{GrindEQ__39_}
\begin{equation}\label{GrindEQ__42_}
    \dot{\rho }_{_{DE}} +3H\left(1+w\right)\rho _{_{DE}} =-\tilde{\rho }_{m0} \frac{dg}{da} \dot{a}\left(\frac{a_{0} }{a} \right)^{3} .
\end{equation}
Assuming that $w=const$ the solution of equation
\eqref{GrindEQ__42_} takes the form
\begin{equation}\label{GrindEQ__43_}
\begin{gathered}
\rho _{_{DE}} =\left(\rho_{_{DE0}} +\tilde{\rho }_{m0} g_{0}
\right)\left(\frac{a_{0} }{a} \right)^{3\left(1+w\right)} -\\
-\tilde{\rho }_{m0} \left(\frac{a_{0} }{a} \right)^{3}
g+3w\tilde{\rho }_{m0} a_{0}^{3} a^{-3\left(1+w\right)} \int _{a_{0}
}^{a}daga^{3w-1} .
\end{gathered}
\end{equation}
Rewrite the second Friedmann equation in the terms of $g(a)$
\begin{equation}
\label{GrindEQ__44_}
\begin{array}{l} {\frac{\ddot{a}}{a} =-\frac{1}{6} \left\{\tilde{\rho }_{m0}
\left(1+g\right)\left(\frac{a_{0} }{a} \right)^{3} +\rho _{b0} \left(\frac{a_{0} }{a} \right)^{3}
+\left(1+3w\right)\right. \times } \\ {\times \left[\left(\rho_{_{DE0}} +\tilde{\rho }_{m0}
g_{0} \right)\left(\frac{a_{0} }{a} \right)^{3\left(1+w\right)} \right.
\left. \left. -\tilde{\rho }_{m0} \left(\frac{a_{0} }{a} \right)^{3}
g+3w\tilde{\rho }_{m0} a_{0}^{3} a^{-3\left(1+w\right)} \int _{a_{0} }^{a}daga^{3w-1}  \right]\right\}.} \end{array}
\end{equation}
In order to solve the equation \eqref{GrindEQ__44_} one have to
explicitly define the function $g(a).$ For the lack of knowledge
about nature of the dark energy, as well as the dark matter, the
function $g(a)$ cannot be derived from the first principles.
Therefore for the model under consideration we introduce the
interaction such that the resulting dynamics in the model
corresponds to the observed data.

\begin{figure}[ht]
\includegraphics[width=.45\textwidth]{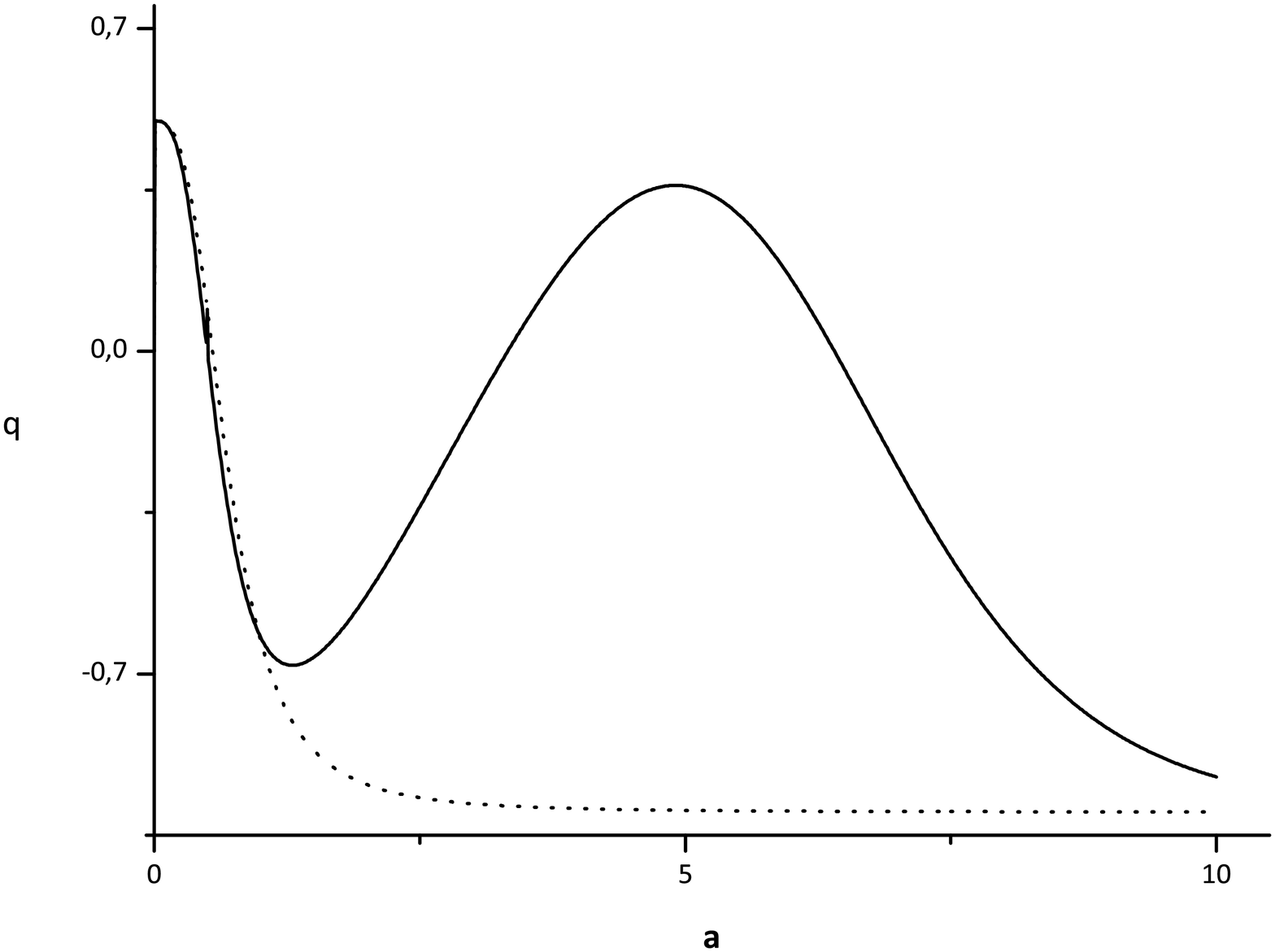}
\includegraphics[width=.45\textwidth]{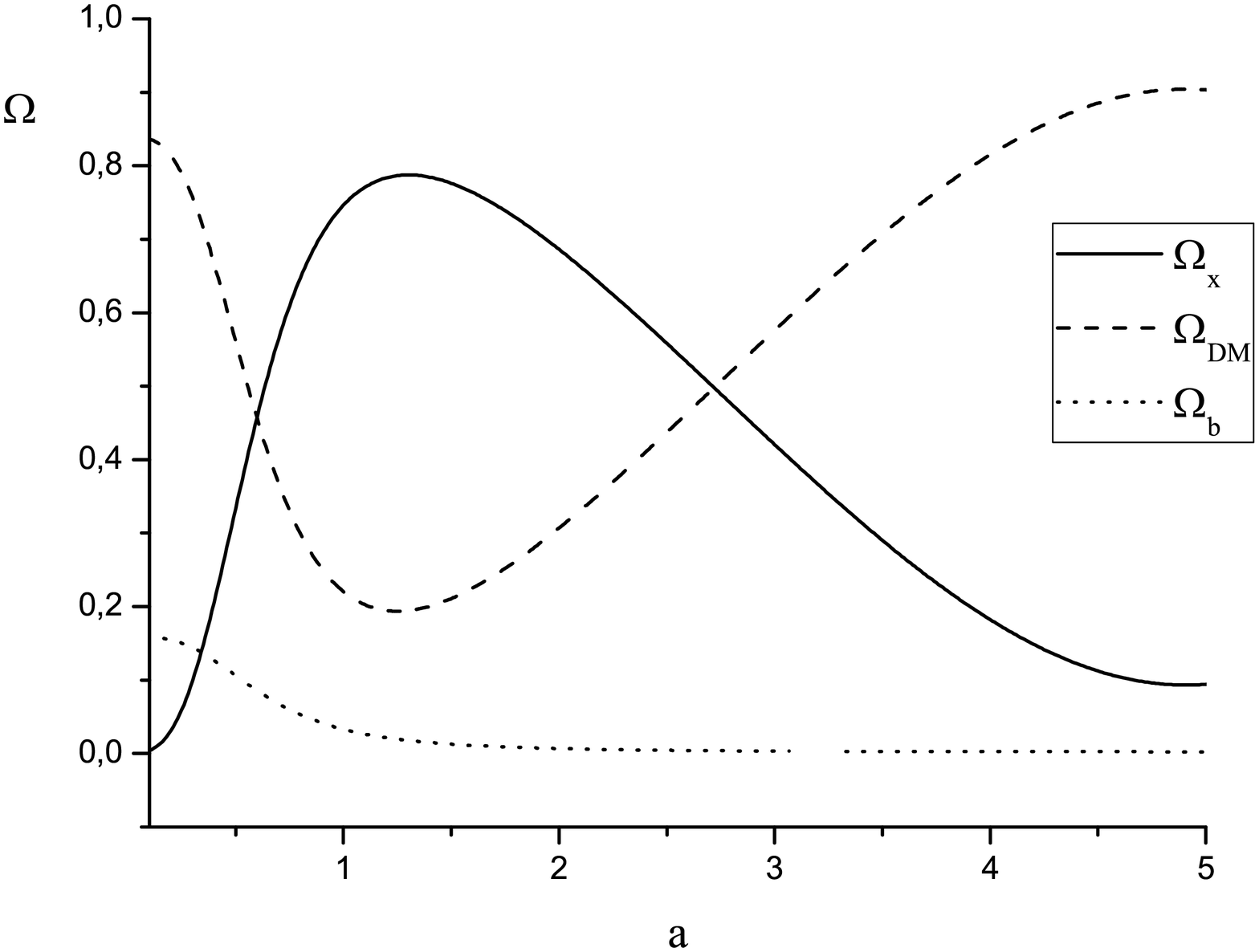}
\caption{The dependencies of relative densities on the scale factor
for $n=7$ and $\sigma =1.5$ (on the left). On the right: the
dependencies of deceleration parameter on the scale factor in the
model with interacting dark energy and dark matter $q(a)$(solid
line) with $n=7$ and $\sigma =1.5$ in comparison to SCM (dashed
line).} \label{fig:W(a)}
\end{figure}

Consider interaction with the function $g(a)$ in the form
$g\left(a\right)=a^{n} \exp \left(-a^{2} /\sigma ^{2} \right),$
where $n$ is natural number, and $\sigma $ is positive real number.
The transient acceleration implies that the dark energy density
starts to diminish, i.e. experiences decay $\frac{dg}{da} >0.$ This
condition requires that $n$ and $\sigma $ satisfy the inequality
$n\sigma ^{2} >2.$ The dependencies of relative densities on the
scale factor for $n=7$ and $\sigma =1.5$ are shown on Fig.
\ref{fig:W(a)}.
The considered model which allows to produce the transient
acceleration with certain choice of the interaction parameters,
however for large values of scale factor as well for small ones it
is indistinguishable from SCM.
\subsubsection{Decaying cosmological constant and transient acceleration phase}
As a simple example of the transient acceleration we consider a
model \cite{cosmolog} with decaying cosmological constant:
\begin{equation} \label{ec}
\dot{\rho}_{dm} + 3\frac{\dot{a}}{a}\rho_{dm} = -
\dot{\rho}_\Lambda\;,
\end{equation}
where $\rho_{dm}$ and $\rho_\Lambda$ are energy densities of dark
matter and cosmological constant $\Lambda$ respectively. On early
stages of Universe expansion, when $\rho_\Lambda$ is sufficiently
small, such decay has almost no influence on the cosmic evolution.
On later stages, the more the dark energy contribution grows, the
more its decay affects the standard dependence for the dark matter
energy density $\rho_{dm} \propto a^{-3}$ on the scale factor $a.$
Let us assume that such deviation can be described by a scale factor
function $\epsilon(a),$ then
\begin{equation} \label{dm}
\rho_{dm} = \rho_{dm, 0}a^{-3 + \epsilon(a)}\;,
\end{equation}
where in the present epoch $a_0 = 1.$ Other material fields
(radiation and baryons) evolve independently and conserve. Therefore
the dark energy density takes the form:
\begin{equation}\label{decayv}
\rho_{\Lambda} =  \rho_{dm,0} \int_{a}^{1}{\epsilon(\tilde{a}) +
\tilde{a}\epsilon' \ln(\tilde{a}) \over \tilde{a}^{4 - \epsilon(a)}}
d\tilde{a} + {\rm{X}},
\end{equation}
where prime denotes the derivative with respect to the scale factor,
and ${\rm{X}}$ -- is the integration constant. Neglecting the
contribution of radiation, the first Friedmann equation takes the
form
\begin{equation}
\label{friedmann} {{H}}= H_0\left[\Omega_{b,0}{a}^{-3} +
\Omega_{dm,0}\varphi(a) + {\Omega}_{{\rm{X,0}}}\right]^{1/2}.
\end{equation}
The function $\varphi(a)$ reads
\begin{equation} \label{f(a)}
\varphi(a) = a^{-3 +\epsilon(a)} + \int_{a}^{1}{\epsilon(\tilde{a})
+ \tilde{a}\epsilon' \ln(\tilde{a}) \over \tilde{a}^{4 -
\epsilon(a)}} d\tilde{a},
\end{equation}
where ${\Omega}_{{\rm{X,0}}}$ denotes relative contribution of the
constant ${\rm{X}}$ into the total relative density. In order to
proceed further, some assumptions on the explicit form of the
function $\epsilon(a)$ should be done. In the present review we
follow the original paper \cite{cosmolog} on $\Lambda$-dark matter
interaction and consider the simplest case
\begin{eqnarray}
\label{Parametrization_a}
\epsilon(a)& = & \epsilon_0a^\xi;\quad  \nonumber \\
& = &\epsilon_0(1+z)^{-\xi},
\end{eqnarray}
where $\epsilon_0$ and $\xi$ can take both positive and negative
values. From the above cited expression (\ref{decayv}) it follows
that
\begin{equation}\label{decayv2}
\rho_{\Lambda} =  \rho_{m0}\epsilon_{0} \int_{a}^{1}{[1 +
\ln(\tilde{a}^{\xi})] \over \tilde{a}^{4 - \xi -
\epsilon_{0}\tilde{a}^{\xi}}} d\tilde{a} + {\rm{X}}.
\end{equation}
Remark that the case $\epsilon_0 = 0$ corresponds to SCM, i.e.
${\rm{X}} \equiv {\rho}_{\Lambda 0}.$

Using the above cited formulae it is easy to obtain the dependence
of the relative densities $\Omega_b(a),$ $\Omega_{dm}(a)$ and
$\Omega_{\Lambda}(a)$:
\begin{subequations}
\begin{equation} \label{8a}
\Omega_{b}(a) = \frac{a^{-3}}{{\rm{A}} + a^{-3} +
{\rm{B^{-1}}}\varphi(a)};
\end{equation}
\begin{equation} \label{8b}
\Omega_{dm}(a) = \frac{a^{-3 + \epsilon(a)}}{{\rm{D}} +
{\rm{B}}a^{-3} + \varphi(a)};
\end{equation}
\begin{equation} \label{8c}
\Omega_{{\rm{\Lambda}}}(a) = \frac{{\rm{D}} + \varphi(a) - a^{-3 +
\epsilon(a)}}{{\rm{D}} + {\rm{B}}a^{-3}  + \varphi(a)};
\end{equation}
\end{subequations}
 where ${\rm{A}} = {\Omega_{{\rm{X}},0}}/{\Omega_{b,0}},$ ${\rm{B}} = {\Omega_{b,0}}/{\Omega_{dm,0}}$
and ${\rm{D}} = {\Omega_{{\rm{X}},0}}/{\Omega_{dm,0}}.$

\begin{figure}[t]
\centerline{\psfig{figure=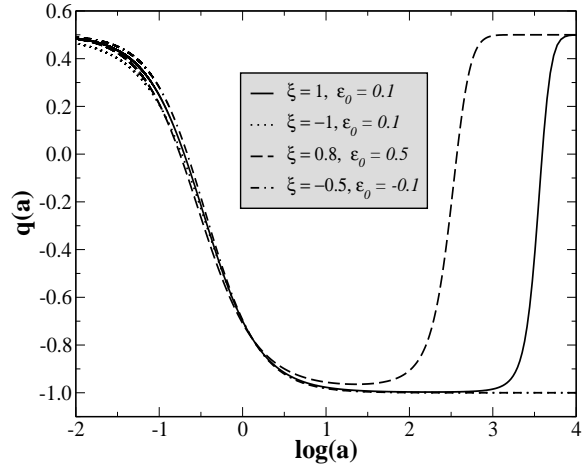,width=2.5truein,height=3.1truein,angle=-90}}
\caption{The deceleration parameter as function of $\log(a)$ for
some selected values of $\epsilon_0$ and $\xi.$ } \label{fig:qzw}
\end{figure}

In such simple model, appropriate choice of the parameters
$\epsilon_0$ and $\xi$ enables to obtain practically any kind of
Universe dynamics. In the context of the present review it is
especially interesting to consider the case when $\epsilon_0 > 0$
and $\xi$ takes large positive values ($\xi \gtrsim 0.8$). The
figure \ref{fig:qzw} shows the dependence of the scale factor for
the case $\xi = 1.0$ and $\epsilon_0 = 0.1.$ Note that with such
parameters in the present time $a\sim 1$ the Universe expands with
acceleration, but unlike the case of SCM, the dark energy domination
will not last forever and at $a\gg 1$ the Universe enters into new
era of non-relativistic matter domination with $a \to \infty.$ Such
type of dynamical behavior is impossible for most models with
$\Lambda(t)$ or those with interacting quintessence, widely
discussed in literature, but it is an attribute of the so-called
thawing \cite{thaw} and hybrid \cite{cqg} potentials, which follow
from string or M-theory \cite{fischler} (see also
\cite{ed})\footnote{Authors of the paper \cite{fischler} present
arguments for the fact that forever expanding Universe, which is
common feature of most quintessence models (including the standard
$\Lambda$CDM-model), contradicts the predictions of
string/M-theories, because those models contain cosmological
horizon, which make impossible to apply the usual procedure of
S-matrix formalism, describing the interaction of the particles.}.

In order to give clearer idea of transient acceleration phenomenon
we find the explicit form of the deceleration parameter
$q=-a\ddot{a}/\dot{a}^2,$ in the considered model
\begin{equation}
q(a) = \frac{3}{2}\frac{\Omega_{b,0}a^{-3} +
\Omega_{dm,0}a^{\epsilon(a) - 3}}{\Omega_{b,0}{a}^{-3} +
\Omega_{dm,0}\varphi(a) + {\Omega}_{\rm{X},0}} -1.
\end{equation}
Its dependence on $\log(a)$ for selected values of $\xi$ and
$\epsilon_0$ is shown on Fig.\ref{fig:qzw}. Note that large values
of the parameter $\xi$ corresponds to the Universe dominated by
matter in the past ($q(a) \rightarrow 1/2$ for $a \gg  1$),
afterwards (at $a_{acc} < 1$) a long-during era of accelerated
expansion takes place, which, unlike SCM, does not last forever,
ending at some $a_{dec} > 1.$
\subsection{Transient acceleration: holographic limitations}
The novel concept of the holographic Universe considered in the
previous section turns out to impose strict limitations to Universe
dynamics. In order to see that we consider the matter entropy
contained within the observed horizon in the case when the matter is
composed of non-relativistic substance. Each particle of the
substance carries one unit of information, and therefore $S_m =K_B
N,$ where $N=(4\pi/3)r^3_An,$ and $n$ is the particle concentration,
for which $n=n_0a^{-3}.$ Thus one obtains for entropy the following
\begin{equation}
S_{m} =  k_{B}\,  \frac{4 \pi}{3}\, \tilde{r}_{A}^{3} \, n_{0}\,
a^{-3} \propto a^{3/2} \, . \label{eq:Sm}
\end{equation}
and it follows that $S''_{m} \propto \frac{3}{4\, \sqrt{a}}
> 0.$ It is interesting to note that as the Universe expands the particle
number contained inside the horizon increases too, which provides
the entropy growth and therefore $S''_{m} \, + \, S''_{A} > 0 .$ It
is easy to see that the Universe tends to maximum entropy provided
its horizon decreases with time. In the simplest case the horizon
equals to Hubble radius and the above condition is satisfied if the
Universe expands with acceleration. In such case the deceleration
parameter is negative ($q<0$), thus
$$
\frac{d}{dt}(R_{H} ) = c(1 + q) < c,
$$
and Hubble sphere has velocity which is less then light speed by the
quantity $cq,$ and thus it is left behind those galaxies. Therefore
the particles initially placed inside the Hubble sphere, gradually
leave it.

From the latter fact that a conclusion follows that in order to obey
the second law of thermodynamics, the universe must be dominated by
dark energy with state equation parameter satisfying the condition $
-1 \leq w \leq- 2/3 .$ Another alternative is to modify the
gravitation theory in order to obtain the accelerated dynamics
without the dark energy.

Likewise, since every isolated system is expected to evolve in a
such a way that it tends to thermodynamic equilibrium the curvature
of the ${\cal A}(a)$ function should never spontaneously change from
negative to positive values. In order to specify the set of possible
cosmological values, or equivalently to impose certain limitations
on the values taken by the state equation parameter $w,$ it is
necessary to use the laws which are not immediately connected to
cosmology. In the context of holographic cosmology it is reasonable
to use the generalized law of the black hole thermodynamics
--- the generalized square law (GSL). According to the latter the total entropy of the system $S$
must obey the condition $S'' < 0.$ As will be shown below, (see
\cite{1012.0474v1}) application of this criterion on later stages of
Universe evolution helps considerable specify the range of values
allowable for the state equation parameter $w.$

\begin{figure}[htb]
\centering
\includegraphics[width=10cm]{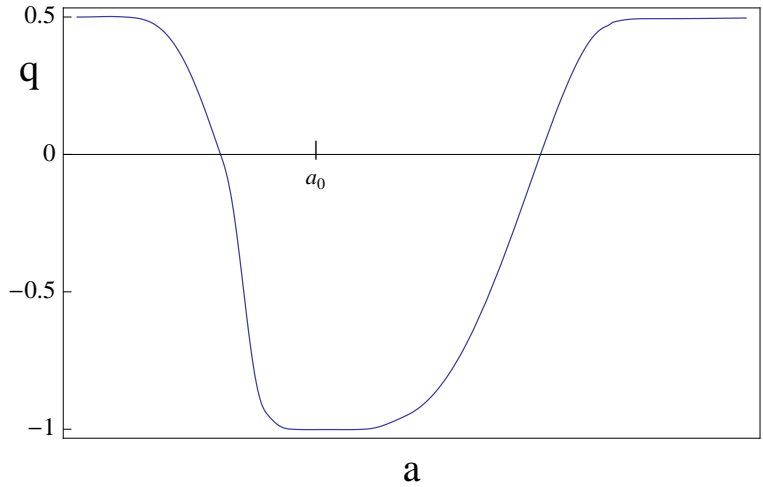}
\caption{The deceleration parameter dependence in some models with
transient acceleration. } \label{fig:q(t)}
\end{figure}

For further convenience we express the horizon area $\cal A$ and its
derivatives through the deceleration parameter $q \equiv
-\ddot{a}/(a \, H^{2}),$ which can be presented in the above
mentioned form:
\begin{equation}
q = -\left(1 \, + \, \frac{a \, H'}{H}\right) \, . \label{decp}
\end{equation}
For the case of spatially flat Universe ${\cal A} = 4 \pi H^{-2},$
therefore ${\cal A}' = 8 \pi (1+q)/H^{2} $ and
\begin{equation}
 {\cal A}'' = 2 {\cal A}\, \left[ (1\, + \, q) \, (1\, + \, 2q) \, + \, \frac{q'}{a}\right] \, .
\label{2primeA2}
\end{equation}
It follows that for the case $q \geq -1/2$ the second order
derivative of the horizon area, ${\cal A}'',$ cannot change from
positive to negative values while $q' < 0.$ It excludes the
dependence for $q$ shown on Fig.\ref{fig:q(t)}. Such dependence of
the deceleration parameter appears in the cosmological models where
the current accelerated expansion is transient, and after a short
dark matter dominated period the dark matter starts dominating
again, which leads to the decelerated Universe expansion era.
\cite{jcap-julio2010,bose-majumdar}.
\subsection{Transient acceleration in models with holographic dark energy}
Current literature usually considers the models where the required
dynamics of Universe is provided by one or another, and always only
one, type of dark energy. As was multiply mentioned above, in order
to explain the observed dynamics of Universe, the action for
gravitational field is commonly complemented, besides the
conventional matter fields (both matter and baryon), by either the
 cosmological constant, which plays role of physical vacuum in SCM,
 or more complicated dynamical objects --- scalar fields,
$K$-essence and so on. In the context of holographic cosmology, the
latter term is usually neglected, restricting to contribution of the
boundary terms. Nevertheless such restriction has no theoretical
motivation.

In the present subsection we consider the cosmological model which
contains both volume and surface terms\cite{JCAPphase,CosmicAccel}. The role of former is played
by homogeneous scalar field in exponential potential, which
interacts with dark matter. The boundary term responds to
holographic dark energy in form of \eqref{Omega_q}. Such model is
shown to produce the transient acceleration phase --- the evolution
stage when the accelerated expansion of Universe changes to the
decelerated one, and afterwards the Universe comes to the stage of
eternal accelerated expansion, thus removing the limitations imposed
in the previous subsection.

To describe the dynamical properties of the Universe it is
convenient to to transform to dimensionless variables as the
following:
\begin{eqnarray}
x=\frac{\dot{\varphi}}{\sqrt{6}M_{Pl}H},\quad
y=\frac{1}{M_{Pl}H}\sqrt{\frac{V(\varphi)}{3}},\quad
z=\frac{1}{M_{Pl}H}\sqrt{\frac{\rho_{m}}{3}},\quad
u=\frac{1}{M_{Pl}H}\sqrt{\frac{\rho_{_{q}}}{3}}.\label{var2}
\end{eqnarray}
The evolution of scalar field is described by the Klein-Gordon
equation, which in the case of interaction between the scalar field
and matter takes the following form:
\begin{equation}
\ddot{\varphi}+3 H \dot{\varphi} + \frac{dV}{d\varphi}=-
\frac{Q}{\dot{\varphi}}. \label{eq:kgeqn}
\end{equation}
In the present section we consider the case when the interaction
parameter $Q$ is a linear combination of energy density for scalar
field and dark energy
\begin{equation}
  \label{Q1}
  Q= 3H(\alpha\rho_\varphi+\beta \rho_m),
\end{equation}
where $\alpha,$ $\beta$ are constant parameter. For given model,
regardless the explicit form of the scalar field potential
$V(\varphi),$ the system of dynamical equations takes the following
form
\begin{eqnarray}
x'&=& \frac{3x}{2}g(x,z,u)-3x + \sqrt{\frac{3}{2}}\lambda y^2-\gamma,\nonumber\hfill\\
 \label{sys_xyz}
y'&=&  \frac{3y}{2}g(x,z,u)- \sqrt{\frac{3}{2}}\lambda xy, \hfill\\
z'&=& \frac{3z}{2}g(x,z,u)-\frac{3}{2}z  + \gamma\frac{x}{z},\nonumber\hfill\\
u'&=& \frac{3u}{2}g(x,z,u)-\frac{u^2}{n},\nonumber\hfill
\end{eqnarray}
where
\begin{equation}
g(x,z,u) = 2x^2+ z^2+ \frac{2}{3n}u^3,~\lambda\equiv
-\frac{1}{V}\frac{dV}{d\varphi} M_{Pl}.
\end{equation}
and
\begin{eqnarray}
  \nonumber
    Q&=&9H^3M_{Pl}^2\left[\alpha(x^2+y^2)+\beta z^3\right];\\
\gamma& = &\frac{\alpha(x^2+y^2)+\beta z^3}{x} \label{Q_gamma_xyz}.
\end{eqnarray}
As was mentioned above, here we consider the simplest case of
exponential potential
\begin{equation}\label{V_exp}
    V=V_0\exp\left(\sqrt{\frac{2}{3}}\frac{\mu\varphi}{M_{Pl}}\right),
\end{equation}
where $\mu$ is constant. Taking into account the expression
\eqref{Q_gamma_xyz}, the system of equations \eqref{sys_xyz} reads
\begin{eqnarray}
x'&=& \frac{3x}{2}\left[g(x,z,u)- \frac{\alpha(x^2+y^2)+\beta z^2}{x^2}\right]-3x - \mu y^2,\nonumber\hfill\\
 \label{sys_xyz_V}
y'&=&  \frac{3y}{2}g(x,z,u)+\mu xy, \hfill\\
z'&=& \frac{3z}{2}\left[g(x,z,u)+\frac{\alpha(x^2+y^2)+\beta z^2}{z^2}\right]-\frac{3}{2}z ,\nonumber\hfill\\
u'&=& \frac{3u}{2}g(x,z,u)-\frac{u^2}{n}.\nonumber\hfill
\end{eqnarray}
In the considered model the deceleration parameter takes the form
\begin{equation}
q=-1+\frac 32\left[2x^2+z^2+\frac{2}{3n}u^3\right].
\end{equation}
Note that the cosmological parameters have no explicit dependence on
the interaction form, and only determine the evolution dynamical
variables. This fact essentially complicates analysis of such
system. Let us make few remarks on the system of equations
(\ref{sys_xyz_V}). First consider the case $y = 0,$ which
corresponds to free scalar field. It easy to obtain some restricting
relations on the interaction parameters, which follow from the
requirement for energy density to be real
\begin{equation}
\label{cond}
    2\sqrt{\frac{\beta}{\alpha}}>1+\alpha+\beta,
\end{equation}
which in turn requires $0>\beta>\alpha,|\alpha|+|\beta|<1.$ This
unstable critical point corresponds to Universe filled by dark
matter. The case of multiple such points is also possible but
irrelevant. To conclude we note that any of such critical points is
easily shown to exist also in the interval $x_0<0.$ For the case
$z\neq 0$ with the restrictions imposed by (\ref{cond}), one obtains
\begin{equation}
    x_c=\left[\left(a+\sqrt{{\beta}/{\alpha}}\right)^{1/2}+a\right]z_c,
\end{equation}
where
$a=\left(2\sqrt{{\beta}/{\alpha}}-(1+\alpha+\beta)\right)^{1/4}.$
\subsubsection{Case $Q=0$}
In the present section we consider in more details the
interaction-free case with scalar field and dark matter. The
critical points of the system (\ref{sys_xyz_V}) in the case
$(\alpha=\beta=0)$ are given in the table \ref{tab:2}. The phase
space generated by the system of equations (\ref{sys_xyz_V})
contains six physically relevant critical points, the last of which
is an attractor. The first critical point $(1,0,0,0)$ is unstable
and corresponds to Universe dominated by scalar field with extremely
rigid state equation $(w_\varphi = 1),$ the second critical point is
also unstable and responds to the evolution period when the scalar
field is dynamically equivalent to cosmological constant.

The next third point $(0,0,1,0)$ is of no interest because it
corresponds to Universe composed solely of dark matter which is also
unstable. The fourth critical point $(0,0,0,1)$ corresponds to
Universe solely composed of holographic dark energy in form of
\eqref{Omega_q}, and was considered above in details. The most
physical interest is presented by the last sixth critical point
which is an attractor. It corresponds to the Universe filled by
scalar field and holographic dark energy. This critical point is
completely determined by the scalar field potential parameter $\mu$
and magnitude of $n:$
\begin{equation}
    \begin{array}{cccccl}
   x_{*}&=&\frac{2}{3n\mu}u_{*}, \quad& y_{*}&=&\sqrt{1-\left(1+\frac{4}{9n^2\mu^2}\right)u_{*}^2},\\
    z_{*}&=&0, \quad                        & u_{*}&=&\frac{3}{2n\mu^2}\left(-1+\sqrt{1+\frac{4n^2\mu^4}{9}}\right).
\end{array}
\end{equation}
The obtained property $x_*\propto u_*$ is typical for the so-called
tracking solutions \cite{Tracking_Solutions}. Remark also there is
also so-called background interaction between the scalar field and
dark energy, caused by the fact that the scalar field dynamics is
affected by the holographic dark energy which has negative pressure
and influences on the Universe expansion rate by means of Hubble
parameter, which enters into the Klein-Gordon equation for the
scalar field.
\begin{table}
\caption{Critical points for the autonomous system of equations
(\ref{sys_xyz_V})} \label{tab:2}
\begin{tabular}{ccccc}
\hline\noalign{\smallskip}
$(x_c,y_c,z_c,u_c)$ & Stability & $q$ & $w_{\varphi}$& $w_{tot}$\\
 coordinates &  character  & & & \\
\noalign{\smallskip}\hline\noalign{\smallskip}
$\;(1,0,0,0)$ &  unstable  & $2$ &$ 1$& $1$\\
$\;(0,1,0,0)$   & unstable  & $-1$ &$-1$& $-1$\\
 $\;(0,0,1,0)$ &unstable &$\frac{1}{2}$\quad &$\nexists$&0\\
 $\;(0,0,0,1)$ & stable  & $-1+\frac{1}{n}$ &$ \nexists$& $-1+\frac{2}{3n}$\\
 $(-\frac{3}{2\mu},\frac{3}{2\mu},\sqrt{1-\frac{3}{2\mu^2}},0)$ & unstable &\quad$\frac{1}{2}$ &$\nexists$&0\\
 $(x_*,y_*,0,u_*)$ & attractor &$q_*<0$&$w_{\varphi*}$&$w_{tot*}$\\
\noalign{\smallskip}\hline
\end{tabular}
\end{table}
In the attractor point the dark matter density turns to zero. To fit
that model with the observations one has to strictly define the
initial conditions such that the accelerated expansion of Universe
starts before then the considered evolution phase begins.
\subsubsection{Review of the case $Q=3H\alpha\rho_\varphi$}
\begin{figure}[t]
\centering
\includegraphics[width=0.45\textwidth]{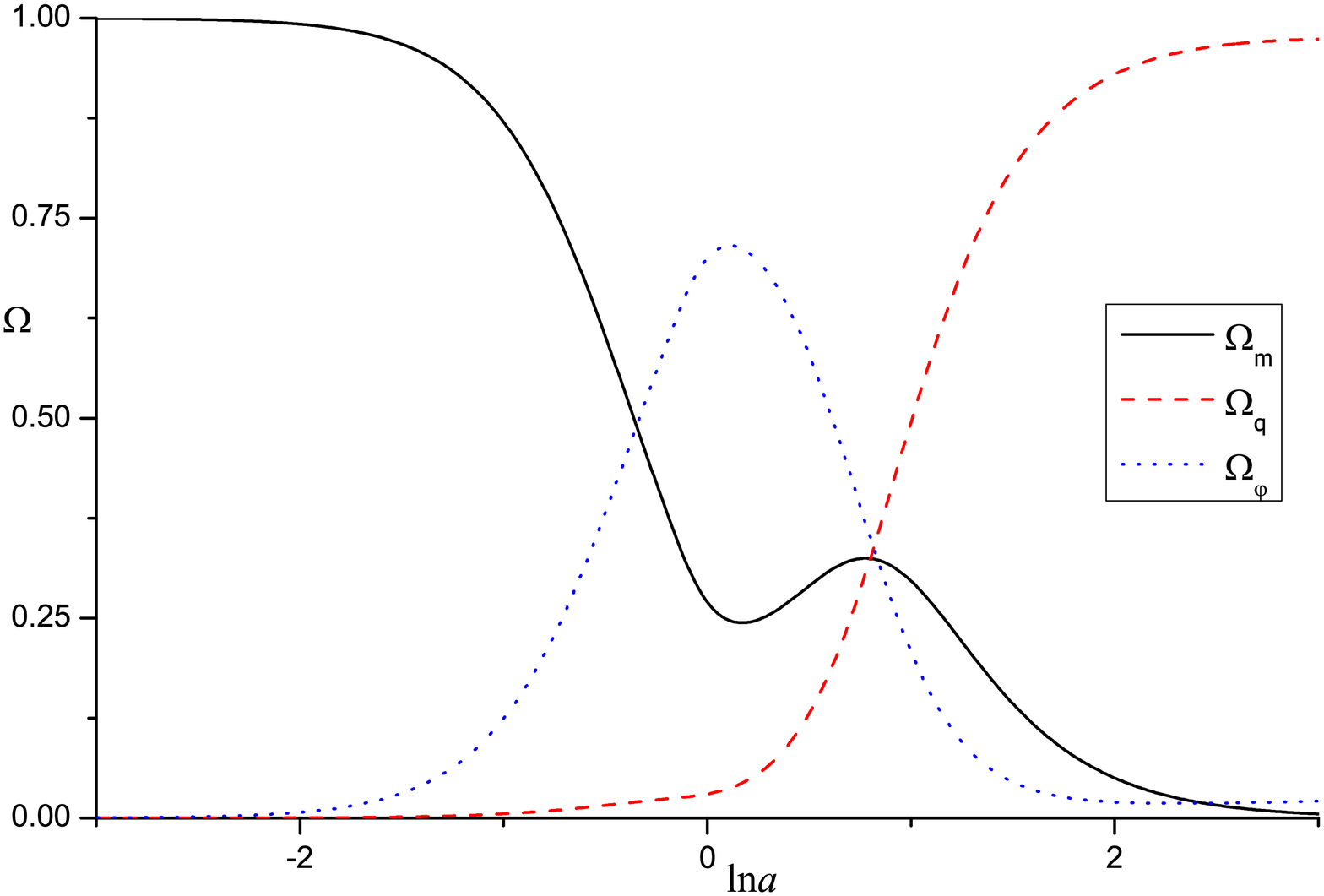}
\includegraphics[width=0.45\textwidth]{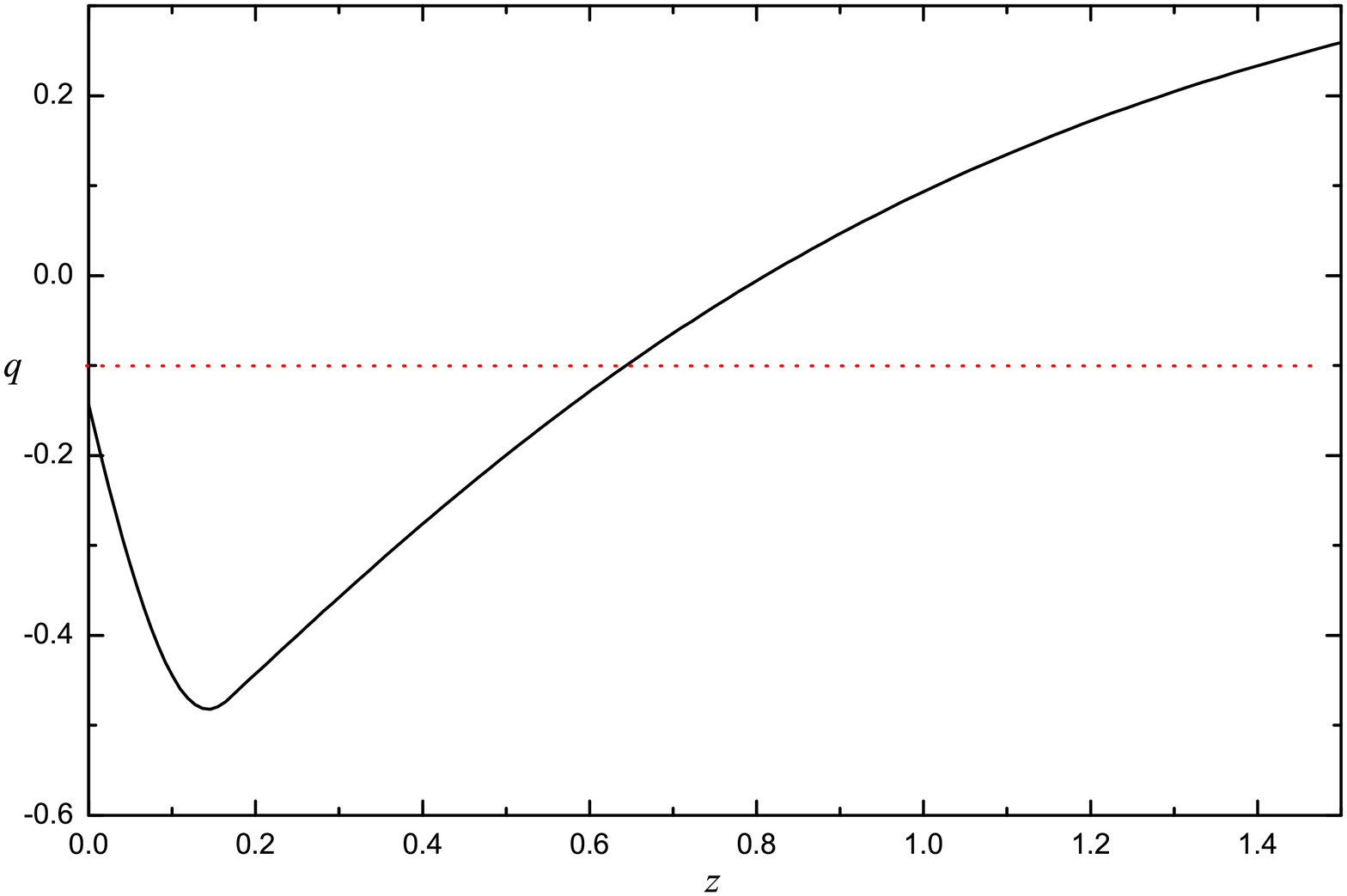}
\caption{Behavior of $\Omega_{\varphi}$ (dot line), $\Omega_{q}$
(dash  line) and $\Omega_{m}$ (solid line) as a function of  $N=\ln
a$ for $n=3,\;\alpha = 0.005$ and $\mu = -5$ (left side). Evolution
of deceleration parameter for this model (right side).}
\label{fig:4}
\end{figure}

In the above case, the phenomenon of transient acceleration that
occurs in such a Universe does not match the observations. In order
to fix that problem we consider a model where scalar field interacts
with dark matter. In the present section we consider the case with
the interaction parameter of the form (\ref{Q1}) with $ \beta = 0.$
The Figure \ref{fig:4} shows the dependencies $\Omega_q, \Omega_{m}$
and $\Omega_{\varphi}$ for the case when $\alpha=0.005,$ $\mu=-5$
and $n = 3.$
From the explicit form of the equations and interaction character it
is easy to see that neither type nor position of the above obtained
critical points change when the interaction disappears. The latter
affects only behavior of the dynamical variables, which corresponds
to different trajectories in the phase space between the critical
points. This corresponds to the fact that the interaction parameters
enter only in the Hubble parameter derivatives of second order and
higher with respect to time.

With such values of interaction parameters the transient
acceleration begins almost in the current era. Like in the
conventional cosmological models with dark energy in form of scalar
field, the latter starts to dominate and causes the accelerated
universe expansion phase. Along with the Universe expansion the
contribution of $\Omega_q$ increases, which leads to the fact that
the background (space) starts increasing faster than the field and
becomes asymptotically free. Such field is known to have the
so-called super-strict state equation and forces the Universe to
decrease its expansion rate. Soon however, when the contribution
$\Omega_q$ increases enough that the scalar field cannot any more
prevent the expansion, Universe starts to speed up again.
\subsection{Observational evidence}
\begin{figure}[t]
\centering
\psfig{figure=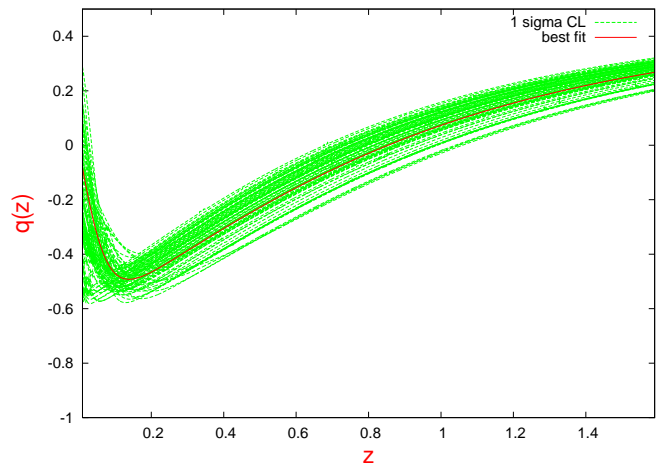,width=0.35\textwidth,angle=-90}
\caption{ The deceleration parameter dependence $q(z)$ reconstructed
from independent observational data, including the brightness curves
for SNe Ia, cosmic microwave background temperature anisotropy and
baryon acoustic oscillations (BAO), SN and parametrization
(\ref{eq:step}). The red solid line shows the best fit on the
confidence level $1\sigma$ CL \cite{starobinsky}.}
\label{fig_Srarob}
\end{figure}

Starobinsky \cite{starobinsky} with co-authors, based on independent
observational data, including the brightness curves for SNe Ia,
cosmic microwave background temperature anisotropy and baryon
acoustic oscillations (BAO), were able to show (see
Fig.\ref{fig_Srarob}), that the acceleration of Universe expansion
reached its maximum value and now decreases. In terms of the
deceleration parameter it means that the latter reached its minimum
value and started to increase. Thus the main result of the analysis
is the following: SCM is not unique though the simplest explanation
of the observational data, and the accelerated expansion of Universe
presently dominated by dark energy is just a transient phenomenon.
The term ''transient acceleration'' is increasingly often used with
respect to the Universe dynamics.

Remark, that the paper \cite{starobinsky} also showed that with the
use of CPL parametrization
 \begin{equation}
\label{cplwz} w(z)=w_0+\frac{w_a \, z}{1+z}.
\end{equation}
for the state equation parameter, it is impossible avoid
contradictions in attempt to combine the data obtained from
observations of near supernovae of type SN1a with those of the
relict radiation anisotropy. One of the way out of that
contradiction is to reject the above mentioned parametrization and
invent a new one. Starobinsky group suggested a novel
parametrization which can combine those data sets:
 \begin{equation}
    w(z)=- \frac{1+ \tanh\left[(z-z_t)\Delta\right]}{2}.
\label{eq:step}
\end{equation}
Within such approximation $w = -1$ on early times of Universe
evolution and increases up to its maximum value $w\sim 0$ for small
values of $z.$

Figure \ref{fig_Srarob} shows the dependence of the deceleration
parameter $q$ reconstructed by the parametrization (\ref{eq:step}).

In the year 2010 in frames of Supernova Cosmology Project (SCP), the
most recent data set on supernovae outbursts was published
\cite{amanullah}. It contains 557 events and is largest up to date.
Besides that the data set concerning supernovae with moderate red
shift $(z<0.3)$ was remarkably extended. Today there are several
works \cite{LiWuYu,LiWuYu1}, making analysis of the above mentioned
data in order to test the transient acceleration hypothesis. All the
authors share common opinion that the ultimate answer will be given
only by remade more accurate measurements. Moreover, obtain a
convincing result one will probably have to remake all the data
treatment procedure. Thus it was shown in \cite{LiWuYu,LiWuYu1} that
there is a contradiction between the data obtained from (SNe Ia+BAO)
for small redshifts and those from (CMB) for high ones. The
contradiction is in the fact that the analysis of two data sets
separately one obtains opposite results. For example, using only
(SNe Ia+BAO) data, one obtains evidence in support of the fact that
the Universe expansion rate reached the maximum value at $(z\sim
0.3)$ and presently starts to decrease. At the same time, if the
data are complemented by the CMB data set, the results of analysis
drastically change and no deviations from $\Lambda$CDM can be
revealed.
\begin{figure}[t]
\centering
\includegraphics[width=0.45\textwidth]{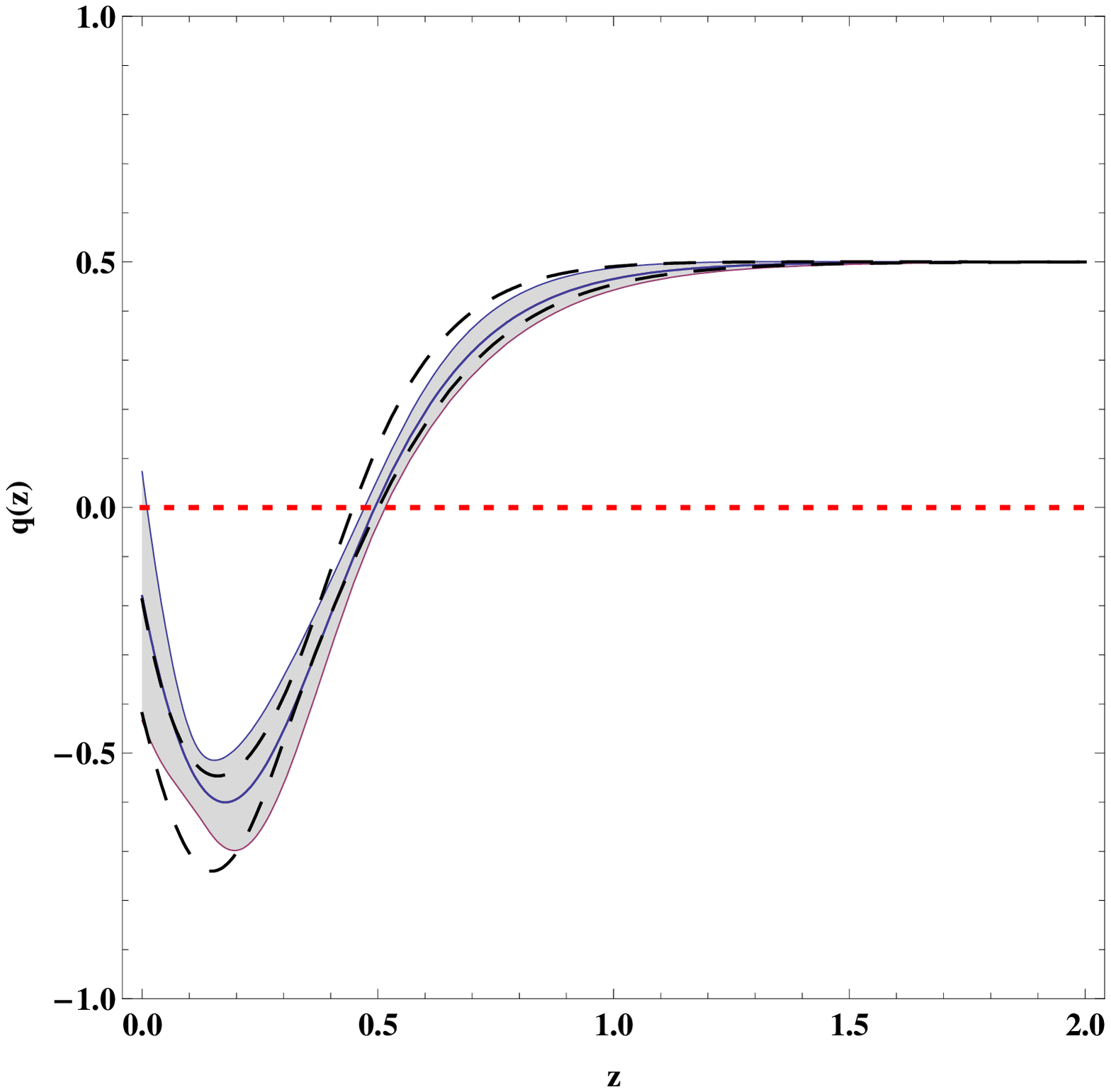}
\includegraphics[width=0.45\textwidth]{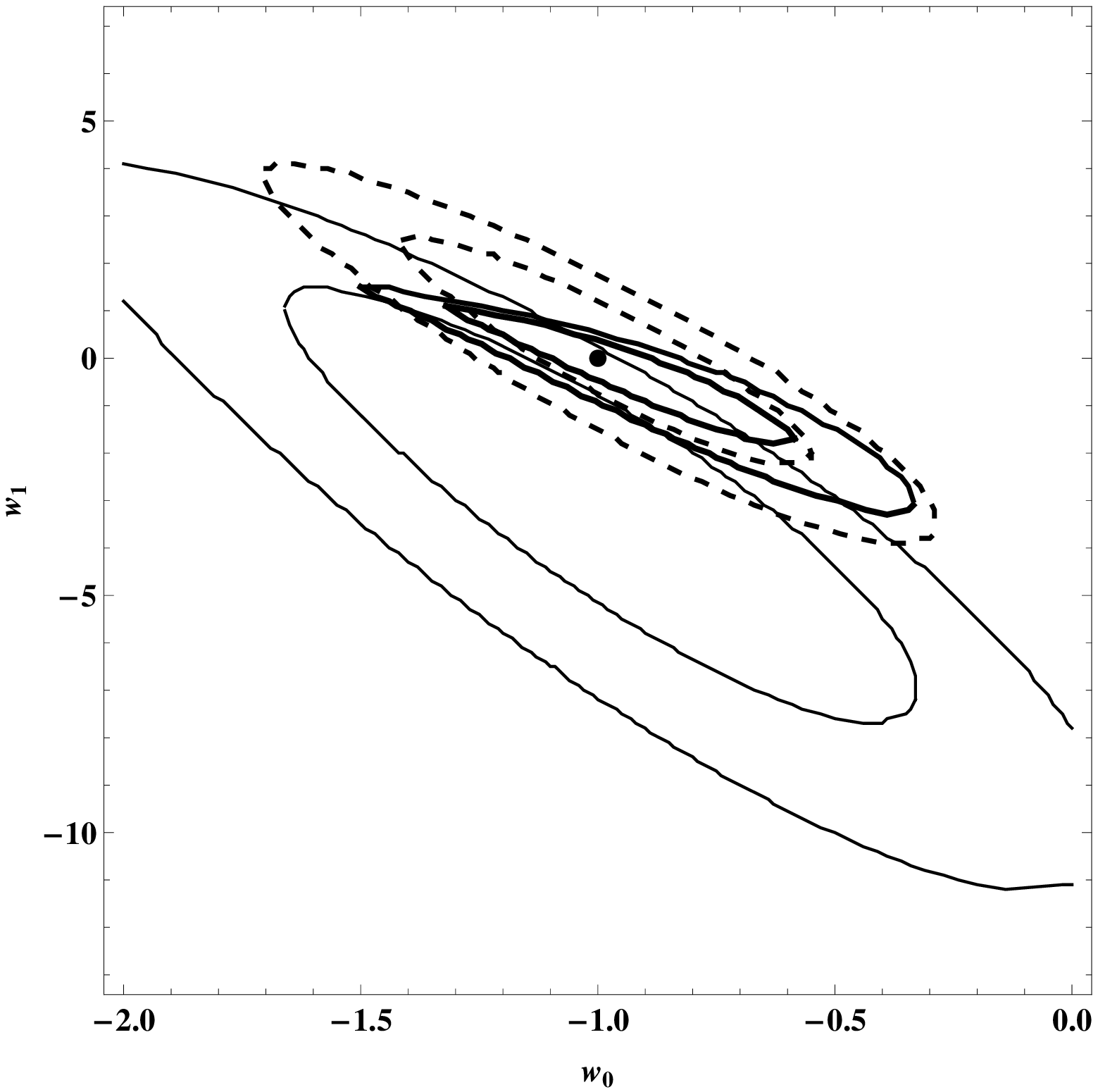}
\caption{On the left: the deceleration parameter dependence $q(z)$
reconstructed from results of analysis of Union2+BAO data sets with
confidence level $2\sigma.$ The grey region and the one between the
two dashed lines corresponds to presence and absence of systematic
errors in the SNe Ia observations. On the right: the confident regions
for $68.3\%$ and $95\%$ levels are shown for $w_0$ and $w_1$ in the
parametrization CPL with $w=w_0+w_1z/(1+z).$ Dashed, solid, thin
solid lines corresponds to the results of Union2S, Union2S+BAO and
Union2S+BAO+CMB respectively. The point $w_0=-1,$ $w_1=0$
corresponds to spatially flat $\Lambda$CDM model(SCM). }
\label{fig:TrAcCh}
\end{figure}
Therefore the result of reconstruction the evolutionary dependence
of dark energy and answer to the question whether our Universe will
expand with deceleration or accelerated expansion will continue
forever (like in SCM), strongly depends on data obtained from SNe Ia
observation, its quality, the particular method of reconstruction
for the cosmological parameters such as $q(z), w(z)$ and
$\Omega_{DE}$), as well as on explicit form of the state equation
parametrization. For detailed answer to that question we have to
wait more precise observational data and seek less model-dependent
ways of their analysis.
\section*{Acknowledgements}

We are grateful   V.A. Cherkaskiy for  qualitative translation  of this review. Without his extraordinary help this review could not be appear. Work is supported in part by the Joint DFFD-RFBR Grant \# F40.2/040.

\end{document}